\documentclass[a4paper, 12pt]{article}

\usepackage{graphicx}

\usepackage[T1]{fontenc}
\usepackage[utf8]{inputenc}
\usepackage[english]{babel}
\usepackage{geometry}                		
\usepackage{graphicx}				

\graphicspath{ {./images/} }

\usepackage[normalem]{ulem}			
\usepackage{pgfplots}	       

\usepackage{amssymb}
\usepackage{amsmath}
\usepackage{amsthm}
\usepackage{mathtools}
\usepackage{dsfont}
\newcommand{\indic}{\mathds{1}}     
\usepackage[pagebackref,bookmarks=false]{hyperref}				
\makeatletter
\renewcommand*{\backref}[1]{}
\renewcommand*{\backrefalt}[4]{%
  \ifcase #1 %
  \or [#2]%
  \else [#2]%
  \fi
}
\makeatother

\usepackage[poorman,noabbrev,nosort]{cleveref}
\crefname{assumption}{assumption}{assumptions}
\Crefname{assumption}{Assumption}{Assumptions}
\crefrangelabelformat{assumption}{#3#1#4--#5#2#6}
\crefname{plemma}{lemma}{lemmas}
\Crefname{plemma}{Lemma}{Lemmas}
\crefname{passumption}{assumption}{assumptions}
\Crefname{passumption}{Assumption}{Assumptions}
\crefrangelabelformat{passumption}{#3#1#4--#5#2#6}

\usepackage[dvipsnames]{xcolor} 
\hypersetup{colorlinks, citecolor=blue, filecolor=blue, linkcolor=blue, urlcolor=blue}

\usepackage{cancel}					
\usepackage{booktabs}				
\usepackage{tabularx}				
\usepackage{rotating}				
\usepackage{threeparttable}         
\usepackage{threeparttablex}
\usepackage{longtable}
\usepackage{array}

\usepackage[title]{appendix}			

\usepackage{csvsimple}				

\usepackage{subfiles}				

\usepackage{pdfpages}

\usepackage{fancyhdr}

\usepackage{indentfirst}

\usepackage{quoting}								
\quotingsetup{font=small}

\usepackage{tikz} 
\usepackage{pgfplots}
\pgfplotsset{compat=1.5}

\usepackage{enumerate}

\usepackage{caption}
\usepackage{newfloat}

\newtheorem{theorem}{Theorem}[]

\newtheorem{lemma}{Lemma}[]
\newtheorem{remark}{Remark}[]

\makeatletter
\newcommand{\primechars}[1]{%
  \ifcase#1\relax
  \or '\or ''\or '''\or ''''\else $^{(#1)}$\fi
}
\makeatother

\newtheorem{plemma}{Lemma}[lemma]

\newtheorem{proposition}{Proposition}[]

\theoremstyle{definition}
\newtheorem{example}{Example}[]

\usepackage{xparse} 

\NewDocumentEnvironment{examplecont}{m o}{%
  \par\medskip
  \noindent\textbf{\Cref{#1}---continued%
  \IfValueT{#2}{\ (#2)}.}\ %
  \ignorespaces
}{%
  \hfill$\triangle$
  \par\medskip
}

\newtheorem{assumption}{Assumption}

\makeatletter

\newcommand{\assumption@prefix}{}

\renewcommand{\theassumption}{%
  \assumption@prefix\arabic{assumption}%
}

\newcommand{\assumptionprefix}[1]{%
  \expandafter\xdef
    \csname assumption@saved@\assumption@prefix\endcsname
    {\arabic{assumption}}%
  \edef\assumption@prefix{#1}%
  \@ifundefined{assumption@saved@\assumption@prefix}{%
    \setcounter{assumption}{0}%
  }{%
    \setcounter{assumption}{%
      \csname assumption@saved@\assumption@prefix\endcsname
    }%
  }%
}

\renewcommand{\theHassumption}{%
  \assumption@prefix.\arabic{assumption}%
}

\makeatother

\newtheorem{passumption}{Assumption}[assumption]

\newtheorem{algorithm}{Algorithm}[]

\makeatletter
    \let\oldalgorithm\algorithm
    \let\endoldalgorithm\endalgorithm
    \renewenvironment{algorithm}[1][]
      {\oldalgorithm[#1]}
      {\hfill$\triangle$\endoldalgorithm}
    \let\oldexample\example
    \let\endoldexample\endexample
    \renewenvironment{example}
      {\oldexample}
      {\hfill$\triangle$\endoldexample}
\makeatother

\makeatletter
\renewenvironment{proof}[1][\proofname]%
  {%
    \par
    \pushQED{\qed}%
    \normalfont
    \topsep=\medskipamount
    \trivlist
    \item[\hskip\labelsep{\bfseries\itshape #1\@addpunct{.}}]%
    \ignorespaces
  }%
  {%
    \popQED
    \endtrivlist
    \@endpefalse
  }
\makeatother

\usepackage{natbib}

\title{\textsc{Debiasing Functionals of Conditional Distributions}
\thanks{Escanciano
acknowledges research support by Spanish grant PID2025-170712NB-I00
(MCI/AEI/FEDER, UE). Gazzei acknowledges financial support through a fellowship from Fundaci\'on Ram\'on Areces.}
}
\vspace{12pt}
\author{Juan Carlos Escanciano\\\textit{Universidad Carlos III de Madrid}
\and Gabriele Gazzei\\\textit{Universidad Carlos III de Madrid}}
\date{\today}

\begin{document}

\maketitle

\vspace{-12pt}

\begin{abstract}
Many economic parameters depend on estimated conditional distributions. We construct locally robust moments for these parameters by representing their first-step derivative and centering the representer conditionally on the covariates. We develop complementary conditional-measure and CDF linearizations, and derive corrections for control functions, counterfactual functionals, conditional rank-rank regression, and transformations of conditional $L$-functionals. Under the CDF linearization, a Riesz-regression criterion estimates the correction automatically; its analysis distinguishes representer estimation from integration against the estimated conditional law. We establish asymptotic oracle equivalence, asymptotic normality, and consistent variance estimation for cross-fitted estimators under continuity and remainder conditions, without requiring functional weak convergence of the first-step estimators. An exact linear representation yields double robustness; for rank-rank regression this property is componentwise. Simulations report reductions in bias and RMSE relative to plug-in estimation, with double cross-fitting improving the finite-sample performance.
\end{abstract}

\newpage
\section{Introduction}\label{sec:introduction}

Many economically relevant parameters depend on conditional cumulative distribution functions (CCDFs). Examples include control functions in models with endogeneity \citep{chesher2003identification,imbens2009identification}, functionals of counterfactual distributions \citep{chernozhukov2013inference}, conditional rank-rank regressions for intergenerational mobility \citep{chernozhukov2025conditional}, and transformations of conditional quantiles (conditional L-statistics). Estimating these parameters requires accounting for estimation of the conditional distribution.

Distributional methods have a long tradition in applied econometrics, including wage-distribution decompositions \citep{dinardo1996labor}, recentered-influence-function regressions \citep{firpo2009unconditional}, and distributional policy effects \citep{rothe2012partial}. Conditional distributions can be estimated by distribution regression \citep{foresi1995conditional} or conditional quantile methods \citep{koenker1978regression}, among other procedures. Even under a correctly specified parametric first step, its estimation generally contributes to second-step uncertainty. Flexible nonparametric and machine-learning methods introduce further difficulties because their regularization bias and convergence rates can affect inference on the target parameter.

A common route to plug-in inference applies the functional
delta method to an estimated conditional distribution \citep{chernozhukov2013inference}.
This requires a suitable weak limit for the entire
first-step estimation error. In fully nonparametric
settings with continuously distributed covariates, this
requirement presents a substantive obstacle: pointwise
asymptotic normality does not ensure convergence of the
joint process over outcomes and covariates. Related
impossibility results for kernel density estimators
\citep{nishiyama2011impossibility,stupfler2016weak}
and the failure of functional convergence for standardized
series estimators \citep[Comment~4.7]{belloni2015some}
demonstrate that the difficulty goes beyond slow
convergence rates. We conjecture that this failure is
widespread among fully nonparametric conditional-CDF
estimators at their natural stochastic rates when a
uniform process limit is required.

In this paper we study parameters that depend on averages
of conditional-distribution functionals. Such averaging
can permit root-$n$ inference even when the estimated
conditional distribution has no suitable functional
weak limit \citep{newey1994asymptotic,rothe2010nonparametric}.
Our approach exploits this distinction by correcting
the first-order effect of conditional-distribution
estimation directly, without requiring a process limit
for the first step.
We construct moments that remove the first-order population effect of conditional-distribution estimation. The approach follows the first-step influence-function construction of \citet{chernozhukov2022locally} and is related to classical orthogonalization \citep{neyman1959optimal}, semiparametric influence functions \citep{newey1994asymptotic,ichimura2022influence}, and debiased estimation \citep{robins1994estimation,chernozhukov2018double}. The correction is a conditionally centered representer of the derivative of the population moment. Combined with cross-fitting, it permits root-$n$ inference under mean-square continuity, small-bias, and interaction-remainder conditions without requiring functional weak convergence of the first-step estimator.

Our target is a finite-dimensional parameter identified by unconditional moments. This differs from the conditional targets studied by \citet{chernozhukov2026conditional}, whose influence-function analysis concerns perturbations of a conditional law and can support inference at nonparametric rates. The distinction is between the target and the nuisance: in our setting the entire conditional distribution is a nuisance, while averaging in the identifying restriction can yield a regular target. A related application-specific analysis is \citet{hahn2019three}, who derive influence-function components for semiparametric control-function estimators. Our construction recovers their conditional-CDF component and applies to other moments as well.

Closely related work by \citet{chetverikov2025weighted} develops
influence-function adjustments for linear regressions of
conditional-distribution functionals, including weighted-average
quantile regression. Our analysis considers general unconditional
moments and develops complementary conditional-measure and CDF
representations, together with automatic representer estimation
under the CDF formulation.

The paper makes four contributions. First, we give two complementary linearizations of the population moment. The conditional-measure formulation covers derivatives represented by integration against a conditional probability measure. A stronger CDF formulation yields the measure representer through a conditional tail integral. The two formulations have different domains of applicability: individual conditional quantiles can have a measure representer without an $L^2$ CDF representer.

Second, under the CDF formulation, we develop automatic estimation of the correction using Riesz regression \citep{chernozhukov2022automatic,chernozhukov2024automatic}. The identified object is the projection of the CDF representer onto the space of admissible directions. We give a sieve-estimation bound and account separately for the error introduced when integrating the fitted representer against an estimated conditional law. This integration error also enters the conditions for inference.

Third, we derive corrections for control functions, counterfactual functionals, conditional rank-rank regression, and transformations of conditional $L$-functionals. The general construction permits discrete and mixed conditional laws. Additional continuity and density conditions arise when needed by a particular functional, such as a quantile, or by a structural interpretation, such as a uniform control variable.

Fourth, we state conditions for local robustness, double robustness, and valid inference. Double robustness follows from an exact linear representation of the population-moment difference. For rank-rank regression it is a componentwise property, holding when the other generation's distribution is correct. For general nonlinear transformations of conditional $L$-functionals, inference instead requires a sufficiently small nonlinear remainder. We establish asymptotic linearity, normality, and consistent variance estimation for a cross-fitted GMM estimator, and provide separate sufficient conditions for consistency.

Monte Carlo experiments illustrate the method in conditional rank-rank regression. The reported XGBoost results show reductions in bias and RMSE relative to plug-in estimation, with double cross-fitting providing finite sample improvements in accuracy for stronger-dependence designs. A separate experiment holds the parental nuisances correctly specified and varies specification of the child CDF and representer, illustrating the componentwise double-robustness result.

\Cref{sec:motivation} introduces the setting and examples. \Cref{sec:FSIF,sec:FSIFexamples} derive the influence functions, and \Cref{sec:local&doublerobustness} studies robustness. \Cref{sec:representer-estimation} develops representer estimation, \Cref{sec:asymptotictheory} gives the asymptotic theory, and \Cref{sec:simulation} presents simulations. Proofs, an empirical-process counterexample, and additional simulation results appear in the Appendix.
\section{Setting and Motivating Examples}\label{sec:motivation}

Let $W=(Y,X,Z)\sim F_0$ denote the observed data, where $Y$ is an outcome, $X$ is a vector of covariates, and $Z$ collects any additional observed variables. Our main first-step object is the conditional CDF (CCDF)
\[
\gamma_0(y,x):=F_{Y|X}(y|x)
=\mathbb{P}(Y\leq y\mid X=x).
\]
Let $\mathcal{G}$ denote a class of measurable conditional probability kernels, represented by their CDFs $\gamma:\mathbb R\times\mathcal X\to[0,1]$, with $\gamma_0\in\mathcal G$. For each $x$, $\gamma(\cdot,x)$ is a right-continuous CDF, and $\gamma(dy,x)$ denotes its induced probability measure. All random variables take values in standard Borel spaces, so regular conditional distributions exist. We use fixed measurable versions of the kernels wherever they are evaluated.

Write $P f=\int f(w)F_0(dw)$, let $P_A$ denote the marginal law of a subvector $A$ under $F_0$, and write $F_X=P_X$ when using measure notation. The notation $F_{Y|X}(y,x)$ is equivalent to $F_{Y|X}(y\mid x)$. For vector-valued functions, $\|f\|_{L^2(P_A)}=(\int\|f\|^2\,dP_A)^{1/2}$, with the Euclidean norm inside the integral; $\|f\|_2$ uses the relevant true marginal law. Matrix norms are operator norms unless a Frobenius norm $\|\cdot\|_F$ is specified. All moments and conditional integrals used below are assumed to exist on their stated domains.

We consider estimation and inference for a finite-dimensional parameter $\theta_0\in\Theta\subseteq\mathbb{R}^q$, identified by the moment restrictions
\begin{equation}\label{eq:originalmoment}
    \mathbb{E}[g(W,\gamma_0,\theta_0)]=0,
\end{equation}
where $g(w,\gamma,\theta)\in\mathbb{R}^p$ is a known moment function, $p\geq q$, and expectations are taken under $F_0$. We assume that $\theta_0$ is the unique solution to these restrictions over $\Theta$. The dependence of $g$ on $\gamma_0$ creates a generated-nuisance problem when the conditional distribution is unknown and must be estimated.

For expositional simplicity, we focus on moments involving a single CCDF. This entails no essential loss of generality on the first-order construction: moments depending on multiple conditional distributions can be handled by combining the corresponding first-step influence functions, as discussed below.

\begin{remark}[Multiple conditional distributions]
The framework extends to moments depending on a finite collection of CCDFs, $\gamma_0=(\gamma_{0,1},\ldots,\gamma_{0,J})$, potentially involving different outcomes and conditioning variables. By linearity of the first-order derivative, the overall first-step effect is the sum of the effects associated with each conditional distribution. Hence, the locally robust correction is obtained by adding the corresponding componentwise corrections, provided the required differentiability and representability conditions hold. The single-CDF formulation is therefore adopted only to simplify notation.
\end{remark}

The following examples illustrate economically relevant parameters that depend on conditional distributions and motivate our general approach to estimation and inference.

For notational simplicity, we formulate the general results for
$\gamma_0=F_{Y|X}$. The same results apply to the conditional
distribution $F_{A|B}$ of any observable scalar variable $A$
given a vector $B$, by replacing $(Y,X)$ with $(A,B)$.
In particular, in the control-function application given next, the first-step nuisance is $F_{X|Z}$, so the relevant variables are $(A,B)=(X,Z)$.

\begin{example}[Control functions]\label{ex:controlfunctions}
The control-function approach provides identification and estimation of nonparametric and semiparametric models with endogeneity. Consider the triangular model
\[
Y=f(X,\varepsilon),
\qquad
X=r(Z,U),
\]
where $X$ is a continuous endogenous variable, $Z$ is an instrument, and $\varepsilon$ and $U$ are unobserved disturbances. Assume that $Z$ is independent of $(\varepsilon,U)$, that $U$ is continuously distributed (without loss of generality, uniformly distributed on $[0,1]$), and that $r(z,\cdot)$ is strictly increasing. Under the usual support conditions, \citet{imbens2009identification} show that
\[
V=\gamma_0(X,Z),
\qquad
\gamma_0(x,z)=F_{X|Z}(x|z),
\]
is a valid control variable. In particular, the average structural function
\[
m(x)=\mathbb{E}[f(x,\varepsilon)]
\]
is identified by
\[
m(x)=\mathbb{E}[\mu(x,V)],
\qquad
\mu(x,v)=\mathbb{E}[Y\mid X=x,V=v].
\]

Estimating the average structural function with an unrestricted
response regression $\mu$ involves three steps: estimating the
conditional distribution $\gamma_0$, estimating $\mu$ using the
generated control, and averaging the estimated response regression \citep{hahn2019three}.
Our two-step theory does not by itself account for estimation of
the second-step regression.

A direct application of our framework is obtained by specifying
the response regression parametrically:
\[
\mu(x,v)=\ell(x,v;\eta_0),
\]
where $\ell$ is known and $\eta_0\in\mathbb R^d$ has fixed
dimension. Let $b(x,v)\in\mathbb R^k$, with $k\geq d$, be a
known vector of instruments such that
\[
\mathbb E\!\left[
b(X,V)\{Y-\ell(X,V;\eta)\}
\right]=0
\]
identifies $\eta_0$. For a fixed structural evaluation point
$x^\star$, write $\beta_0=m(x^\star)$ and include the response
coefficients and the structural target in the parameter vector
$\theta_0=(\eta_0',\beta_0)'$. The identifying moments are then
\begin{equation}\label{eq:cf-parametric-stack}
g(W,\gamma,\theta)
=\varrho\big(W,\gamma(X,Z),\theta\big),
\qquad
\varrho(W,v,\theta)=
\begin{pmatrix}
b(X,v)\{Y-\ell(X,v;\eta)\}\\
\ell(x^\star,v;\eta)-\beta
\end{pmatrix}.
\end{equation}
The only nonparametric first-step object in this system is
$\gamma_0$. Applying our conditional-CDF correction to every
component of the stacked moment and estimating $\theta$ jointly
accounts for estimation of the response coefficients through
the full GMM Jacobian. Thus, although the sample equations may
be solved sequentially, the problem fits our two-step framework.

More generally, the same argument applies to known moments
$\varrho(W,V,\theta)$, provided all unknown finite-dimensional
response coefficients are included in $\theta$.

For regular targets with a nonparametric response regression,
an alternative is to first construct a moment locally robust
to estimation of that regression, following
\citet{escanciano2023generated}, and then apply our conditional-CDF
correction to the resulting moment. Under the corresponding
differentiability conditions, orthogonality to the second step
eliminates the indirect effect of estimating the control that
operates through the response regression. Our construction
corrects the remaining effect of conditional-CDF estimation.
The correction must therefore be derived from the entire
second-step-robust moment, including its adjustment term.
Three-step debiased inference can be established by combining
the results in this paper with those of
\citet{escanciano2023generated}, including the additional
continuity and remainder conditions associated with the
estimated second step.
\end{example}

\begin{example}[Functionals of counterfactuals]\label{ex:counterfactuals}
    Counterfactual distributions are widely used to study distributional effects of changes in observed characteristics, labor-market institutions, policies, or other economic environments (see, among others, \citet{dinardo1996labor}, \citet{machado2005counterfactual}, \citet{firpo2009unconditional}, and \citet{chernozhukov2013inference}). They provide a natural framework for distributional decompositions and for comparing an observed outcome distribution with the distribution that would prevail under an alternative distribution of covariates.

    Let $F_X^*$ denote a counterfactual distribution of the covariates and define the counterfactual outcome distribution as
    $$
        F_Y^* = \int_{\mathcal X} F_{Y|X}(\cdot,x)F_X^*(dx).
    $$
    The parameter of interest may then be a scalar functional of this distribution,
    $$
        \theta_0 = \varphi(F_Y^*),
    $$
    including, for example, counterfactual means, variances, quantiles, inequality measures such as the Gini coefficient, and other distributional statistics.

    Since $F_Y^*$ depends on the conditional distribution $F_{Y|X}$, this class of parameters falls directly into our framework. Writing $\gamma_0=F_{Y|X}$, the identifying moment can be expressed as
    $$
        g(W,\gamma,\theta) = \varphi\left( \int_{\mathcal X} \gamma(\cdot,x)F_X^*(dx) \right) - \theta.
    $$

    \citet{chernozhukov2013inference} develop a general theory for estimation and inference on counterfactual distributions and their functionals when the conditional distribution is estimated by distribution regression or other first-step procedures satisfying an appropriate weak-convergence requirement (see the Introduction for discussion). \citet{rothe2010nonparametric} instead studies nonparametric kernel estimators of distributional policy effects and derives their asymptotic distribution using undersmoothing. These results illustrate an important difficulty of counterfactual inference: estimation error in the entire conditional distribution propagates into the counterfactual distribution and subsequently through the functional $\varphi(\cdot)$.

    The locally robust approach developed below provides an alternative way to account for this first-step estimation effect. In particular, it permits the use of flexible nonparametric and machine-learning estimators of the conditional distribution under suitable rate and remainder conditions, without requiring the same weak-convergence conditions on the entire first-step process or relying on undersmoothing.
\end{example}

\begin{example}[Conditional rank-rank regression]\label{ex:rankrank}
    Rank-rank regressions are widely used in the empirical literature on intergenerational mobility to measure the association between the positions of children and parents in their respective income distributions. More recently, \citet{chernozhukov2025conditional} propose using conditional rather than unconditional ranks, thereby measuring intergenerational rank dependence after accounting for observed covariates.

    Let $Y_c$ and $Y_p$ denote the income of child and parents, respectively, and let $X$ denote a vector of observed covariates.
    Then, the conditional ranks are measured by
    $$
        F_{Y_c|X}(Y_c,X), \quad F_{Y_p|X}(Y_p,X).
    $$
    If $Y_c$ and $Y_p$ are continuously distributed conditional on $X$, the conditional probability integral transform gives
    $$
        F_{Y_s|X}(Y_s,X) \sim U(0,1) \quad \text{for } s\in\left\{c,p\right\}.
    $$
    Hence,
    $$
        \mathbb{E}\left[F_{Y_s|X}(Y_s,X)\right] = \frac{1}{2}, \quad \operatorname{Var}\left(F_{Y_s|X}(Y_s,X)\right) = \frac{1}{12} \quad \text{for } s\in\left\{c,p\right\}.
    $$
    The slope coefficient from the linear projection of the child conditional rank $F_{Y_c|X}(Y_c,X)$ on the parental conditional rank $F_{Y_p|X}(Y_p,X)$ is therefore
    $$
        \theta_0 = \frac{ \mathbb{E}\left[ F_{Y_c|X}(Y_c,X) F_{Y_p|X}(Y_p,X) \right] -1/4 }{1/12}.
    $$
    Thus, writing $\gamma_0=(\gamma_{0,c},\gamma_{0,p})=\left(F_{Y_c|X},F_{Y_p|X}\right)$, the parameter is identified by the moment condition
    $$
        g(W,\gamma,\theta)=\gamma_c(Y_c,X)\gamma_p(Y_p,X)-1/4-\theta/12.
    $$
    Conditional rank-rank regression is therefore a natural application of our framework with multiple first-step conditional distributions. Estimation error in either CCDF propagates into the product of the estimated conditional ranks, and the interaction between the two first steps must also be controlled.

    \citet{chernozhukov2025conditional} derive asymptotic results for conditional rank-rank coefficients using distribution-regression estimators of the conditional distributions. Their theory relies on first-step estimators satisfying the corresponding weak-convergence requirements. The locally robust approach developed below instead accounts explicitly for the first-step estimation effects through influence-function corrections and can accommodate more flexible nonparametric and machine-learning estimators under suitable rate and remainder conditions.
\end{example}

\begin{example}[Conditional $L$-functionals]
\label{ex:l-functionals}

Many economically relevant distributional parameters can be expressed as $L$-functionals of conditional quantiles. Let
\[
\gamma^{-1}(s,x):=\inf\{y:\gamma(y,x)\geq s\},
\qquad s\in(0,1),
\]
and define
\[
L_\gamma(x)
=\int_0^1 a\left(\gamma^{-1}(s,x)\right)K(ds),
\]
where $a$ is a known transformation, $K$ is a finite signed measure on $(0,1)$, and the displayed integral is finite for the conditional laws under consideration. Different choices of $a$ and $K$ generate conditional means, quantiles, trimmed means, expected shortfall, generalized Lorenz ordinates, and $L$-moments. Point masses of $K$ yield individual quantiles and their linear combinations, whereas absolutely continuous measures yield weighted averages of quantiles. More generally, we consider moments of the form
\[
g(W,\gamma,\theta)
=\mathcal{M}\big(W,L_\gamma(X),\theta\big),
\]
where $\mathcal{M}$ is a known, possibly nonlinear map. This formulation encompasses several important economic applications.

\medskip
\noindent\textit{Best linear approximations and tail risk.}
The best linear approximation of a conditional $L$-functional is identified by
\[
g(W,\gamma,\theta)
=p(X)\big\{L_\gamma(X)-p(X)'\theta\big\},
\]
where $p(X)\in\mathbb R^q$ is a specified vector of covariate transformations, with finite second moments and nonsingular $\mathbb E[p(X)p(X)']$. Here $p(\cdot)$ denotes a function, distinct from the scalar moment dimension $p$. With $a(y)=y$ and $K(ds)=\psi(s)ds$, this includes the weighted-average quantile regression parameter of \citet{chetverikov2025weighted}. Choosing $K$ to concentrate weight on the lower or upper tail yields regressions for conditional expected shortfall, allowing researchers to study how tail outcomes vary with observed characteristics. Related regression models for conditional distributional measures are studied by \citet{hossjer2023use}.

\medskip
\noindent\textit{Weighted averages and social welfare.}
Another class of parameters is identified by
\[
g(W,\gamma,\theta)
=\omega(X)L_\gamma(X)-\theta,
\]
where $\omega(X)$ is a known weighting function. For $a(y)=y$ and $K=\delta_s$, this yields the weighted average conditional quantile
\[
\theta_0
=\mathbb{E}\left[
\omega(X)F_{Y|X}^{-1}(s,X)
\right].
\]
More generally, choosing $K(ds)=w(s)ds$, where $w(s)\geq 0$ integrates to one, yields rank-weighted social welfare measures,
\[
\theta_W=\mathbb{E}[L_{\gamma_0}(X)],
\qquad
L_\gamma(x)
=\int_0^1\gamma^{-1}(s,x)w(s)ds.
\]
Greater weight on lower quantiles reflects stronger concern for disadvantaged individuals, as in generalized Gini social welfare criteria \citep{weymark1981generalized}. Our framework also accommodates nonlinear welfare transformations. For a strictly increasing, concave utility function $u$ with an inverse on the relevant range, let
\[
L_\gamma(x)
=\int_0^1 u\left(\gamma^{-1}(s,x)\right)w(s)ds.
\]
The parameter
\[
\theta_{CE}
=\mathbb{E}\left[
u^{-1}\big\{L_{\gamma_0}(X)\big\}
\right]
\]
is the average equally distributed equivalent income across covariate groups, expressing inequality-adjusted welfare in income units.

\medskip
\noindent\textit{Distributional treatment effects.}
Let $D\in\{0,1\}$ denote treatment status and $Y(1),Y(0)$ the potential outcomes. Under consistency $Y=DY(1)+(1-D)Y(0)$, conditional unconfoundedness $(Y(1),Y(0))\perp D\mid X$, and overlap $0<p_d(X):=P(D=d\mid X)<1$, their conditional distributions are identified by
\[
\gamma_{d,0}(y,x)
=F_{Y|X,D=d}(y|x),
\qquad d\in\{0,1\}.
\]
For $a(y)=y$ and a common measure $K$, consider
\[
\theta_\Delta
=\mathbb{E}\left[
\omega(X)\left\{
L_{\gamma_{1,0}}(X)
-
L_{\gamma_{0,0}}(X)
\right\}
\right].
\]
This parameter summarizes treatment effects across economically relevant regions of the conditional outcome distribution. Taking $K=\delta_s$ gives the average conditional quantile treatment effect at rank $s$, while assigning weight to the lower or upper tail captures treatment effects on tail outcomes. With $a(y)=u(y)$, the same construction measures treatment-induced changes in inequality-sensitive welfare. These parameters complement conventional average treatment effects by capturing distributional effects that may be obscured by changes in the mean; see \citet{kallus2023robust} for related orthogonal methods for conditional distributional treatment effects.

In the treatment-effect case, the distribution of covariates in each arm differs from the target distribution $P_X$. The correction therefore includes inverse propensity weights. Under the conditional-quantile regularity developed in \Cref{ass:Lhadam-diff}, put $q_{d,0}(s,x)=\gamma_{d,0}^{-1}(s,x)$ and let $f_{d,0}$ be the corresponding conditional density at these quantiles. Define
\[
\ell_{d,0}(y,x)=\int_0^1
\frac{a'(q_{d,0}(s,x))}{f_{d,0}(q_{d,0}(s,x),x)}
\{s-\indic\{y\leq q_{d,0}(s,x)\}\}K(ds).
\]
The first-step correction for the weighted arm-$d$ average is
\[
\phi_{d,0}(W)=\frac{\indic\{D=d\}}{p_d(X)}\omega(X)\ell_{d,0}(Y,X),
\]
and the correction for the difference is $\phi_{1,0}-\phi_{0,0}$. These expressions require the relevant integrability and square-integrability conditions; overlap alone does not imply them. They follow by conditioning on $(X,D)$ and reweighting the arm-specific covariate law to $P_X$. For candidate arm laws $\eta=(\eta_1,\eta_0)$, the identifying moment is $g(W,\eta,\theta)=\omega(X)\{L_{\eta_1}(X)-L_{\eta_0}(X)\}-\theta$. Estimation of the propensity scores contributes to representer error and must satisfy the corresponding remainder conditions.

These applications illustrate both nonlinear dependence on the inverse-CDF map and moments involving multiple conditional distributions. The regularity conditions for the existence of the corresponding first-step influence functions are developed below.
\end{example}

\section{First-Step Influence Functions}
\label{sec:FSIF}

We derive first-step influence functions (FSIFs) for identifying moments whose nuisance parameter is a CCDF. Following \citet{ichimura2022influence}, consider a perturbation $F_\tau=(1-\tau)F_0+\tau H$, with $\tau\in[0,1]$, and let $\gamma_\tau=F_{Y|X,\tau}$ denote the corresponding CCDF. We restrict attention to perturbations $H$ for which $H_X\ll F_X$, the induced conditional law is admissible for sufficiently small $\tau$, and the derivatives below exist. For the measure representer at each $\theta$, impose the path integrability requirement
\[
\mathbb E_H\left[\|\alpha_0(Y,X,\theta)\|
 +\int\|\alpha_0(y,X,\theta)\|\gamma_0(dy,X)\right]<\infty.
\]
For the CDF formulation below, it suffices to impose
$\mathbb E_H[\int\|\delta_0(y,X,\theta)\|\gamma_0(dy,X)]<\infty$.
These requirements are part of admissibility, in addition to any application-specific restrictions on the path. Derivatives at zero are right derivatives along the mixture path. Holding the distribution in the outer expectation fixed at $F_0$, the first-step effect is
\[
D_\theta(H)
:=
\left.\frac{d}{d\tau}
\mathbb{E}[g(W,\gamma_\tau,\theta)]
\right|_{\tau=0}.
\]
An FSIF is a mean-zero, square-integrable function $\phi_0(W,\theta)$ satisfying
\begin{equation}\label{eq:infdef}
D_\theta(H)=\int\phi_0(w,\theta)H(dw)
\end{equation}
for all admissible $H$. Adding the FSIF to the original identifying moment is the standard construction of locally robust moments. Our objective is to obtain this correction explicitly when the first-step nuisance is a CCDF.

We provide two complementary linearizations of the first-step effect. The first operates on conditional probability measures.

\begin{assumption}[Conditional-measure linearization]
\label{ass:linearization}
For each $\theta\in\Theta$, there exists a measurable function $\alpha_0(y,x,\theta)$ satisfying
$\mathbb{E}[\|\alpha_0(Y,X,\theta)\|^2]<\infty$ such that, for every admissible $H$,
\begin{equation}\label{eq:linearization1}
D_\theta(H)
=
\left.\frac{d}{d\tau}
\iint \alpha_0(y,x,\theta)
F_{Y|X,\tau}(dy,x)F_X(dx)
\right|_{\tau=0}.
\end{equation}
\end{assumption}

The following result provides the corresponding FSIF.

\begin{lemma}\label{lemma:correctionterm}
Under \Cref{ass:linearization}, the FSIF is
\[
\phi_0(w,\theta)
=
\alpha_0(y,x,\theta)
-
\mathbb{E}[\alpha_0(Y,X,\theta)\mid X=x].
\]
\end{lemma}

The correction is obtained by centering the representer conditional on $X$. This centering isolates the effect of perturbing the conditional distribution from changes in the marginal distribution of $X$.

An alternative linearization works directly with the CCDF, rather than the conditional probability measure. This is particularly convenient when the identifying moment is differentiable with respect to CDF evaluations.

\begin{passumption}[CDF linearization]
\label{ass:linearizationcdf}
For each $\theta\in\Theta$, there exists a measurable function $\delta_0(y,x,\theta)$ satisfying
$\mathbb{E}[\|\delta_0(Y,X,\theta)\|^2]<\infty$ such that, for every admissible $H$,
\begin{equation}\label{eq:linearization2}
D_\theta(H)
=
\left.\frac{d}{d\tau}
\iint \delta_0(y,x,\theta)
F_{Y|X,\tau}(y,x)
F_{Y|X}(dy,x)F_X(dx)
\right|_{\tau=0}.
\end{equation}
\end{passumption}

The CDF-based representer determines the conditional-measure representer through the following transformation.

\begin{plemma}\label{lemma:correctiontermprime}
Under \Cref{ass:linearizationcdf}, the FSIF is
\[
\phi_0(w,\theta)
=
\alpha_0(y,x,\theta)
-
\mathbb{E}[\alpha_0(Y,X,\theta)\mid X=x],
\]
where
\[
\alpha_0(y,x,\theta)
=
\mathbb{E}\left[
\delta_0(Y,X,\theta)
\indic\{y\leq Y\}
\mid X=x
\right].
\]
\end{plemma}

Thus, both linearizations lead to the same form of correction. The CDF-based assumption implies the conditional-measure representation under the stated regularity conditions, but the converse need not hold. In particular, the existence of a square-integrable CDF-based representer is a stronger requirement than the existence of a conditional-measure representer.

\begin{remark}[Relation between the two linearizations]
The conditional-measure linearization is more general than its CDF-based counterpart. Indeed, under suitable integrability conditions, \Cref{ass:linearizationcdf} implies \Cref{ass:linearization}, with
\[
\alpha_0(y,x,\theta)
=
\int_{[y,\infty)}
\delta_0(t,x,\theta)F_{Y|X}(dt\mid x).
\]
The converse need not hold, since the CDF-based representation requires a version of the measure representer in the range of this integral operator. Measure representers are only identified up to additive functions of $x$, which disappear under conditional centering. This distinction matters for functionals involving pointwise CDF evaluations or conditional quantiles, whose derivatives may admit square-integrable conditional-measure representers but no square-integrable CDF-based representer. Thus, the conditional-measure approach provides the general formulation, while the CDF-based approach offers a convenient construction when the identifying moment is differentiable with respect to CCDF evaluations.
\end{remark}

The resulting locally robust moment is
\begin{equation}\label{eq:LRmoment}
\psi(W,\gamma_0,\alpha_0,\theta)
=
g(W,\gamma_0,\theta)+\phi_0(W,\theta).
\end{equation}
For estimation, its extension to generic nuisance values is
\begin{equation}\label{eq:moment}
\psi(W,\gamma,\alpha,\theta)
=
g(W,\gamma,\theta)
+
\alpha(Y,X,\theta)
-
\int\alpha(y,X,\theta)\gamma(dy,X),
\end{equation}
We use the notation
\[
\phi(W,\gamma,\alpha,\theta)
=\alpha(Y,X,\theta)-\int\alpha(y,X,\theta)\gamma(dy,X)
\]
for the correction at generic nuisance values; thus $\psi=g+\phi$ and
$\phi_0(W,\theta)=\phi(W,\gamma_0,\alpha_0,\theta)$.

By construction, the correction has mean zero at $\gamma_0$ for any square-integrable $\alpha$. Moreover, under the regularity conditions established below, the moment satisfies
\begin{equation}\label{eq:LRproperty}
\left.\frac{d}{d\tau}
\mathbb{E}[\psi(W,\gamma_\tau,\alpha_0,\theta)]
\right|_{\tau=0}=0,
\qquad
\forall H \text{ admissible},\quad \forall\theta\in\Theta.
\end{equation}
Thus, the first-order effect of estimating the CCDF is eliminated. The precise conditions for local robustness are developed in \Cref{thm:LR}.

\section{First-Step Influence Functions in Applications}
\label{sec:FSIFexamples}

\subsection{Control functions}\label{subsec:controlfunction}

Consider the control-function setting of \Cref{ex:controlfunctions}, where
\[
V=\gamma_0(X,Z),\qquad
\gamma_0(x,z)=F_{X|Z}(x|z),
\]
and the identifying moment takes the general form
\[
g(W,\gamma,\theta)
=\varrho\big(W,\gamma(X,Z),\theta\big).
\]
Thus, the first-step nuisance is the CCDF of $X$ given $Z$, and our general results apply with $(Y,X)$ replaced by $(X,Z)$.

\begin{proposition}[Control-function correction]
\label{prop:controlfunction}
Suppose that $\varrho(W,v,\theta)$ is differentiable in $v$, differentiation can be interchanged with expectation along admissible paths, and
$\mathbb{E}[\|\partial_v\varrho(W,V,\theta)\|^2]<\infty$.
Then the CDF-based linearization holds with
\[
\delta_0(x,z,\theta)
=
\mathbb{E}\left[
\partial_v\varrho(W,V,\theta)\mid X=x,Z=z
\right],
\]
and the FSIF is
\[
\phi_0(w,\theta)
=
\alpha_0(x,z,\theta)
-
\mathbb{E}[\alpha_0(X,Z,\theta)\mid Z=z],
\]
where
\[
\alpha_0(x,z,\theta)
=
\mathbb{E}\left[
\partial_v\varrho(W,V,\theta)
\indic\{x\leq X\}\mid Z=z
\right].
\]
\end{proposition}

The representer $\alpha_0(x,z,\theta)$ aggregates the sensitivity
of the identifying moment to changes in the generated control.
By conditional Fubini,
\[
\mathbb E[\alpha_0(X,Z,\theta)\mid Z=z]
=
\mathbb E\!\left[
\partial_v\varrho(W,V,\theta)V\mid Z=z
\right].
\]
Consequently, the first-step correction can also be written as
\begin{equation}\label{eq:phicontrolfunction}
\phi_0(w,\theta)
=
\mathbb E\!\left[
\partial_v\varrho(W,V,\theta)
\{\indic\{x\leq X\}-V\}
\mid Z=z
\right].
\end{equation}

For the stacked parametric response model in
\Cref{eq:cf-parametric-stack}, differentiation with respect to
the control, holding $\theta=(\eta',\beta)'$ fixed, gives
\[
\partial_v\varrho(W,v,\theta)
=
\begin{pmatrix}
\partial_v b(X,v)\{Y-\ell(X,v;\eta)\}
-b(X,v)\partial_v\ell(X,v;\eta)\\
\partial_v\ell(x^\star,v;\eta)
\end{pmatrix}.
\]
Substituting this derivative into
\Cref{eq:phicontrolfunction} yields the conditional-CDF
correction for every component of the stacked moment.
The response coefficients $\eta$ and the structural target
$\beta$ are then estimated from the corrected system.
The correction removes the first-order effect of estimating
$\gamma_0$, while the full GMM Jacobian accounts for estimation
of the finite-dimensional response coefficients.

Continuity of $F_{X|Z}$ is not required for the FSIF calculation,
which also applies to discrete and mixed conditional laws.
The structural interpretation of $V$ in
\Cref{ex:controlfunctions}, however, uses a continuous endogenous
variable. With a nonparametric response regression, an additional
correction for estimation of that regression is required.

The following example connects our conditional-CDF correction
with the influence-function analysis of control-function
estimators in \citet{hahn2019three}, who consider parameters
$\mathbb E[H(X,m(X))]$ with known $H$, and the second-step
orthogonalization of \citet{escanciano2023generated}.
For a nonparametric response regression, we instead consider
the integrated average structural function under a fixed
known probability measure $Q$,
\begin{equation}\label{eq:cf-integrated-target}
\beta_{Q,0}
=\int m(t)Q(dt)
=\mathbb E\!\left[\int\mu(t,V)Q(dt)\right],
\end{equation}
where $Q$ is a fixed known probability measure on structural
values of $X$. Let $P_{X,V}$ denote the joint law of $(X,V)$
and $P_V$ the marginal law of $V$. Suppose that
$\mathbb E[Y^2]<\infty$ and
\[
Q\otimes P_V\ll P_{X,V},
\qquad
a_{2,0}(x,v)
:=\frac{d(Q\otimes P_V)}{dP_{X,V}}(x,v)
\in L^2(P_{X,V}).
\]
The function $a_{2,0}$ represents the derivative with respect
to the response regression: for every
$k\in L^2(P_{X,V})$,
\begin{equation}\label{eq:cf-second-step-representer}
\mathbb E\!\left[\int k(t,V)Q(dt)\right]
=
\mathbb E[a_{2,0}(X,V)k(X,V)].
\end{equation}
The integrated target permits this square-integrable
representer. By contrast, when $X$ is continuously distributed,
point evaluation $Q=\delta_{x^\star}$ fails the stated
absolute-continuity condition. The pointwise ASF considered
in \Cref{eq:cf-parametric-stack} is accommodated by the
parametric restriction on the response regression.

Let $h$ denote a candidate response regression, whose true
value is $\mu$, and let $a_2$ be a candidate second-step
representer. Define
\begin{equation}\label{eq:cf-second-step-robust}
\varrho_2(W,v,h,a_2,\beta)
=
\int h(t,v)Q(dt)-\beta
+a_2(X,v)\{Y-h(X,v)\}.
\end{equation}
At the true control $v=V$, the moment in
\Cref{eq:cf-second-step-robust} has expectation zero at
$(h,a_2,\beta)=(\mu,a_{2,0},\beta_{Q,0})$ and is locally
robust with respect to both $h$ and $a_2$.
The corresponding population derivatives vanish by the
representer identity in
\Cref{eq:cf-second-step-representer} and the restriction
$\mathbb E[Y-\mu(X,V)\mid X,V]=0$, respectively.

Under the differentiability conditions for the functional
chain rule, this second-step orthogonality eliminates the
indirect effect of estimating the control that operates
through the response regression and its representer
\citep{escanciano2023generated}. The remaining direct effect
of estimating $\gamma_0$ must be computed from the entire
moment $\varrho_2$, including its residual adjustment.

We now derive the remaining conditional-CDF correction.
In the triangular model of \Cref{ex:controlfunctions},
\[
\mathbb E[Y\mid X,Z]=\mu(X,V).
\]
Consequently, writing $V_\gamma=\gamma(X,Z)$, the population
residual adjustment satisfies
\[
\mathbb E[a_{2,0}(X,V_\gamma)\{Y-\mu(X,V_\gamma)\}]
=
\mathbb E[a_{2,0}(X,V_\gamma)
\{\mu(X,V)-\mu(X,V_\gamma)\}].
\]
This identity allows us to differentiate the population
moment without differentiating the second-step representer.

Suppose that $\mu$ is differentiable in its second argument,
that the corresponding regression evaluations are
differentiable in mean square along admissible CCDF paths,
and that $a_{2,0}(X,V_\gamma)$ is continuous in mean square
at $\gamma_0$. Assume also that differentiation can be
interchanged with $Q$-integration and expectation. Define
\begin{equation}\label{eq:cf-control-sensitivity}
s_0(x,v)
=
\int\partial_v\mu(t,v)Q(dt)
-a_{2,0}(x,v)\partial_v\mu(x,v),
\end{equation}
and suppose that $\mathbb E[s_0(X,V)^2]<\infty$.
For an admissible path $\gamma_\tau$ through $\gamma_0$,
with derivative $\dot\gamma$, these conditions give
\[
\left.\frac{d}{d\tau}
\mathbb E\!\left[
\varrho_2(W,\gamma_\tau(X,Z),\mu,a_{2,0},\beta_{Q,0})
\right]\right|_{\tau=0}
=
\mathbb E[s_0(X,V)\dot\gamma(X,Z)].
\]
The second term in \Cref{eq:cf-control-sensitivity} is the
contribution of the residual adjustment. Omitting it would
give the derivative of the uncorrected moment.

The CDF-based representer is therefore
\[
\delta_{\gamma,0}(x,z)
=s_0\big(x,\gamma_0(x,z)\big).
\]
Applying \Cref{lemma:correctiontermprime}, with $(Y,X)$
replaced by $(X,Z)$, yields
\[
\alpha_{\gamma,0}(x,z)
=
\mathbb E\!\left[
s_0(X,V)\indic\{x\leq X\}\mid Z=z
\right],
\]
and the remaining first-step influence function is
\begin{equation}\label{eq:cf-cdf-correction}
\phi_{\gamma,0}(w)
=
\alpha_{\gamma,0}(x,z)
-\int\alpha_{\gamma,0}(t,z)\gamma_0(dt,z).
\end{equation}

Combining the two corrections gives the moment
\begin{equation}\label{eq:cf-three-step-score}
\begin{split}
\psi_{\rm CF}(W,\gamma,h,a_2,\alpha,\beta)
&=\varrho_2(W,\gamma(X,Z),h,a_2,\beta)\\
&\quad+\alpha(X,Z)
-\int\alpha(t,Z)\gamma(dt,Z).
\end{split}
\end{equation}
At the true nuisance values
$(\gamma_0,\mu,a_{2,0},\alpha_{\gamma,0})$, this moment
has expectation zero at $\beta_{Q,0}$ and is locally robust
with respect to the conditional distribution, the response
regression, and both correction nuisances.

Equation~\eqref{eq:cf-three-step-score} illustrates how our
conditional-CDF correction combines with the second-step
orthogonalization of \citet{escanciano2023generated}.
A full asymptotic analysis of the resulting three-step
estimator is beyond the scope of this paper.
\subsection{Counterfactuals}\label{subsec:counterfactuals}
\assumptionprefix{C}

Consider the counterfactual distribution
\[
F_Y^*(t)=\int \gamma_0(t,x)F_X^*(dx),
\]
where $F_X^*$ is a specified counterfactual distribution of covariates. For a functional $\varphi$, the parameter $\theta_0=\varphi(F_Y^*)$ is identified by
\[
g(W,\gamma,\theta)
=
\varphi\left(\int\gamma(\cdot,x)F_X^*(dx)\right)-\theta.
\]
We derive the FSIF associated with estimating $\gamma_0$, treating $F_X^*$ as fixed. Two conditions are sufficient.

\begin{assumption}\label{ass:abs-cont}
$F_X^*\ll F_X$, with Radon--Nikodym derivative
$r_0=dF_X^*/dF_X$ satisfying
$\mathbb{E}[r_0(X)^2]<\infty$.
\end{assumption}

\begin{assumption}\label{ass:Chadam-diff}
The functional $\varphi:\mathcal{F}\subset\mathbb{B}\to\mathbb{R}$, where $\mathcal F$ is a class of CDFs and $\mathbb B$ is a Banach space, is Hadamard differentiable at $F_Y^*$ tangentially to a linear subspace $\mathcal T_C\subseteq\mathbb B$. Its derivative admits the representation
\[
\varphi'_{F_Y^*}[\eta]
=
\int a_0(t)\eta(t)\nu(dt),
\qquad \eta\in\mathcal{T}_C,
\]
where $\nu$ is a signed measure and
$\int|a_0(t)|\,|\nu|(dt)<\infty$.
\end{assumption}

The absolute-continuity condition allows us to express the counterfactual distribution as
\[
F_Y^*(t)
=
\mathbb{E}[r_0(X)\gamma_0(t,X)].
\]
The second assumption permits differentiation of the functional along perturbations of this distribution.

\begin{proposition}[Functional of the counterfactual distribution]
\label{prop:functionalcounterfactuals}
Suppose that \Cref{ass:abs-cont,ass:Chadam-diff} hold. Assume further that, along every admissible path, the counterfactual distribution induced by $\gamma_\tau$ is differentiable in $\mathbb{B}$, with derivative in $\mathcal{T}_C$, and that differentiation and integration can be interchanged. Then the FSIF is
\begin{equation}\label{eq:phicounterfactuals1}
\phi_0(y,x)
=
\alpha_0(y,x)
-
\mathbb{E}[\alpha_0(Y,X)\mid X=x],
\end{equation}
where
\[
\alpha_0(y,x)
=
r_0(x)\int a_0(t)\indic\{y\leq t\}\nu(dt).
\]
Moreover, $\mathbb{E}[\alpha_0(Y,X)^2]<\infty$.
\end{proposition}

The representer combines two components: $r_0(x)$ reweights the observed covariate distribution to the counterfactual population, while the integral captures the sensitivity of the functional to changes in the conditional outcome distribution.

Two familiar counterfactual parameters illustrate the construction. First, consider the counterfactual mean
\[
\varphi(F_Y^*)=\int y\,F_Y^*(dy).
\]
For true and candidate conditional laws supported on a common compact interval $[\underline y,\overline y]$, integration by parts gives
\[
\varphi'_{F_Y^*}[\eta]
=
-\int_{\underline y}^{\overline y}\eta(t)dt.
\]
Thus, $a_0(t)=-1$ and $\nu$ is Lebesgue measure on the outcome support, yielding
\[
\phi_0(y,x)
=
r_0(x)\left\{y-\mathbb{E}[Y\mid X=x]\right\}.
\]

Second, consider the counterfactual $\tau$-quantile,
\[
q_\tau^*=(F_Y^*)^{-1}(\tau),
\qquad \tau\in(0,1).
\]
If the quantile is uniquely defined and $F_Y^*$ has a positive, continuous density at $q_\tau^*$, the derivative is
\[
\varphi'_{F_Y^*}[\eta]
=
-\frac{\eta(q_\tau^*)}{f_Y^*(q_\tau^*)}.
\]
Hence, taking
\[
a_0(t)=-\frac{1}{f_Y^*(q_\tau^*)},
\qquad
\nu(dt)=\delta_{q_\tau^*}(dt),
\]
we obtain
\[
\phi_0(y,x)
=
\frac{r_0(x)}{f_Y^*(q_\tau^*)}
\left\{
\gamma_0(q_\tau^*,x)
-
\indic\{y\leq q_\tau^*\}
\right\}.
\]

The quantile example illustrates the usefulness of the conditional-measure linearization: the derivative involves a Dirac measure and need not admit the square-integrable CDF-based representation of \Cref{ass:linearizationcdf}.

\subsection{Measures of Intergenerational Mobility}
\label{subsec:conditionalRRR}

Consider the conditional rank-rank regression in \Cref{ex:rankrank}, with two CCDF nuisances,
$\gamma_0=(\gamma_{0,c},\gamma_{0,p})=(F_{Y_c|X},F_{Y_p|X})$.
Since the identifying moment is linear in each CCDF when the other is held fixed, its first-step derivative is the sum of the two componentwise derivatives. Applying \Cref{lemma:correctiontermprime} yields the following result.

\begin{proposition}[Conditional rank-rank correction]
\label{prop:rankrank}
For $s\in\{c,p\}$, write $-s$ for the other generation and let $\gamma_{s,\tau}$ denote the CCDF induced by an admissible perturbation $F_\tau=(1-\tau)F_0+\tau H$, with $H_X\ll F_X$. Then the FSIF is
\begin{equation}\label{eq:phirankrank}
\phi_0(w)
=
\sum_{s\in\{c,p\}}
\left\{
\alpha_{0,s}(y_s,x)
-
\mathbb{E}[\alpha_{0,s}(Y_s,X)\mid X=x]
\right\},
\end{equation}
where
\[
\alpha_{0,s}(y_s,x)
=
\mathbb{E}\left[
\delta_{0,s}(Y_s,X)
\indic\{y_s\leq Y_s\}\mid X=x
\right],
\]
and
\[
\delta_{0,s}(y_s,x)
=
\mathbb{E}\left[
\gamma_{0,-s}(Y_{-s},X)
\mid Y_s=y_s,X=x
\right].
\]
\end{proposition}

The representer $\delta_{0,s}(y_s,x)$ is the expected conditional rank of one generation given the income of the other generation and observed characteristics. Accordingly, $\alpha_{0,s}(y_s,x)$ measures the corresponding rank-weighted upper-tail probability. The two corrections account separately for estimation of the child and parental CCDFs, while their interaction contributes only to the second-order remainder.

The representers are bounded, so square-integrability holds automatically. Moreover, the required interchange of differentiation and integration follows by dominated convergence for admissible perturbations with $H_X\ll F_X$. Continuity of the conditional distributions is not required for the FSIF, although it is needed for the uniform-rank normalization underlying the regression coefficient in \Cref{ex:rankrank}.

\subsection[Transformations of Conditional L-Functionals]{Transformations of Conditional $L$-Functionals}
\label{subsec:Lfunctionals}
\assumptionprefix{L}

Consider the conditional $L$-functionals introduced in
\Cref{ex:l-functionals}, with identifying moments
\[
g(W,\gamma,\theta)
=\mathcal{M}\big(W,L_\gamma(X),\theta\big),
\qquad
L_\gamma(x)
=\int_0^1 a\left(\gamma^{-1}(s,x)\right)K(ds).
\]
The influence function of an $L$-functional follows from the
derivative of the quantile function; see
\citet[p.~321]{van2000asymptotic}. The same principle applies
conditionally on $X$, but the nonlinear inverse-CDF map
requires regularity at the quantiles relevant for $K$.

Define
\[
b_0(x,\theta)
=
\mathbb{E}\left[
\partial_\ell\mathcal{M}
\big(W,L_{\gamma_0}(X),\theta\big)
\mid X=x
\right],
\]
where $\partial_\ell\mathcal{M}$ denotes differentiation with
respect to the second argument.

\begin{assumption}\label{ass:Lhadam-diff}
For each $\theta\in\Theta$, the following conditions hold:
(i) $a$ is continuously differentiable at the relevant
quantiles, and $K$ is a finite signed measure on $(0,1)$;
(ii) for $F_X$-almost every $x$ and $|K|$-almost every $s$,
the CCDF is continuously differentiable in a neighborhood of
$\gamma_0^{-1}(s,x)$, with positive and finite conditional
density, and the usual quantile derivative holds along
admissible paths;
(iii) $\mathcal{M}(W,\ell,\theta)$ is differentiable in $\ell$,
and differentiation can be interchanged with the integrations
defining the population moment and $L_\gamma$;
(iv) $\partial_\ell\mathcal M(W,L_{\gamma_0}(X),\theta)$ is integrable, so $b_0$ is well defined, and
\[
\mathbb E\left[\|b_0(X,\theta)\|\int_0^1
\frac{|a'(\gamma_0^{-1}(s,X))|}
     {f_{Y|X}(\gamma_0^{-1}(s,X),X)}|K|(ds)\right]<\infty;
\]
(v) the representer in \Cref{prop:l-functionals} is square-integrable.
Here $f_{Y|X}(y,x)=\partial_y\gamma_0(y,x)$ at the quantiles where the derivative is assumed to exist. The path integrability requirements of \Cref{sec:FSIF} are maintained.
\end{assumption}

\begin{proposition}[Conditional $L$-functional correction]
\label{prop:l-functionals}
Under \Cref{ass:Lhadam-diff}, the FSIF is
\begin{equation}\label{eq:philstatisticsprime}
\phi_0(y,x,\theta)
=
\alpha_0(y,x,\theta)
-
\mathbb{E}[\alpha_0(Y,X,\theta)\mid X=x],
\end{equation}
where
\[
\begin{aligned}
\alpha_0(y,x,\theta)
={}&b_0(x,\theta)
\int_0^1
\frac{a'\left(\gamma_0^{-1}(s,x)\right)}
{f_{Y|X}\left(\gamma_0^{-1}(s,x),x\right)}
\\[-2pt]
&\quad\times
\left\{
s-\indic\left\{y\leq\gamma_0^{-1}(s,x)\right\}
\right\}K(ds).
\end{aligned}
\]
Moreover,
$\mathbb{E}[\alpha_0(Y,X,\theta)\mid X]=0$,
so $\phi_0(y,x,\theta)=\alpha_0(y,x,\theta)$.
\end{proposition}

The representer combines the sensitivity of the conditional
$L$-functional to changes in the quantiles with the sensitivity
of the identifying moment to changes in $L_{\gamma_0}(X)$.
For best linear approximations, $b_0(X,\theta)=p(X)$,
recovering the influence-function adjustment underlying
\citet{chetverikov2025weighted}. For the welfare-equivalent income in \Cref{ex:l-functionals}, assuming $u$ is continuously differentiable with strictly positive derivative at the relevant incomes,
\[
b_0(X,\theta)
=
\frac{1}{u'\left(u^{-1}\{L_{\gamma_0}(X)\}\right)}.
\]

Importantly, the proposition allows $K$ to have point masses
at interior quantiles, including individual conditional
quantiles and interquantile ranges. Such functionals generally
do not admit the square-integrable CDF-based representation
of \Cref{ass:linearizationcdf}, but are covered by the
conditional-measure linearization. The density conditions
above are sufficient for the quantile-based derivation, not
necessary for regularity of every $L$-functional.

\section{Local and Double Robustness}\label{sec:local&doublerobustness}

\subsection{Local robustness}\label{subsec:orthogonality}

Local robustness is a derivative property at the true nuisance values. Its domain must be specified: cancellation along admissible perturbations does not establish orthogonality in directions outside their span.
Let $\mathbb G$ be a normed space containing the CDF differences under consideration. Let $\mathcal H_0$ be a class of admissible perturbations with $H_X=F_X$, and define
\[
\mathcal T=\overline{\operatorname{span}}^{\|\cdot\|_{\mathbb G}}
\{H_{Y|X}-\gamma_0:H\in\mathcal H_0\}.
\]
For these paths, $\gamma_\tau=\gamma_0+\tau(H_{Y|X}-\gamma_0)$ exactly. Derivatives on $\mathcal T$ below mean continuous linear extensions of derivatives along admissible paths; arbitrary elements of $\mathcal T$ need not themselves define probability distributions.

\begin{lemma}[Conditional centering]\label{lemma:thm2CEINR}
Fix $\theta$ and let $\alpha$ be measurable with
\[
\mathbb{E}\left[\|\alpha(Y,X,\theta)\|\right]<\infty.
\]
Then
\begin{equation}\label{eq:zero-mean}
\mathbb{E}[\phi(W,\gamma_0,\alpha,\theta)]=0.
\end{equation}
\end{lemma}

\begin{theorem}[Local robustness]\label{thm:LR}
Suppose \Cref{ass:linearization} holds and, for each $\theta$, the map
$\gamma\mapsto\mathbb{E}[\psi(W,\gamma,\alpha_0,\theta)]$
admits a continuous linear derivative on $\mathcal T$ agreeing with its derivatives along $\mathcal H_0$. Then
\[
\partial_\gamma
\mathbb{E}[\psi(W,\gamma_0,\alpha_0,\theta)][h]=0,
\qquad h\in\mathcal T.
\]
The derivative with respect to the representer is also zero at $\gamma_0$ along every integrable representer direction.
\end{theorem}

The theorem follows by cancellation of the two first-order effects on each generating direction and continuity on their closed linear span. No equality between $\mathcal T$ and the ambient nuisance space is required. In particular, a restricted collection of perturbations need not generate every CDF difference that could arise from an estimator. The remainder conditions in \Cref{sec:asymptotictheory} are imposed on the actual fitted nuisances.

\subsection{Double robustness}\label{subsec:doublerobustness}

Double robustness is stronger than local robustness. For an admissible class $\mathcal G$ and an integrable representer class $\mathcal A$, it requires
\begin{equation}\label{eq:DR}
\mathbb{E}[\psi(W,\gamma_0,\alpha,\theta_0)]=0,
\qquad
\mathbb{E}[\psi(W,\gamma,\alpha_0,\theta_0)]=0,
\quad \gamma\in\mathcal G,\ \alpha\in\mathcal A.
\end{equation}
The first equality follows from conditional centering. The second requires an exact population identity, not merely a first-order expansion.

\begin{theorem}[Double robustness]\label{thm:doublerobustness}
Suppose that, for every $\gamma\in\mathcal G$ and $\theta\in\Theta$, all terms below are integrable and
\begin{equation}\label{eq:exact-linear-representation}
\mathbb{E}[g(W,\gamma,\theta)-g(W,\gamma_0,\theta)]
=
\mathbb{E}\left[
\int\alpha_0(y,X,\theta)(\gamma-\gamma_0)(dy,X)
\right].
\end{equation}
Then
\[
\mathbb{E}[\psi(W,\gamma,\alpha_0,\theta)]
=
\mathbb{E}[\psi(W,\gamma_0,\alpha,\theta)]
=
\mathbb{E}[g(W,\gamma_0,\theta)],
\]
for every integrable $\alpha\in\mathcal A$. In particular, \Cref{eq:DR} holds.
A sufficient CDF-based condition for \Cref{eq:exact-linear-representation} is
\begin{equation}\label{eq:exact-cdf-representation}
\mathbb{E}[g(W,\gamma,\theta)-g(W,\gamma_0,\theta)]
=
\mathbb{E}\left[
\delta_0(Y,X,\theta)(\gamma-\gamma_0)(Y,X)
\right],
\end{equation}
with $\delta_0\in L^2(P_{Y,X})$ and $\alpha_0$ as in \Cref{lemma:correctiontermprime}.
\end{theorem}

An affine population moment on a convex class satisfies the exact identity when its derivative representation holds along every segment joining $\gamma_0$ to a member of that class. Orthogonality on a smaller tangent space alone is insufficient. These population identities imply consistency under misspecification only when the corresponding sample moments converge uniformly and the limiting moment continues to identify $\theta_0$. They do not imply valid standard errors without additional assumptions.

\begin{examplecont}{ex:controlfunctions}[Control functions]
\begin{proposition}\label{prop:DRcontrolfunction}
Suppose
\[
\varrho(W,v,\theta)=h_0(W,\theta)+h_1(W,\theta)v,
\qquad
\mathbb{E}[\|h_0(W,\theta)\|]<\infty,
\quad
\mathbb{E}[\|h_1(W,\theta)\|^2]<\infty.
\]
Then \Cref{eq:DR} holds for every admissible CCDF $\gamma$ of $X$ given $Z$, with
\[
\delta_0(x,z,\theta)
=\mathbb{E}[h_1(W,\theta)\mid X=x,Z=z],
\quad
\alpha_0(x,z,\theta)
=\mathbb{E}[h_1(W,\theta)\indic\{x\leq X\}\mid Z=z].
\]
\end{proposition}
For a nonlinear dependence on the control $v$, local robustness still holds under \Cref{prop:controlfunction}, but double robustness requires additional structure.
\end{examplecont}

\begin{examplecont}{ex:counterfactuals}[Counterfactuals]
Write $F_\gamma^*(t)=\int\gamma(t,x)F_X^*(dx)$.
\begin{proposition}\label{prop:DRcounterfactuals}
Suppose \Cref{ass:abs-cont} holds and, for every admissible $\gamma$,
\[
\varphi(F_\gamma^*)-\varphi(F_Y^*)
=\int a_0(t)\{F_\gamma^*(t)-F_Y^*(t)\}\nu(dt),
\qquad \int|a_0|\,d|\nu|<\infty.
\]
Then \Cref{eq:DR} holds with the representer in \Cref{prop:functionalcounterfactuals}.
\end{proposition}
This identity holds for the bounded-support counterfactual mean discussed above. A counterfactual quantile is generally nonlinear and does not satisfy it.
\end{examplecont}

\begin{examplecont}{ex:rankrank}[Conditional rank-rank regression]
\begin{proposition}\label{prop:DRrankrank}
Fix $s\in\{c,p\}$ and keep $\gamma_{-s}=\gamma_{0,-s}$ and $\alpha_{-s}=\alpha_{0,-s}$. The corrected rank-rank moment has expectation zero at $\theta_0$ if either $\gamma_s=\gamma_{0,s}$ or $\alpha_s=\alpha_{0,s}$, with the other component arbitrary subject to integrability.
\end{proposition}
This is a componentwise property. With both representers correct but both CDFs potentially incorrect, the remaining bias is
\[
\mathbb{E}[\psi(W,\gamma,\alpha_0,\theta_0)]
=
\mathbb{E}\left[
(\gamma_c-\gamma_{0,c})(Y_c,X)
(\gamma_p-\gamma_{0,p})(Y_p,X)
\right],
\]
which need not vanish. Thus correctness of both representers alone does not imply double robustness with respect to unrestricted simultaneous misspecification of both CDFs.
\end{examplecont}

\begin{examplecont}{ex:l-functionals}[Conditional $L$-functionals]
\begin{remark}\label{rmk:DRl-functionals}
The inverse-CDF map and the transformation $\mathcal M$ are generally nonlinear, so there is no double-robustness claim for this entire class. Special choices can nevertheless yield an affine population moment. For example, with $a(y)=y$, $K(ds)=ds$, and $\mathcal M(W,L,\theta)=L-\theta$, the conditional $L$-functional is the conditional mean, and the measure representer $\alpha_0(y,x)=y$ gives exact cancellation whenever the relevant first moments are finite. In nonlinear cases, inference instead requires control of the population remainder.
\end{remark}
\end{examplecont}
\section{Estimation of the Representer}
\label{sec:representer-estimation}

The locally robust moment in \Cref{eq:moment} depends on the unknown representer $\alpha_0$, which must be estimated. We consider two approaches. The first replaces the unknown components of the analytical expressions derived above by consistent estimators. The second estimates the representer automatically, without requiring its analytical expression, by exploiting the Riesz representation of the first-step derivative. The latter approach extends Riesz regression methods developed for regression-function nuisances \citep{chernozhukov2022automatic,chernozhukov2024automatic} to CCDF first steps.

\subsection{Plug-in Estimation}
\label{subsec:representer-plugin}
\assumptionprefix{}

When the analytical expression of $\alpha_0$ is available, a plug-in estimator can be constructed by replacing its unknown components with suitable nonparametric or machine-learning estimates. For example, the counterfactual correction involves a density ratio, while the conditional $L$-functional correction generally involves conditional densities evaluated at quantiles. Consistency follows under appropriate convergence and stability conditions for these components. In particular, inverse-density terms require additional regularity to prevent small estimated denominators from generating large estimation errors.

Collect these components and their training-sample estimators in
$$
    \eta_0=(\eta_{1,0},\ldots,\eta_{J,0}),
    \qquad
    \widehat\eta_{-\ell}
    =(\widehat\eta_{1,-\ell},\ldots,\widehat\eta_{J,-\ell}),
$$
where $\widehat\eta_{-\ell}$ uses only observations outside $I_\ell$. Cross-fitting is specified in \Cref{subsec:estimation}.
Suppressing the parameter argument, write
$\alpha_0(y,x)=\alpha(y,x;\eta_0)$ and define
$$
    \widehat\alpha_{-\ell}(y,x)
    =\alpha(y,x;\widehat\eta_{-\ell}).
$$
When the CDF-based representer is available, define $\widehat\delta_{-\ell}$ analogously.
If a preliminary parameter estimate enters the analytical formula, include $\theta_0$ and $\widetilde\theta_{-\ell}$ in the corresponding collections. Each component is equipped with a norm $\|\cdot\|_j$ appropriate to its role in the formula. Population norms involving estimated functions hold the training sample fixed and integrate over an independent observation.

Consistency of the component estimators transfers to the representer under continuity of the analytical formula in these norms. \Cref{ass:representer-plugin} imposes a quantitative version of this requirement, allowing the component rates to be propagated to the representer. For formulas involving inverse densities, verifying this condition requires control of small estimated denominators. When $\alpha_0$ is recovered by conditional integration of $\delta_0$, additional conditions control the error from estimating the integrating law.

\begin{assumption}[Plug-in regularity]\label{ass:representer-plugin}
For each fold $\ell$, the following conditions hold, with $J$ fixed,
$C<\infty$, and $\kappa_j\in(0,1]$.
\begin{enumerate}
    \item[(i)]
    For a neighborhood $\mathcal U$ of $\eta_0$,
    \[
        \mathbb{P}(\widehat\eta_{-\ell}\in\mathcal U)\to1,
        \qquad
        \|\widehat\eta_{j,-\ell}-\eta_{j,0}\|_j
        =O_P(r_{j,n}),
        \qquad r_{j,n}\to0.
    \]

    \item[(ii)]
    For direct estimation of $\alpha_0$, its analytical formula
    satisfies, for every $\eta\in\mathcal U$,
    \[
        \|\alpha(\cdot,\cdot;\eta)-\alpha_0\|_2
        \le C\sum_{j=1}^J
        \|\eta_j-\eta_{j,0}\|_j^{\kappa_j}.
    \]
    Impose the corresponding bound for $\delta_0$ when it is
    estimated directly.

    \item[(iii)]
    When $\alpha_0$ is recovered from $\delta_0$ by conditional
    integration, replace the continuity condition for $\delta_0$
    by
    \[
        \left(
        \mathbb{E}_X\!\left[
            \sup_y\|\delta(y,X;\eta)-\delta_0(y,X)\|^2
        \right]
        \right)^{1/2}
        \le C\sum_{j=1}^J
        \|\eta_j-\eta_{j,0}\|_j^{\kappa_j}.
    \]
    Moreover, $\delta_0(\cdot,x)$ is componentwise right-continuous,
    \[
        B_0 := \sup_x
            \sup_y\|\delta_0(y,x)\| < \infty, \quad
        V_0 := \sup_x \sum_{k=1}^p
            \operatorname{TV}_{\mathbb R}
            \big(\delta_{0,k}(\cdot,x)\big)
        <\infty,
    \]
    and $\widehat\gamma_{-\ell}$ is a valid conditional CDF with
    \[
        \mathbb{E}_X\!\left[
            \sup_y
            |\widehat\gamma_{-\ell}(y,X)-\gamma_0(y,X)|^2
        \right]
        =O_P(r_{\gamma,n}^2),
        \qquad r_{\gamma,n}\to0.
    \]
    \item[(iv)]
    All plug-in expressions are measurable and well defined.
\end{enumerate}
\end{assumption}

The following proposition gives the resulting mean-square rates for direct substitution and for recovery through conditional integration.

\begin{proposition}[Plug-in estimation]\label{prop:representer-plugin}
Under \Cref{ass:representer-plugin}, direct plug-in estimators satisfy
\[
    \|\widehat\alpha_{-\ell}-\alpha_0\|_2^2
    =O_P\!\left(\sum_{j=1}^J r_{j,n}^{2\kappa_j}\right),
    \qquad
    \|\widehat\delta_{-\ell}-\delta_0\|_2^2
    =O_P\!\left(\sum_{j=1}^J r_{j,n}^{2\kappa_j}\right),
\]
whenever the corresponding analytical formula is available.
If instead
\[
    \widehat\alpha_{-\ell}(y,x)
    =
    \int_{[y,\infty)}
    \widehat\delta_{-\ell}(t,x)\widehat\gamma_{-\ell}(dt,x),
\]
with $\alpha_0$ given by the corresponding population integral, then
\[
    \|\widehat\alpha_{-\ell}-\alpha_0\|_2^2
    =
    O_P\!\left(
        \sum_{j=1}^J r_{j,n}^{2\kappa_j}
        +r_{\gamma,n}^2
    \right).
\]
\end{proposition}

The rate of a directly estimated representer depends on the accuracy
of its estimated components and the sensitivity of its analytical
expression. Their stochastic estimation errors are already contained
in the rates $r_{j,n}$. Recovery from $\widehat\delta_{-\ell}$ introduces
the additional conditional-CDF error through $r_{\gamma,n}$.
Direct estimation of $\alpha_0$ remains available when its analytical
formula exists without a square-integrable CDF-based representer.
For inference, the resulting rates must be combined with the score
continuity, interaction, and nonlinear-remainder conditions in
\Cref{sec:asymptotictheory}.

\subsection{Automatic Estimation via Riesz Regression}
\label{subsec:representer-automatic1}

We now develop automatic estimation based on the CDF linearization in \Cref{ass:linearizationcdf}. Unlike the conditional-measure approach, this representation identifies a linear functional on a space of CCDF perturbations, making Riesz regression applicable.
Estimating the representer directly can avoid explicit ratios of estimated nuisance quantities, which may amplify estimation error and destabilize the resulting locally robust estimator when the corresponding population denominators are close to zero. This is particularly relevant for conditional $L$-functionals whose analytical representers involve inverse conditional densities.

Let $\mathcal T_2$ be the closed linear span in $L^2(P_{Y,X})$ of the admissible CDF directions used for the derivative representation. It is the $L^2$ counterpart of the space in \Cref{subsec:orthogonality}; no equality with the full $L^2$ space is assumed. For this section, $\theta$ is fixed (at $\theta_0$ for the inference application), and $p$ is fixed. We assume that the derivative of the population moment admits a continuous linear extension to this space. For each component $j=1,\ldots,p$, define
\[
\Lambda_{\theta,j}(h)
=
\left.
\frac{d}{dt}
\mathbb{E}[g_j(W,\gamma_0+th,\theta)]
\right|_{t=0},
\qquad h\in\mathcal T_2,
\]
where the derivative is understood through this extension. Under \Cref{ass:linearizationcdf},
\begin{equation}\label{eq:automatic1}
\Lambda_{\theta,j}(h)
=
\mathbb{E}[\delta_{0,j}(Y,X,\theta)h(Y,X)],
\qquad h\in\mathcal T_2.
\end{equation}

Only the projection of $\delta_{0,j}$ onto $\mathcal T_2$ is identified by these restrictions. Let
\begin{equation}\label{eq:riesz-projection}
\delta_{\mathcal T_2,j}
=
\Pi_{\mathcal T_2}\delta_{0,j}.
\end{equation}
By the Riesz representation theorem, $\delta_{\mathcal T_2,j}$ is the unique element of $\mathcal T_2$ satisfying \Cref{eq:automatic1} for every $h\in\mathcal T_2$.

The key observation is that $\delta_{\mathcal T_2}$ can be estimated without knowing the analytical expression of $\delta_0$. For a fixed $\theta$, define
\[
\mathcal{Q}_\theta(\delta)
=
\mathbb{E}\left[
\sum_{j=1}^p
\left\{
\delta_j(Y,X)^2
-
2\partial_\gamma g_j(W,\gamma_0,\theta)[\delta_j]
\right\}
\right],
\]
where $\partial_\gamma g_j(W,\gamma_0,\theta)[h]$ denotes an integrable linear extension of the pointwise derivative. We assume its expectation equals $\Lambda_{\theta,j}(h)$ for the candidate directions; existence of a population derivative alone does not supply a computable pointwise extension. By \Cref{eq:automatic1},
\[
\mathcal{Q}_\theta(\delta)
=
\|\delta-\delta_{\mathcal T_2}\|_2^2
-
\|\delta_{\mathcal T_2}\|_2^2,
\qquad \delta\in\mathcal T_2^p.
\]
Consequently,
\[
\delta_{\mathcal T_2}
=
\arg\min_{\delta\in\mathcal T_2^p}
\mathcal{Q}_\theta(\delta).
\]
This characterization provides an automatic estimation criterion involving only the derivative of the identifying moment and evaluations of candidate representers.

For implementation, let $b_n(y,x)=(b_{1n}(y,x),\ldots,b_{K_n n}(y,x))'$ be a dictionary whose elements belong to $\mathcal T_2$, and approximate $\delta_{\mathcal T_2}$ by
\[
\delta(y,x)=R'b_n(y,x),
\qquad R\in\mathbb{R}^{K_n\times p}.
\]
For each evaluation fold $I_\ell$, define the training-sample matrices
\[
\widehat G
=
\mathbb{P}_{-\ell}[b_n(Y,X)b_n(Y,X)']
\]
and
\[
\widehat M_{kj}
=
\mathbb{P}_{-\ell}
\left[
\partial_\gamma
g_j(W,\widehat\gamma_{-\ell},
\widetilde\theta_{-\ell})[b_{kn}]
\right],
\]
where $\mathbb P_{-\ell}$ denotes the average over the training observations outside $I_\ell$ and $\widetilde\theta_{-\ell}$ estimates the fixed target $\theta$. The derivative extension is also assumed to be defined at the fitted nuisance values. The Riesz regression estimator is
\begin{equation}\label{eq:estimation-automatic1}
\widehat R
=
\arg\min_R
\left\{
\operatorname{tr}(R'\widehat G R)
-
2\operatorname{tr}(R'\widehat M)
+
\lambda_n\|R\|_F^2
\right\},
\end{equation}
and
\[
\widehat\delta(y,x)
=
\widehat R'b_n(y,x).
\]
The objective can be estimated on a subsample independent of that used to train $\widehat\gamma_{-\ell}$, or under suitable empirical-process conditions. This accommodates nonlinear identifying moments through their derivatives evaluated at preliminary nuisance estimates.

To recover the representer entering the FSIF, define the conditional integration operator
\[
(\mathcal{K}\delta)(y,x)
=
\int_{[y,\infty)}
\delta(t,x)\gamma_0(dt,x).
\]
The tail-integral construction gives a correction representer for the directions in $\mathcal T_2$,
\[
\alpha_{\mathcal T_2}
=
\mathcal{K}\delta_{\mathcal T_2}.
\]
We estimate it by
\begin{equation}\label{eq:alpha-automatic}
\widehat\alpha(y,x)
=
\int_{[y,\infty)}
\widehat\delta(t,x)\widehat\gamma(dt,x).
\end{equation}
Thus, automatic estimation proceeds by Riesz regression followed by integration with respect to the estimated conditional distribution.

The following result gives sufficient conditions for consistency. Let $\delta_n^*$ denote the $L^2(P_{Y,X})$ projection of $\delta_{\mathcal T_2}$ onto the sieve space generated by $b_n$.

\begin{proposition}[Automatic estimation via Riesz regression]
\label{prop:automatic1}
Suppose that:
(i) the derivative in \Cref{eq:automatic1} is continuous on $\mathcal T_2$;
(ii) the eigenvalues of $G=\mathbb{E}[b_n(Y,X)b_n(Y,X)']$ are uniformly bounded away from zero and infinity;
(iii) $\|\widehat G-G\|_{\mathrm{op}}=o_P(1)$; and
(iv) $\lambda_n\geq0$ and $\lambda_n=o(1)$. All fitted integration operators are assumed to be well defined on the fitted dictionary.

Then, with probability approaching one,
\[
\|\widehat\delta-\delta_{\mathcal T_2}\|_2
\leq
\|\delta_n^*-\delta_{\mathcal T_2}\|_2
+
C\left\{
\|\widehat M-M\|_F
+
\|\widehat G-G\|_{\mathrm{op}}
+
\lambda_n
\right\},
\]
where $C<\infty$ is independent of $n$ and may depend on the eigenvalue bounds and $\|\delta_{\mathcal T_2}\|_2$, and
\[
M_{kj}
=
\mathbb{E}\left[
\partial_\gamma g_j(W,\gamma_0,\theta)[b_{kn}]
\right].
\]
Moreover,
\[
\|\widehat\alpha-\alpha_{\mathcal T_2}\|_2
\leq
\|\widehat\delta-\delta_{\mathcal T_2}\|_2
+
\|(\widehat{\mathcal{K}}-\mathcal{K})
\widehat\delta\|_2,
\]
where $\widehat{\mathcal{K}}$ replaces $\gamma_0$ by $\widehat\gamma$ in $\mathcal{K}$.
Consequently, if the approximation error, the estimation errors in $\widehat M$ and $\widehat G$, and the conditional integration error all converge to zero, then
\[
\|\widehat\alpha-\alpha_{\mathcal T_2}\|_2=o_P(1).
\]
\end{proposition}

The additional conditional integration error is specific to recovering the CCDF correction from its CDF-based Riesz representer. In particular, consistency of $\widehat\delta$ alone does not guarantee consistency of $\widehat\alpha$. When $\delta_0\in\mathcal T_2$, the projected representer coincides with $\delta_0$ and $\alpha_{\mathcal T_2}=\alpha_0$. More generally, $\alpha_{\mathcal T_2}$ provides the first-step correction along the tangent directions in $\mathcal T_2$.

Finally, the automatic method requires a square-integrable CDF-based representer. This condition is stronger than the conditional-measure linearization and may fail for functionals involving individual conditional quantiles. Such cases remain covered by the analytical FSIF construction, but require a different automatic estimation argument.

\section{Asymptotic Theory}\label{sec:asymptotictheory}

We establish asymptotic linearity and valid inference for the cross-fitted estimator. The argument specializes the general locally robust GMM theory of \citet{chernozhukov2022locally}. The application-specific requirements concern continuity of the score and two population remainders: the nonlinear remainder from estimating the conditional distribution and the interaction between the conditional-distribution and representer errors. Cross-fitting controls centered sampling errors without requiring weak convergence of the first-step estimators or Donsker restrictions on their function classes. Appendix~\ref{subsec:lack-SE} discusses why consistency of a conditional-CDF estimator alone does not justify an empirical-process argument without cross-fitting.

\subsection{Cross-fitted estimation}\label{subsec:estimation}

Let $W_1,\ldots,W_n$ be i.i.d.\ with distribution $F_0$. Partition the observations independently of their values into a fixed number $L\geq2$ of folds $I_1,\ldots,I_L$, with $n_\ell=|I_\ell|$ and $n_\ell/n\to\pi_\ell\in(0,1)$. For each fold, estimate $\gamma_0$, $\alpha_0$, and, when needed, a preliminary parameter $\widetilde\theta_{-\ell}$ using only observations outside $I_\ell$. These estimates are measurable with respect to the training-sample sigma-field $\mathcal W_\ell^c$. All expectations and population norms involving fitted functions below hold the training sample fixed and integrate over an independent observation. All displayed conditional integrals are assumed well defined on the fitted nuisance class. Write $Pf=\int f(w)F_0(dw)$ and $\|f\|_2=(P\|f\|^2)^{1/2}$, using the relevant marginal distribution when $f$ depends on fewer variables.

To simplify notation, set
\[
\alpha_0(y,x):=\alpha_0(y,x,\theta_0),\qquad
\widehat a_\ell(y,x)=\widehat\alpha_{-\ell}(y,x,\widetilde\theta_{-\ell}).
\]
The preliminary estimate is held fixed when optimizing over $\theta$. Define
\[
\begin{split}
\widehat\psi_{i\ell}(\theta)
&=g(W_i,\widehat\gamma_{-\ell},\theta)
  +\widehat a_\ell(Y_i,X_i)
  -\int\widehat a_\ell(y,X_i)\widehat\gamma_{-\ell}(dy,X_i),\\
\widehat\psi(\theta)
&=\frac1n\sum_{\ell=1}^L\sum_{i\in I_\ell}\widehat\psi_{i\ell}(\theta).
\end{split}
\]
Write $\widehat\psi_\ell(W,\theta)$ for the corresponding score function on fold $\ell$. The locally robust GMM estimator is
\begin{equation}\label{eq:LRestimator}
\widehat\theta\in\arg\min_{\theta\in\Theta}
\widehat\psi(\theta)'\widehat\Upsilon_n\widehat\psi(\theta),
\end{equation}
where $\widehat\Upsilon_n$ is symmetric positive definite with probability approaching one, does not depend on $\theta$, and converges in probability to a symmetric positive definite matrix $\Upsilon$. A measurable minimizer is assumed to exist. The nuisance estimates are assumed to define admissible conditional probability measures and finite correction integrals with probability approaching one. The oracle score is
\[
\psi_0(W)=g(W,\gamma_0,\theta_0)+\alpha_0(Y,X)-\mathbb E[\alpha_0(Y,X)\mid X],
\qquad P\psi_0=0.
\]

\subsection{Nuisance remainders and asymptotic linearity}

For each fold, let $e_\ell=\widehat a_\ell-\alpha_0$ and define
\[
\begin{split}
u_\ell(W)&=g(W,\widehat\gamma_{-\ell},\theta_0)-g(W,\gamma_0,\theta_0),\\
d_\ell(X)&=\int \alpha_0(y,X)(\widehat\gamma_{-\ell}-\gamma_0)(dy,X),\\
c_\ell(Y,X)&=e_\ell(Y,X)-\int e_\ell(y,X)\gamma_0(dy,X),\\
\widehat\Delta_{-\ell}(X)&=-\int e_\ell(y,X)(\widehat\gamma_{-\ell}-\gamma_0)(dy,X).
\end{split}
\]
Here $Pc_\ell=0$ exactly from \Cref{lemma:thm2CEINR}, regardless of the fitted representer. The score difference therefore admits the exact decomposition
\begin{equation}\label{eq:score-decomposition}
\widehat\psi_\ell(W,\theta_0)-\psi_0(W)
=u_\ell(W)-d_\ell(X)+c_\ell(Y,X)+\widehat\Delta_{-\ell}(X).
\end{equation}
For an admissible conditional distribution $\gamma$, define its population linearization remainder by
\begin{equation}\label{eq:population-remainder}
\begin{split}
\mathcal R(\gamma)
&=P\{g(W,\gamma,\theta_0)-g(W,\gamma_0,\theta_0)\}
  -\mathbb E\!\left[\int \alpha_0(y,X)(\gamma-\gamma_0)(dy,X)\right]\\
&=P\psi(W,\gamma,\alpha_0,\theta_0).
\end{split}
\end{equation}
Consequently, the conditional population bias of the feasible score is
\begin{equation}\label{eq:conditional-score-bias}
P\widehat\psi_\ell(W,\theta_0)
=\mathcal R(\widehat\gamma_{-\ell})+P\widehat\Delta_{-\ell}.
\end{equation}
This identity separates the nonlinear first-step remainder from the interaction remainder without imposing a density representation on the estimated conditional law.

\assumptionprefix{AT}
\begin{assumption}[Mean-square continuity]\label{ass:meansquareconsistency}
For every fold,
\[
\text{(i) }\|u_\ell\|_2=o_P(1),\qquad
\text{(ii) }\|d_\ell\|_2=o_P(1),\qquad
\text{(iii) }\|c_\ell\|_2=o_P(1).
\]
\end{assumption}
Conditional centering is an $L^2$ contraction, so $\|e_\ell\|_2=o_P(1)$ suffices for part~(iii). This condition includes any effect of evaluating the representer at $\widetilde\theta_{-\ell}$. Neither raw $L^2$ convergence of the representer nor $L^2$ convergence of CDF evaluations alone guarantees convergence of integrals against the estimated conditional measure; part~(ii) and the following interaction condition address these distinct requirements.

\begin{assumption}[Interaction remainder]\label{ass:rateremainder}
For every fold, at least one of the following conditions holds:
\[
\begin{array}{ll}
\text{(i)}&\sqrt n\,\|P\widehat\Delta_{-\ell}\|=o_P(1),
\quad \|\widehat\Delta_{-\ell}\|_2=o_P(1);\\[3pt]
\text{(ii)}&\sqrt n\,P\|\widehat\Delta_{-\ell}\|=o_P(1);\\[3pt]
\text{(iii)}&n_\ell^{-1/2}\displaystyle\sum_{i\in I_\ell}
\widehat\Delta_{-\ell}(X_i)=o_P(1).
\end{array}
\]
\end{assumption}
Parts~(i) and (ii) are population sufficient conditions for part~(iii), by conditional Chebyshev's and Markov's inequalities, respectively. The empirical formulation also accommodates application-specific cancellations.

\begin{assumption}[Nonlinear remainder]\label{ass:rateandcontinuity}
For every fold, at least one of the following conditions holds:
(i) $\mathcal R(\gamma)=0$ on an admissible class containing $\widehat\gamma_{-\ell}$ with probability approaching one;
(ii) for an application-specific norm $\|\cdot\|_{\mathbb G}$,
\[
\|\mathcal R(\gamma)\|\leq C\|\gamma-\gamma_0\|_{\mathbb G}^2
\]
in a neighborhood of $\gamma_0$, and
$\|\widehat\gamma_{-\ell}-\gamma_0\|_{\mathbb G}=o_P(n^{-1/4})$;
or (iii) $\sqrt n\,\|\mathcal R(\widehat\gamma_{-\ell})\|=o_P(1)$ directly.
\end{assumption}
Part~(i) follows from an exact affine representation of the population moment, as in the double-robustness results in \Cref{subsec:doublerobustness}. Affinity together with orthogonality along every segment connecting $\gamma_0$ to an admissible candidate is sufficient. Orthogonality only along a restricted tangent space does not establish this identity for arbitrary candidates. Part~(ii) requires a quadratic remainder bound in addition to first-order differentiability. The general condition is part~(iii), which also permits products of errors in different conditional distributions.

\begin{proposition}[Asymptotic linearity of the sample moment]\label{prop:asymptoticlinearity}
Under \Cref{ass:meansquareconsistency,ass:rateremainder,ass:rateandcontinuity},
\begin{equation}\label{eq:asymptoticlinearity}
\sqrt n\,\widehat\psi(\theta_0)
=\frac1{\sqrt n}\sum_{i=1}^n\psi_0(W_i)+o_P(1).
\end{equation}
\end{proposition}
Conditional on each training sample, the centered sums of $u_\ell$, $d_\ell$, and $c_\ell$ are negligible by mean-square continuity. The two remainder assumptions control the remaining terms. Dependence between different training samples causes no difficulty because $L$ is fixed; independence across folds is not assumed. A proof is given in Appendix~\ref{app:asymptotictheory}.

\subsection{Asymptotic normality and variance estimation}

Let
\[
J_0=\mathbb E[\partial_{\theta'}g(W,\gamma_0,\theta_0)],\qquad
\widehat J_n(\theta)=\frac1n\sum_{\ell=1}^L\sum_{i\in I_\ell}
\partial_{\theta'}g(W_i,\widehat\gamma_{-\ell},\theta).
\]
The feasible Jacobian involves only $g$, since the correction is held fixed in $\theta$. The population correction has expectation zero at $\gamma_0$ for every $\theta$, therefore under interchange of differentiation and expectation it contributes zero to the population Jacobian as well. Define
\[
A_0=(J_0'\Upsilon J_0)^{-1}J_0'\Upsilon,
\qquad \Psi=P[\psi_0\psi_0'],\qquad V=A_0\Psi A_0'.
\]
The standard consistency and Jacobian conditions are collected in \Cref{ass:consistency,ass:jacobian} in the Appendix.

\begin{theorem}[Asymptotic normality]\label{thm:asymptoticnormality}
Suppose $\theta_0$ is an interior point of $\Theta$, $\widehat\theta\to_P\theta_0$, $P\|\psi_0\|^2<\infty$, and \Cref{ass:meansquareconsistency,ass:rateremainder,ass:rateandcontinuity,ass:jacobian} hold. Then
\begin{equation}\label{eq:parameter-linear-representation}
\sqrt n(\widehat\theta-\theta_0)
=-A_0\frac1{\sqrt n}\sum_{i=1}^n\psi_0(W_i)+o_P(1),
\end{equation}
and hence
\begin{equation}\label{eq:asymptoticnormality}
\sqrt n(\widehat\theta-\theta_0)\rightsquigarrow N(0,V).
\end{equation}
\end{theorem}
Consistency can be verified using \Cref{prop:consistency}. For moments affine in $\theta$ with a nonsingular constant Jacobian, it also follows directly from the explicit estimator and \Cref{eq:asymptoticlinearity}.

For variance estimation, use scores evaluated at the training-sample preliminary estimates,
\[
\widehat s_{i\ell}=\widehat\psi_{i\ell}(\widetilde\theta_{-\ell}),\qquad
\widehat\Psi=\frac1n\sum_{\ell=1}^L\sum_{i\in I_\ell}
\widehat s_{i\ell}\widehat s_{i\ell}',
\]
and set
\[
\widehat A=(\widehat J'\widehat\Upsilon_n\widehat J)^{-1}
\widehat J'\widehat\Upsilon_n,
\qquad \widehat J=\widehat J_n(\widehat\theta),\qquad
\widehat V=\widehat A\widehat\Psi\widehat A'.
\]

\begin{assumption}[Variance continuity]\label{ass:varconsistency}
For every fold,
\[
\|g(W,\widehat\gamma_{-\ell},\widetilde\theta_{-\ell})
-g(W,\widehat\gamma_{-\ell},\theta_0)\|_2=o_P(1),
\qquad \|\widehat\Delta_{-\ell}\|_2=o_P(1).
\]
\end{assumption}

\begin{proposition}[Variance estimation]\label{prop:variance-consistency}
Under the conditions of \Cref{thm:asymptoticnormality} and \Cref{ass:varconsistency}, $\widehat\Psi\to_P\Psi$ and $\widehat V\to_P V$.
\end{proposition}
The asymptotic covariance of $\widehat\theta$ is estimated by $\widehat V/n$. Wald inference for a contrast requires its limiting variance to be positive. If $\Psi$ is nonsingular, $\Upsilon=\Psi^{-1}$ minimizes the asymptotic variance among GMM estimators based on these moments. In just-identified models the weighting matrix does not affect the first-order variance.

These results impose convergence to the nuisance values used in the oracle score. Double robustness can preserve consistency when one nuisance converges to an incorrect limit, but it does not by itself establish \Cref{eq:parameter-linear-representation} at that limit. Inference under such misspecification requires a separate expansion.

\begin{remark}[Computed scores]\label{rmk:computed-score}
The formulas above use exact conditional integrals. Suppose an implementation changes the fold score by a training-measurable function $\varepsilon_\ell(W)$ that is constant in $\theta$, as for a numerical approximation of the fixed correction. For a consistent implemented estimator, the asymptotic linearity and variance conclusions are unchanged if
\[
\frac1{\sqrt n}\sum_{\ell=1}^L\sum_{i\in I_\ell}\varepsilon_\ell(W_i)=o_P(1),
\qquad
\frac1n\sum_{\ell=1}^L\sum_{i\in I_\ell}\|\varepsilon_\ell(W_i)\|^2=o_P(1).
\]
The first condition preserves asymptotic linearity and the second preserves the score covariance estimate. These conditions also need verification when a diagonal-removal or other centering modification changes the theoretical score. They do not follow from outer cross-fitting alone.
\end{remark}

\subsection{Sufficient rate conditions and automatic estimation}

The interaction condition is expressed in terms of conditional measures because an $L^2$ error in CDF values does not generally control integration against the corresponding signed measure. One sufficient bound is
\begin{equation}\label{eq:interaction-TV-bound}
P\|\widehat\Delta_{-\ell}\|
\leq\mathbb E\!\left[\int\|e_\ell(y,X)\|
|\widehat\gamma_{-\ell}-\gamma_0|(dy,X)\right].
\end{equation}
If both conditional laws have densities $\widehat f_{-\ell}$ and $f_0$ relative to a common fixed measure $\mu$, Cauchy--Schwarz gives
\[
P\|\widehat\Delta_{-\ell}\|
\leq\|e_\ell\|_{L^2(P_X\otimes\mu)}
\|\widehat f_{-\ell}-f_0\|_{L^2(P_X\otimes\mu)}.
\]
Thus a product of these two errors of order $o_P(n^{-1/2})$ suffices for \Cref{ass:rateremainder}(ii). This is an optional sufficient condition, not a requirement that conditional densities be estimated. It does not replace the separate mean-square condition used for variance estimation.

For the automatic estimator of \Cref{sec:representer-estimation}, the tail-integral structure gives a more direct connection to CDF errors. Take $\alpha_0=\mathcal K\delta_{\mathcal T_2}$ and
$\widehat a_\ell=\widehat{\mathcal K}_\ell\widehat\delta_\ell$, suppressing the preliminary parameter in the notation. Define
\[
\begin{split}
h_\ell&=\widehat\gamma_{-\ell}-\gamma_0,
\qquad \eta_\ell=(\widehat{\mathcal K}_\ell-\mathcal K)\widehat\delta_\ell,\\
q_\ell(X)&=\int\eta_\ell(y,X)h_\ell(dy,X),\\
r_{\gamma,\ell}&=\|h_\ell\|_2,
\qquad r_{\delta,\ell}=\|\widehat\delta_\ell-\delta_{\mathcal T_2}\|_2.
\end{split}
\]
Here $h_\ell(y,x)$ denotes the CDF difference and $h_\ell(dy,x)$ its induced signed measure. Fubini's theorem yields
\begin{equation}\label{eq:automatic-interaction}
\widehat\Delta_{-\ell}(X)
=-\mathbb E[(\widehat\delta_\ell-\delta_{\mathcal T_2})(Y,X)
h_\ell(Y,X)\mid X]-q_\ell(X).
\end{equation}
Consequently,
\begin{equation}\label{eq:automatic-rate-bounds}
\begin{split}
\|e_\ell\|_2&\leq r_{\delta,\ell}+\|\eta_\ell\|_2,\\
\|P\widehat\Delta_{-\ell}\|&\leq
r_{\delta,\ell}r_{\gamma,\ell}+P\|q_\ell\|,\\
\|\widehat\Delta_{-\ell}\|_2&\leq r_{\delta,\ell}+\|q_\ell\|_2.
\end{split}
\end{equation}
The last inequality uses $|h_\ell|\leq1$. Hence representer consistency, $\|\eta_\ell\|_2=o_P(1)$, $\|q_\ell\|_2=o_P(1)$, and
\[
\sqrt n\{r_{\delta,\ell}r_{\gamma,\ell}+P\|q_\ell\|\}=o_P(1)
\]
verify the representer and interaction requirements. The remaining conditions on $u_\ell$, $d_\ell$, and $\mathcal R$ must also hold. In particular, consistency of Riesz regression alone is insufficient: the estimated integration operator contributes $q_\ell$.

When only the projection $\delta_{\mathcal T_2}$ is identified, the remainder in \Cref{eq:population-remainder} is defined using $\alpha_0=\mathcal K\delta_{\mathcal T_2}$. Relative to an unprojected representer $\delta_0$, it contains the additional term
$P[(\delta_0-\delta_{\mathcal T_2})h_\ell]$. This term vanishes when $h_\ell\in\mathcal T_2$; otherwise its size must be included in the nonlinear-remainder condition. The rate requirements thus apply to the directions actually generated by the first-step estimator.

\subsection{Applications}

The following conditions specialize the two remainder requirements to the examples. All sampling and weighting-matrix conditions in \Cref{subsec:estimation} are maintained. All fitted functions below are computed outside their evaluation fold. Whenever the representer depends on $\theta$, its fitted value below is evaluated at the training-sample preliminary estimate, as in $\widehat a_\ell$. The interaction $\widehat\Delta_{-\ell}$ is defined as above, with the appropriate outcome and conditioning variables, and is summed over components when several conditional distributions enter the moment.

\begin{examplecont}{ex:controlfunctions}[Control functions]
Consider the scalar moment
\[
g(W,\gamma,\beta)=h(W,\gamma(X,Z))-\beta,
\]
where $h$ is a known function,
$\gamma_0=F_{X|Z}$, and
$\beta_*=\mathbb E[h(W,\gamma_0(X,Z))]$.
We use the correction in \Cref{prop:controlfunction}.
The assumptions and proposition below concern this
two-step scalar specialization.

For the stacked parametric response model in
\Cref{eq:cf-parametric-stack},
\Cref{thm:asymptoticnormality} applies to the full parameter
$\theta=(\eta',\beta)'$, subject to its stated conditions.
The corresponding vector correction and full GMM Jacobian
account jointly for conditional-CDF estimation and estimation
of the finite-dimensional response coefficients.
\assumptionprefix{ACF}
\begin{assumption}\label{ass:ANcontrolfunction}
For every fold: (i) $\|\widehat\gamma_{-\ell}-\gamma_0\|_{L^2(P_{X,Z})}=o_P(1)$,
$\|h(W,\widehat\gamma_{-\ell}(X,Z))-h(W,\gamma_0(X,Z))\|_2=o_P(1)$, and $\|\widehat\alpha_{-\ell}-\alpha_0\|_{L^2(P_{X,Z})}=o_P(1)$;
(ii) either $\sqrt n|\mathcal R(\widehat\gamma_{-\ell})|=o_P(1)$ directly, or the exact representation in \Cref{prop:DRcontrolfunction} holds;
(iii) $n_\ell^{-1/2}\sum_{i\in I_\ell}\widehat\Delta_{-\ell}(Z_i)=o_P(1)$.
For variance estimation, also assume (iv) $\widetilde\beta_{-\ell}\to_P\beta_*$ and (v) $\|\widehat\Delta_{-\ell}\|_2=o_P(1)$.
\end{assumption}
A sufficient condition for part~(ii) is $\|\widehat\gamma_{-\ell}-\gamma_0\|_2=o_P(n^{-1/4})$ and twice continuous differentiability of $h$ in its second argument with
\[
\mathbb E[\sup_{v\in[0,1]}|\partial_v^2h(W,v)|\mid X,Z]\leq C.
\]
Taylor expansion then gives $|\mathcal R(\gamma)|\leq(C/2)\|\gamma-\gamma_0\|_2^2$. The correction-continuity condition follows from the tail representation of $\alpha_0$, square-integrability of its CDF-based representer, and the boundedness of CDF differences.

\begin{proposition}[Control functions]\label{prop:ANcontrolfunction}
Under the conditions of \Cref{prop:controlfunction} for this scalar moment, \Cref{ass:ANcontrolfunction}(i)--(iii), $P\psi_0^2<\infty$, and $\beta_*$ interior to $\Theta$, the locally robust estimator is consistent and
$\sqrt n(\widehat\beta-\beta_*)\rightsquigarrow N(0,P\psi_0^2)$.
Under parts~(iv)--(v), $\widehat V\to_P P\psi_0^2$.
\end{proposition}
The Jacobian is $-1$. For the general control-function moment $\varrho(W,\gamma(X,Z),\theta)$, \Cref{thm:asymptoticnormality} applies with its corresponding Jacobian and remainder.
\end{examplecont}

\begin{examplecont}{ex:counterfactuals}[Counterfactuals]
Let $F_\gamma^*(t)=\int\gamma(t,x)F_X^*(dx)$ and
$L_0[\eta]=\int a_0(t)\eta(t)\nu(dt)$, where $a_0(t)$ is the derivative weight in \Cref{ass:Chadam-diff}. Since $\int|a_0|\,d|\nu|<\infty$, this linear functional is defined on finite CDF differences, including differences outside the original Hadamard tangent space. With the representer in \Cref{prop:functionalcounterfactuals},
\[
\mathcal R(\gamma)
=\varphi(F_\gamma^*)-\varphi(F_Y^*)-L_0[F_\gamma^*-F_Y^*].
\]
\assumptionprefix{AC}
\begin{assumption}\label{ass:ANcounterfactuals}
For every fold: (i) $|\varphi(F_{\widehat\gamma_{-\ell}}^*)-\varphi(F_Y^*)|=o_P(1)$,
$\|\widehat\alpha_{-\ell}-\alpha_0\|_2=o_P(1)$, and
\[
\mathbb E\!\left[r_0(X)^2
\left|\int a_0(t)(\widehat\gamma_{-\ell}-\gamma_0)(t,X)\nu(dt)\right|^2\right]=o_P(1);
\]
(ii) $\sqrt n|\mathcal R(\widehat\gamma_{-\ell})|=o_P(1)$;
(iii) $n_\ell^{-1/2}\sum_{i\in I_\ell}\widehat\Delta_{-\ell}(X_i)=o_P(1)$.
For variance estimation, also assume (iv) $\widetilde\theta_{-\ell}\to_P\theta_0$ and (v) $\|\widehat\Delta_{-\ell}\|_2=o_P(1)$.
\end{assumption}
The weighted condition in part~(i) follows, for example, from
$\mathbb E[r_0(X)^2\|\widehat\gamma_{-\ell}(\cdot,X)-\gamma_0(\cdot,X)\|_\infty^2]=o_P(1)$.
Part~(ii) follows from a quadratic expansion of $\varphi$ and
$\|F_{\widehat\gamma_{-\ell}}^*-F_Y^*\|_{\mathbb B}=o_P(n^{-1/4})$; it holds exactly under \Cref{prop:DRcounterfactuals}. Hadamard differentiability alone does not imply a quadratic remainder.

\begin{proposition}[Counterfactuals]\label{prop:ANcounterfactuals}
Under the conditions of \Cref{prop:functionalcounterfactuals}, \Cref{ass:ANcounterfactuals}(i)--(iii), $P\psi_0^2<\infty$, and $\theta_0$ interior to $\Theta$, the estimator is consistent and
$\sqrt n(\widehat\theta-\theta_0)\rightsquigarrow N(0,P\psi_0^2)$.
Under parts~(iv)--(v), $\widehat V\to_P P\psi_0^2$.
\end{proposition}
This result treats $F_X^*$ as fixed, as in the definition of the counterfactual moment. Its estimation would require the corresponding additional influence-function component.
\end{examplecont}

\begin{examplecont}{ex:rankrank}[Conditional rank-rank regression]
Let $h_{s,\ell}=\widehat\gamma_{s,-\ell}-\gamma_{0,s}$ for $s\in\{c,p\}$. For the moment in \Cref{ex:rankrank} and the representers in \Cref{prop:rankrank}, direct expansion gives
\begin{equation}\label{eq:rankrank-remainder}
\mathcal R(\widehat\gamma_{-\ell})
=\mathbb E[h_{c,\ell}(Y_c,X)h_{p,\ell}(Y_p,X)].
\end{equation}
Thus the nonlinear bias is a product of the two CDF errors.
\assumptionprefix{ACRR}
\begin{assumption}\label{ass:ANconditionalRRR}
For every fold: (i) for $s\in\{c,p\}$,
\[
\|h_{s,\ell}\|_{L^2(P_{Y_s,X})}=o_P(1),\qquad
\|\widehat\alpha_{s,-\ell}-\alpha_{0,s}\|_{L^2(P_{Y_s,X})}=o_P(1);
\]
(ii) $\sqrt n\,\|h_{c,\ell}\|_2\|h_{p,\ell}\|_2=o_P(1)$;
(iii) $n_\ell^{-1/2}\sum_{i\in I_\ell}\widehat\Delta_{-\ell}(X_i)=o_P(1)$, where $\widehat\Delta_{-\ell}=\widehat\Delta_{c,-\ell}+\widehat\Delta_{p,-\ell}$.
For variance estimation, also assume (iv) $\widetilde\theta_{-\ell}\to_P\theta_0$ and (v) $\|\widehat\Delta_{-\ell}\|_2=o_P(1)$.
\end{assumption}
Bounded CDFs and bounded CDF-based representers imply the required mean-square continuity. Cauchy--Schwarz bounds \Cref{eq:rankrank-remainder} by the product in part~(ii), allowing different convergence rates for the two CDFs. Part~(iii) separately controls estimation of their correction representers and can be verified using the bounds in the preceding subsection.

\begin{proposition}[Conditional rank-rank regression]\label{prop:ANrank-rank}
In the setting of \Cref{ex:rankrank}, under \Cref{ass:ANconditionalRRR}(i)--(iii) and $\theta_0$ interior to $\Theta$, the estimator is consistent and
\[
\sqrt n(\widehat\theta-\theta_0)\rightsquigarrow N(0,144P\psi_0^2).
\]
Under parts~(iv)--(v), $\widehat V\to_P144P\psi_0^2$.
\end{proposition}
The factor $144$ follows from $J_0=-1/12$. Square-integrability of the oracle score holds because the conditional ranks and their representers are bounded.
\end{examplecont}

\begin{examplecont}{ex:l-functionals}[Transformations of conditional $L$-functionals]
Retain the general moment
$g(W,\gamma,\theta)=\mathcal M(W,L_\gamma(X),\theta)$
and the representer in \Cref{prop:l-functionals}. Put $L_0=L_{\gamma_0}$ and $\widehat L_\ell=L_{\widehat\gamma_{-\ell}}$. The relevant nonlinear remainder is
\[
\begin{split}
\mathcal R(\gamma)
={}&P\{\mathcal M(W,L_\gamma(X),\theta_0)
-\mathcal M(W,L_0(X),\theta_0)\}\\
&-\mathbb E\!\left[\int\alpha_0(y,X,\theta_0)(\gamma-\gamma_0)(dy,X)\right].
\end{split}
\]
It accounts for nonlinearity of both the inverse-CDF map and $\mathcal M$.
\assumptionprefix{AL}
\begin{assumption}\label{ass:ANl-functionals}
For every fold: (i)
\[
\begin{split}
\|\mathcal M(W,\widehat L_\ell(X),\theta_0)-\mathcal M(W,L_0(X),\theta_0)\|_2&=o_P(1),\\
\left\|\int\alpha_0(y,X,\theta_0)(\widehat\gamma_{-\ell}-\gamma_0)(dy,X)\right\|_2&=o_P(1),\\
\|e_\ell\|_2&=o_P(1);
\end{split}
\]
(ii) $\sqrt n\,\|\mathcal R(\widehat\gamma_{-\ell})\|=o_P(1)$;
(iii) $n_\ell^{-1/2}\sum_{i\in I_\ell}\widehat\Delta_{-\ell}(X_i)=o_P(1)$.
For variance estimation, also assume (iv)
$\|\mathcal M(W,\widehat L_\ell(X),\widetilde\theta_{-\ell})-\mathcal M(W,\widehat L_\ell(X),\theta_0)\|_2=o_P(1)$ and (v) $\|\widehat\Delta_{-\ell}\|_2=o_P(1)$.
\end{assumption}
Part~(ii) follows from an application-specific quadratic bound on the displayed remainder and an $o_P(n^{-1/4})$ rate in its associated nuisance norm. The first-order conditions in \Cref{ass:Lhadam-diff} do not alone supply this bound. No general affine or double-robustness alternative is imposed for this nonlinear class.

\begin{proposition}[Transformations of conditional $L$-functionals]\label{prop:ANl-functionals}
Under \Cref{ass:Lhadam-diff,ass:jacobian}, \Cref{ass:ANl-functionals}(i)--(iii), the sampling and weighting-matrix conditions of \Cref{subsec:estimation}, $\widehat\theta\to_P\theta_0$, $\theta_0$ interior to $\Theta$, and $P\|\psi_0\|^2<\infty$, \Cref{eq:parameter-linear-representation,eq:asymptoticnormality} hold with $V=A_0\Psi A_0'$. Under parts~(iv)--(v), $\widehat V\to_P V$.
\end{proposition}
For $\mathcal M(W,L,\theta)=L-\theta$, this reduces to $J_0=-1$ and $V=P\psi_0^2$. For the best linear approximation,
$\mathcal M(W,L,\theta)=p(X)\{L-p(X)'\theta\}$,
the Jacobian is $-\mathbb E[p(X)p(X)']$, assumed nonsingular. These cases require the corresponding remainder and moment conditions; the Jacobian is not generally $-1$ for transformations or regressions of conditional $L$-functionals.
\end{examplecont}

\section{Simulation Study}\label{sec:simulation}

We illustrate the estimator using conditional rank-rank regression. The main experiment uses XGBoost first steps in a nonlinear, heteroskedastic location--scale design, in the spirit of \citet{wang2026conditional}. A separate Gaussian experiment in Appendix~\ref{app:simulDRCRRR} studies componentwise double robustness by varying the child nuisance specifications while keeping the parental specifications correct.

\subsection{Design and implementation}

The displayed data-generating family has independent covariates
\[
X_j\sim U[-1,1],\qquad j=1,\ldots,5.
\]
Independently of $X$, let
\[
\begin{pmatrix}\varepsilon_c\\\varepsilon_p\end{pmatrix}
\sim N\left(\begin{pmatrix}0\\0\end{pmatrix},
\begin{pmatrix}1&c\\c&1\end{pmatrix}\right),
\qquad c\in\{0.25,0.50,0.75\}.
\]
For $s\in\{c,p\}$, set
\[
Y_s=\mu_s(X)+\sigma_s(X)g_s(\varepsilon_s),
\]
where $g_s$ is the identity map on $\mathbb R$ and the location and scale functions are given by
\[
\begin{split}
\mu_c(X) & = 1.5\sin\left(\pi X_1X_2\right) + 0.5\left( X_3^2-X_4^2 \right),\allowdisplaybreaks\\[3pt]
    \mu_p(X) & = 1.2\cos\left(\pi X_1\right) + 0.6 X_2X_3,\allowdisplaybreaks\\[3pt]
\sigma_c(X) & = \exp\left( 0.1 + 0.18X_4 + 0.15X_1X_5 \right),\allowdisplaybreaks\\[3pt]
    \sigma_p(X) & = \exp\left( 0.08 + 0.15X_3 - 0.12X_4X_5 \right).
\end{split}
\]

Let $\Phi$ denote the standard normal CDF. Strict monotonicity of $g_s$, positivity of $\sigma_s$, and independence of the latent errors from $X$ imply
\[
F_{Y_s|X}(Y_s\mid X)=\Phi(\varepsilon_s).
\]
Thus the conditional ranks have a Gaussian copula with correlation parameter $c$, and the target coefficient is
\[
\theta_0(c)=\frac6\pi\arcsin(c/2).
\]
This identity does not require the conditional outcome distributions themselves to be Gaussian.

For each outcome, boosted regression trees estimate a conditional location and a positive scale. These define the Gaussian working conditional distribution
\[
\widehat\gamma_s(y,x)
=\Phi\left(\frac{y-\widehat\mu_s(x)}{\widehat\sigma_s(x)}\right).
\]

Since $g_s(z)=z$ in this design, the Gaussian conditional-distribution
family is correctly specified, with location and scale varying
nonlinearly with $X$.

We compare shared-sample nuisance estimation with separate-sample nuisance
estimation, following the double cross-fitting approach of
\citet{newey2018cross}. Both procedures use three-fold outer cross-fitting. Let $I_\ell$ denote an evaluation fold and $I_\ell^c$ its training sample. For each generation, XGBoost estimates the conditional location and scale. A threshold-augmented XGBoost regression estimates $\alpha_{0,s}$ using $(y,X)$ as predictors and the estimated conditional rank of the other generation multiplied by $\mathbf{1}\{y\leq Y_s\}$ as the response. These pseudo-ranks are constructed by inner cross-fitting. All preliminary estimates and pseudo-ranks used to estimate a representer are learned within the training group assigned to that representer.

For a training group $G$, write $\widehat\gamma_{s,G}$ and $\widehat\alpha_{s,G}$ for the corresponding nuisance estimates. Given training groups $G$ and $H$, define the score contribution
\[
\begin{split}
\widehat q_{i\ell}(G,H)
={}&
\widehat\gamma_{c,G}(Y_{ci},X_i)
\widehat\gamma_{p,H}(Y_{pi},X_i)\\
&+\widehat\alpha_{c,H}(Y_{ci},X_i)
-\int \widehat\alpha_{c,H}(y,X_i)
      \widehat\gamma_{c,G}(dy,X_i)\\
&+\widehat\alpha_{p,G}(Y_{pi},X_i)
-\int \widehat\alpha_{p,G}(y,X_i)
      \widehat\gamma_{p,H}(dy,X_i),
\qquad i\in I_\ell.
\end{split}
\]
The conditional centering integrals are evaluated under the corresponding fitted Gaussian laws.
Under shared-sample estimation, all nuisance functions use the full outer training sample, giving
\[
\widehat q_{i\ell}^{\mathrm{sh}}
=
\widehat q_{i\ell}(I_\ell^c,I_\ell^c).
\]
Under separate-sample estimation, we partition $I_\ell^c$ into two disjoint groups, $A_\ell$ and $B_\ell$. The child CDF and parental representer are estimated within $A_\ell$, while the parental CDF and child representer are estimated within $B_\ell$. This allocation uses separate observations to estimate the two CDFs and to estimate each CDF and its own representer. We also reverse the allocation and average the resulting scores observation by observation:
\[
\widehat q_{i\ell}^{\mathrm{sep}}
=
\frac{1}{2}
\left\{
\widehat q_{i\ell}(A_\ell,B_\ell)
+
\widehat q_{i\ell}(B_\ell,A_\ell)
\right\}.
\]

For $k\in\{\mathrm{sh},\mathrm{sep}\}$, the locally robust estimator is
\[
\widehat\theta^{LR,k}
=
\frac{12}{n}
\sum_{\ell=1}^{L}\sum_{i\in I_\ell}
\widehat q_{i\ell}^{k}
-3,
\qquad L=3.
\]
Standard errors use the score-based variance estimator described above, with training-sample preliminary estimates of $\theta_0$. For separate-sample estimation, the two score contributions are averaged before computing their contribution to the variance estimate. The comparison is paired across Monte Carlo replications, using the same simulated observations and outer folds for both procedures, with nuisance-learning settings held fixed. The plug-in estimator omits the two correction terms. The reported experiment uses $n\in\{500,1000,5000\}$, $c\in\{0.25,0.50,0.75\}$, and 300 replications per design point. Errors resulting from integration must be small enough to preserve the estimator’s first-order asymptotic distribution and the consistency of its variance estimator, as specified in \Cref{rmk:computed-score}; cross-fitting alone does not guarantee this accuracy.

\subsection{Reported statistics and results}

We report bias, empirical standard deviation, average estimated standard error, root mean squared error (RMSE), and coverage of nominal $95\%$ confidence intervals. The plug-in standard error omits first-step uncertainty, so its coverage is a descriptive benchmark. Appendix~\ref{app:diagCRRR} reports nuisance-error, interaction-remainder, asymptotic-linearity, and variance diagnostics. 

\begin{sidewaystable}[p]
\centering
\begin{threeparttable}
\caption{Conditional rank--rank regression: shared- and separate-sample nuisance estimation}
\label{tab:MLCRRR_main}
\small
\setlength{\tabcolsep}{3pt}
\renewcommand{\arraystretch}{1.05}
\begin{tabular}{@{}rrrl*{10}{c}@{}}
\toprule
\multicolumn{4}{c}{}
& \multicolumn{5}{c}{Locally robust}
& \multicolumn{5}{c}{Plug-in} \\
\cmidrule(lr){5-9}\cmidrule(l){10-14}
$n$ & $c$ & $\theta_0$ & Training
& Bias & Emp.\ SD & Avg.\ SE & RMSE & Cov.\ 95\%
& Bias & Emp.\ SD & Avg.\ SE & RMSE & Cov.\ 95\% \\
\midrule
500 & 0.25 & 0.2394 & Shared
& 0.0533 & 0.0805 & 0.1003 & 0.0964 & 0.977
& 0.0710 & 0.1020 & 0.1685 & 0.1241 & 0.997 \\
& & & Separate
& 0.0239 & 0.0770 & 0.0993 & 0.0805 & 0.987
& 0.0244 & 0.0985 & 0.1661 & 0.1013 & 1.000 \\
\addlinespace[2pt]
500 & 0.50 & 0.4826 & Shared
& 0.0908 & 0.0809 & 0.1001 & 0.1215 & 0.903
& 0.1680 & 0.0958 & 0.1790 & 0.1934 & 0.960 \\
& & & Separate
& 0.0209 & 0.0742 & 0.1003 & 0.0770 & 0.983
& 0.0803 & 0.0902 & 0.1764 & 0.1207 & 1.000 \\
\addlinespace[2pt]
500 & 0.75 & 0.7341 & Shared
& 0.1427 & 0.0699 & 0.0987 & 0.1588 & 0.790
& 0.2713 & 0.0925 & 0.1886 & 0.2866 & 0.860 \\
& & & Separate
& 0.0219 & 0.0670 & 0.1005 & 0.0704 & 1.000
& 0.1399 & 0.0924 & 0.1865 & 0.1676 & 0.993 \\
\addlinespace[5pt]
1000 & 0.25 & 0.2394 & Shared
& 0.0325 & 0.0523 & 0.0618 & 0.0616 & 0.953
& 0.0565 & 0.0598 & 0.1120 & 0.0822 & 0.997 \\
& & & Separate
& 0.0179 & 0.0490 & 0.0643 & 0.0521 & 0.980
& 0.0375 & 0.0622 & 0.1134 & 0.0725 & 1.000 \\
\addlinespace[2pt]
1000 & 0.50 & 0.4826 & Shared
& 0.0691 & 0.0471 & 0.0615 & 0.0836 & 0.863
& 0.1396 & 0.0573 & 0.1197 & 0.1508 & 0.950 \\
& & & Separate
& 0.0270 & 0.0460 & 0.0647 & 0.0532 & 0.980
& 0.1028 & 0.0593 & 0.1211 & 0.1186 & 0.987 \\
\addlinespace[2pt]
1000 & 0.75 & 0.7341 & Shared
& 0.1002 & 0.0443 & 0.0604 & 0.1095 & 0.637
& 0.2136 & 0.0583 & 0.1266 & 0.2214 & 0.700 \\
& & & Separate
& 0.0321 & 0.0460 & 0.0645 & 0.0560 & 0.980
& 0.1519 & 0.0577 & 0.1280 & 0.1625 & 0.953 \\
\addlinespace[5pt]
5000 & 0.25 & 0.2394 & Shared
& 0.0088 & 0.0180 & 0.0210 & 0.0200 & 0.973
& 0.0273 & 0.0204 & 0.0447 & 0.0341 & 1.000 \\
& & & Separate
& 0.0042 & 0.0178 & 0.0222 & 0.0183 & 0.980
& 0.0281 & 0.0221 & 0.0457 & 0.0358 & 1.000 \\
\addlinespace[2pt]
5000 & 0.50 & 0.4826 & Shared
& 0.0170 & 0.0152 & 0.0206 & 0.0228 & 0.943
& 0.0552 & 0.0196 & 0.0480 & 0.0586 & 0.980 \\
& & & Separate
& 0.0064 & 0.0159 & 0.0219 & 0.0171 & 0.990
& 0.0585 & 0.0204 & 0.0490 & 0.0619 & 0.973 \\
\addlinespace[2pt]
5000 & 0.75 & 0.7341 & Shared
& 0.0319 & 0.0128 & 0.0197 & 0.0344 & 0.710
& 0.0914 & 0.0173 & 0.0512 & 0.0930 & 0.697 \\
& & & Separate
& 0.0150 & 0.0136 & 0.0213 & 0.0202 & 0.977
& 0.0953 & 0.0186 & 0.0523 & 0.0971 & 0.647 \\
\bottomrule
\end{tabular}

\begin{tablenotes}[flushleft]
\footnotesize
\item \textit{Notes:}
Results are based on 300 Monte Carlo replications per configuration. Both procedures use XGBoost first steps with three-fold outer cross-fitting and inner-cross-fitted pseudo-ranks. Shared training uses the full outer training sample for all nuisance functions. Separate training uses disjoint groups for $(\widehat\gamma_c,\widehat\alpha_p)$ and $(\widehat\gamma_p,\widehat\alpha_c)$, then swaps the groups and averages the resulting scores observation by observation. The target is $\theta_0(c)=(6/\pi)\arcsin(c/2)$. Emp.\ SD denotes the Monte Carlo standard deviation; Avg.\ SE, the average estimated standard error; RMSE, the root mean squared error; and Cov.\ 95\%, the empirical coverage of nominal 95\% confidence intervals. LR standard errors use training-sample preliminary estimates. Plug-in standard errors omit first-step estimation uncertainty. Only the correctly specified nuisance configuration is considered.
\end{tablenotes}
\end{threeparttable}
\end{sidewaystable}

\Cref{tab:MLCRRR_main} reports lower bias and RMSE for the locally robust (LR) estimator than for the plug-in estimator at every design point under both training procedures. Additional sample separation reduces LR bias and RMSE throughout the experiment, with particularly large improvements when $c=0.75$. In this case, bias falls from $0.1427$ to $0.0219$ at $n=500$, from $0.1002$ to $0.0321$ at $n=1000$, and from $0.0319$ to $0.0150$ at $n=5000$. The corresponding RMSE falls from $0.1588$ to $0.0704$ at $n=500$ and from $0.0344$ to $0.0202$ at $n=5000$.

The comparison highlights the importance of separating nuisance-training samples when products of estimation errors enter the second-order remainder. Outer cross-fitting separates nuisance training from score evaluation, but nuisance estimators trained on the same observations can retain correlated estimation errors. The additional split removes this common-sample source of dependence and can reduce the resulting cross-product bias, even though each nuisance is estimated using fewer observations. Consistent with this interpretation, the diagnostics in Appendix~\ref{app:diagCRRR} show larger individual nuisance errors and CDF--CDF product bounds under separate training, alongside smaller asymptotic-linearity discrepancies. The improvement therefore illustrates a benefit of sample separation that individual error norms alone do not capture.

Sample separation also substantially changes coverage. For $c=0.75$,
LR coverage increases from $0.790$, $0.637$, and $0.710$ under shared training to $1.000$, $0.980$, and $0.977$ under separate training at $n=500$, $1000$, and $5000$, respectively. Across all design points, separate-sample LR coverage ranges from $0.977$ to $1.000$. This coverage is conservative, since average estimated standard errors exceed empirical standard deviations. For $c=0.75$ and $n=5000$, the residual bias is comparable to the empirical standard deviation, and the larger estimated standard errors compensate for it. The diagnostics in Appendix~\ref{app:diagCRRR} suggest that the slow convergence of the representer estimates contributes to this variance inflation.

\section{Conclusions}\label{sec:conclusions}

Conditional-distribution first steps enter many economic parameters through generated controls, counterfactual distributions, conditional ranks, and quantiles. We characterize their first-order effects and obtain locally robust moments by conditional centering of a representer. The conditional-measure formulation accommodates a broader class of functionals than the CDF formulation. When the latter applies, automatic Riesz regression provides an estimation criterion, with a separate error term that must be controlled when the correction is recovered by integration against the fitted conditional law.

Cross-fitting and explicit control of the nonlinear and interaction remainders yield asymptotic linearity, normality, and consistent variance estimation. These conditions concern the actual nuisance estimators and do not follow from first-order differentiability alone. Exact affine representations can additionally yield double robustness. For conditional rank-rank regression, this is a componentwise result, whereas general transformations of conditional $L$-functionals need not have this property.
Consistency under one-sided nuisance misspecification does not by itself ensure valid inference, which may require faster convergence of the correctly specified nuisance estimator and additional adjustments for its first-order contribution.

The simulations show that the locally robust estimator reduces bias and RMSE relative to the plug-in estimator. Additional sample separation provides substantial further improvements, highlighting the importance of controlling cross-product bias when several nuisance functions are estimated.
Deriving primitive rate conditions for specific conditional-distribution learners, extending automatic estimation to functionals without an $L^2$ CDF representer, and adapting existing methods for doubly robust inference to the framework developed here remain useful directions for further work.

\bibliographystyle{ecta}
\bibliography{bibliography}

@techreport{escanciano2023generated,
  author      = {Escanciano, Juan Carlos and
                 P{\'e}rez-Izquierdo, Telmo},
  title       = {Automatic Locally Robust {GMM} with
                 Machine-Learning-Generated Regressors},
  year        = {2023},
  institution = {arXiv},
  number      = {2301.10643},
  note        = {Revised 2026},
  url         = {https://arxiv.org/abs/2301.10643}
}

@article{andrews1994asymptotics,
  title={Asymptotics for semiparametric econometric models via stochastic equicontinuity},
  author={Andrews, Donald WK},
  journal={Econometrica: Journal of the Econometric Society},
  pages={43--72},
  year={1994},
  publisher={JSTOR}
}

@article{andrews1994empirical,
  title={Empirical process methods in econometrics},
  author={Andrews, Donald WK},
  journal={Handbook of Econometrics},
  volume={4},
  pages={2247--2294},
  year={1994},
  publisher={Elsevier}
}

@article{belloni2015some,
  title={Some new asymptotic theory for least squares series: Pointwise and uniform results},
  author={Belloni, Alexandre and Chernozhukov, Victor and Chetverikov, Denis and Kato, Kengo},
  journal={Journal of Econometrics},
  volume={186},
  number={2},
  pages={345--366},
  year={2015},
  publisher={Elsevier}
}

@article{chernozhukov2013inference,
  title={Inference on counterfactual distributions},
  author={Chernozhukov, Victor and Fern{\'a}ndez-Val, Iv{\'a}n and Melly, Blaise},
  journal={Econometrica},
  volume={81},
  number={6},
  pages={2205--2268},
  year={2013},
  publisher={Wiley Online Library}
}

@article{chernozhukov2014gaussian,
  author = {Victor Chernozhukov and Denis Chetverikov and Kengo Kato},
  title = {{Gaussian approximation of suprema of empirical processes}},
  journal = {The Annals of Statistics},
  year = {2014},
  volume = {42},
  number = {4},
  publisher = {Institute of Mathematical Statistics},
  pages = {1564 -- 1597},
  doi = {10.1214/14-AOS1230},
  url = {https://doi.org/10.1214/14-AOS1230},
}

@article{chernozhukov2018double,
  title={Double/debiased machine learning for treatment and structural parameters},
  author={Chernozhukov, Victor and Chetverikov, Denis and Demirer, Mert and Duflo, Esther and Hansen, Christian and Newey, Whitney and Robins, James},
  journal={The Econometrics Journal},
  volume={21},
  number={1},
  pages={C1--C68},
  year={2018},
  publisher={Oxford University Press Oxford, UK},
  doi = {10.1111/ectj.12097},
  issn={1368-4221}
}

@article{chernozhukov2022automatic,
  title={Automatic debiased machine learning of causal and structural effects},
  author={Chernozhukov, Victor and Newey, Whitney K and Singh, Rahul},
  journal={Econometrica},
  volume={90},
  number={3},
  pages={967--1027},
  year={2022},
  publisher={Wiley Online Library}
}

@article{chernozhukov2022locally,
  title={Locally robust semiparametric estimation},
  author={Chernozhukov, Victor and Escanciano, Juan Carlos and Ichimura, Hidehiko and Newey, Whitney K and Robins, James M},
  journal={Econometrica},
  volume={90},
  number={4},
  pages={1501--1535},
  year={2022},
  publisher={Wiley Online Library}
}

@misc{chernozhukov2024automatic,
      title={Automatic Debiased Machine Learning via Riesz Regression}, 
      author={Victor Chernozhukov and Whitney K. Newey and Victor Quintas-Martinez and Vasilis Syrgkanis},
      year={2024},
      eprint={2104.14737},
      archivePrefix={arXiv},
      primaryClass={math.ST},
      url={https://arxiv.org/abs/2104.14737}, 
}

@misc{chernozhukov2025conditional,
      title={Conditional Rank-Rank Regression}, 
      author={Victor Chernozhukov and Iván Fernández-Val and Jonas Meier and Aico van Vuuren and Francis Vella},
      year={2025},
      eprint={2407.06387},
      archivePrefix={arXiv},
      primaryClass={econ.EM},
      url={https://arxiv.org/abs/2407.06387}, 
}

@article{chernozhukov2026conditional,
  author  = {Chernozhukov, Victor and Newey, Whitney K.
             and Syrgkanis, Vasilis},
  title   = {Conditional Influence Functions},
  journal = {Quantitative Economics},
  year    = {2026},
  note    = {Forthcoming. arXiv:2412.18080},
  url     = {https://arxiv.org/abs/2412.18080}
}

@article{chesher2003identification,
  title={Identification in nonseparable models},
  author={Chesher, Andrew},
  journal={Econometrica},
  volume={71},
  number={5},
  pages={1405--1441},
  year={2003},
  publisher={Wiley Online Library}
}

@article{chetverikov2025weighted,
  title={Weighted-average quantile regression},
  author={Chetverikov, Denis and Liu, Yukun and Tsyvinski, Aleh},
  journal={Journal of Econometrics},
  volume={252},
  pages={106115},
  year={2025},
  publisher={Elsevier}
}

@article{dinardo1996labor,
  author  = {DiNardo, John and Fortin, Nicole M. and Lemieux, Thomas},
  title   = {Labor Market Institutions and the Distribution of Wages, 1973--1992: A Semiparametric Approach},
  journal = {Econometrica},
  year    = {1996},
  volume  = {64},
  number  = {5},
  pages   = {1001--1044}
}

@article{dukes2024doubly,
  title={On doubly robust inference for double machine learning in semiparametric regression},
  author={Dukes, Oliver and Vansteelandt, Stijn and Whitney, David},
  journal={Journal of Machine Learning Research},
  volume={25},
  number={279},
  pages={1--46},
  year={2024}
}

@article{farrell2021deep,
  title={Deep neural networks for estimation and inference},
  author={Farrell, Max H and Liang, Tengyuan and Misra, Sanjog},
  journal={Econometrica},
  volume={89},
  number={1},
  pages={181--213},
  year={2021},
  publisher={Wiley Online Library}
}

@article{firpo2009unconditional,
  title={Unconditional quantile regressions},
  author={Firpo, Sergio and Fortin, Nicole M and Lemieux, Thomas},
  journal={Econometrica},
  volume={77},
  number={3},
  pages={953--973},
  year={2009},
  publisher={Wiley Online Library}
}

@article{foresi1995conditional,
  title={The conditional distribution of excess returns: An empirical analysis},
  author={Foresi, Silverio and Peracchi, Franco},
  journal={Journal of the American Statistical Association},
  volume={90},
  number={430},
  pages={451--466},
  year={1995},
  publisher={Taylor \& Francis}
}

@article{hahn2019three,
  title={Three-stage semi-parametric inference: Control variables and differentiability},
  author={Hahn, Jinyong and Ridder, Geert},
  journal={Journal of econometrics},
  volume={211},
  number={1},
  pages={262--293},
  year={2019},
  publisher={Elsevier}
}

@article{hardle1988strong,
  title={Strong uniform consistency rates for estimators of conditional functionals},
  author={H{\"a}rdle, Wolfgang and Janssen, Paul and Serfling, Robert},
  journal={The Annals of Statistics},
  pages={1428--1449},
  year={1988},
  publisher={JSTOR}
}

@article{hossjer2023use,
  title={On the use of L-functionals in regression models},
  author={H{\"o}ssjer, Ola and Karlsson, M{\aa}ns},
  journal={Open Mathematics},
  volume={21},
  number={1},
  pages={20220597},
  year={2023},
  publisher={De Gruyter}
}

@article{ichimura2022influence,
  title={The influence function of semiparametric estimators},
  author={Ichimura, Hidehiko and Newey, Whitney K},
  journal={Quantitative Economics},
  volume={13},
  number={1},
  pages={29--61},
  year={2022},
  publisher={Wiley Online Library}
}

@article{imbens2009identification,
  title={Identification and estimation of triangular simultaneous equations models without additivity},
  author={Imbens, Guido W and Newey, Whitney K},
  journal={Econometrica},
  volume={77},
  number={5},
  pages={1481--1512},
  year={2009},
  publisher={Wiley Online Library}
}

@inproceedings{kallus2023robust,
  title={Robust and agnostic learning of conditional distributional treatment effects},
  author={Kallus, Nathan and Oprescu, Miruna},
  booktitle={International Conference on Artificial Intelligence and Statistics},
  pages={6037--6060},
  year={2023},
  organization={PMLR}
}

@article{koenker1978regression,
  title={Regression quantiles},
  author={Koenker, Roger and Bassett Jr, Gilbert},
  journal={Econometrica: journal of the Econometric Society},
  pages={33--50},
  year={1978},
  publisher={JSTOR}
}

@article{lee2022nonparametric,
  title={Nonparametric weighted average quantile derivative},
  author={Lee, Ying-Ying},
  journal={Econometric Theory},
  volume={38},
  number={3},
  pages={497--535},
  year={2022},
  publisher={Cambridge University Press}
}

@article{liang2021how,
  author  = {Tengyuan Liang},
  title   = {How Well Generative Adversarial Networks Learn Distributions},
  journal = {Journal of Machine Learning Research},
  year    = {2021},
  volume  = {22},
  number  = {228},
  pages   = {1--41},
  url     = {http://jmlr.org/papers/v22/20-911.html}
}

@article{machado2005counterfactual,
  title={Counterfactual decomposition of changes in wage distributions using quantile regression},
  author={Machado, Jos{\'e} AF and Mata, Jos{\'e}},
  journal={Journal of applied Econometrics},
  volume={20},
  number={4},
  pages={445--465},
  year={2005},
  publisher={Wiley Online Library}
}

@article{meinshausen2006quantile,
  title={Quantile regression forests.},
  author={Meinshausen, Nicolai},
  journal={Journal of machine learning research},
  volume={7},
  number={6},
  year={2006}
}

@article{newey1991uniform,
  title={Uniform convergence in probability and stochastic equicontinuity},
  author={Newey, Whitney K},
  journal={Econometrica: Journal of the Econometric Society},
  pages={1161--1167},
  year={1991},
  publisher={JSTOR}
}

@article{newey1994asymptotic,
  title={The asymptotic variance of semiparametric estimators},
  author={Newey, Whitney K},
  journal={Econometrica: Journal of the Econometric Society},
  pages={1349--1382},
  year={1994},
  publisher={JSTOR}
}

@misc{newey2018cross,
      title={Cross-Fitting and Fast Remainder Rates for Semiparametric Estimation}, 
      author={Whitney K. Newey and James R. Robins},
      year={2018},
      eprint={1801.09138},
      archivePrefix={arXiv},
      primaryClass={math.ST},
      url={https://arxiv.org/abs/1801.09138}, 
}

@incollection{neyman1959optimal,
  author    = {Neyman, Jerzy},
  title     = {Optimal Asymptotic Tests of Composite Statistical Hypotheses},
  booktitle = {Probability and Statistics: The Harald Cram{\'e}r Volume},
  editor    = {Grenander, Ulf},
  publisher = {Almqvist \& Wiksell},
  address   = {Stockholm},
  year      = {1959},
  pages     = {213--234}
}

@article{nishiyama2011impossibility,
  author  = {Nishiyama, Yoichi},
  title   = {Impossibility of weak convergence of kernel
             density estimators to a non-degenerate law
             in {$L^2(\mathbb{R}^d)$}},
  journal = {Journal of Nonparametric Statistics},
  year    = {2011},
  volume  = {23},
  number  = {1},
  pages   = {129--135},
  doi     = {10.1080/10485251003678507}
}

@article{portier2025nearest,
  title={Nearest neighbor empirical processes},
  author={Portier, Fran{\c{c}}ois},
  journal={Bernoulli},
  volume={31},
  number={1},
  pages={312--332},
  year={2025},
  publisher={Bernoulli Society for Mathematical Statistics and Probability}
}

@article{robins1994estimation,
  title={Estimation of regression coefficients when some regressors are not always observed},
  author={Robins, James M and Rotnitzky, Andrea and Zhao, Lue Ping},
  journal={Journal of the American statistical Association},
  volume={89},
  number={427},
  pages={846--866},
  year={1994},
  publisher={Taylor \& Francis}
}

@article{rothe2010nonparametric,
  title={Nonparametric estimation of distributional policy effects},
  author={Rothe, Christoph},
  journal={Journal of Econometrics},
  volume={155},
  number={1},
  pages={56--70},
  year={2010},
  publisher={Elsevier}
}

@article{rothe2012partial,
  title={Partial distributional policy effects},
  author={Rothe, Christoph},
  journal={Econometrica},
  volume={80},
  number={5},
  pages={2269--2301},
  year={2012},
  publisher={Wiley Online Library}
}

@article{schmidt2020nonparametric,
author = {Johannes Schmidt-Hieber},
title = {{Nonparametric regression using deep neural networks with ReLU activation function}},
volume = {48},
journal = {The Annals of Statistics},
number = {4},
publisher = {Institute of Mathematical Statistics},
pages = {1875 -- 1897},
year = {2020},
doi = {10.1214/19-AOS1875},
URL = {https://doi.org/10.1214/19-AOS1875}
}

@article{stupfler2016weak,
  title={On the weak convergence of the kernel density estimator in the uniform topology},
  author={Stupfler, Gilles},
  journal={Electronic Communications in Probability},
  year={2016}
}

@article{stute1986almost,
  title={On almost sure convergence of conditional empirical distribution functions},
  author={Stute, Winfried},
  journal={The Annals of Probability},
  volume={14},
  number={3},
  pages={891--901},
  year={1986},
  publisher={Institute of Mathematical Statistics}
}

@incollection{van1996weak,
  author = {Van Der Vaart, Aad W and Wellner, Jon A},
  title = {Weak Convergence and Empirical Processes: with Applications to Statistics},
  booktitle = {Weak Convergence and Empirical Processes: with Applications to Statistics},
  publisher = {Springer},
  year = {1996},
  pages = {16--28},
}

@book{van2000asymptotic,
  title={Asymptotic statistics},
  author={Van der Vaart, Aad W},
  volume={3},
  year={2000},
  publisher={Cambridge university press}
}

@misc{vanderlaan2026doubly,
      title={Doubly robust inference via calibration}, 
      author={Lars van der Laan and Alex Luedtke and Marco Carone},
      year={2026},
      eprint={2411.02771},
      archivePrefix={arXiv},
      primaryClass={stat.ME},
      url={https://arxiv.org/abs/2411.02771}, 
}

@article{wager2018estimation,
  title={Estimation and inference of heterogeneous treatment effects using random forests},
  author={Wager, Stefan and Athey, Susan},
  journal={Journal of the American Statistical Association},
  volume={113},
  number={523},
  pages={1228--1242},
  year={2018},
  publisher={Taylor \& Francis}
}

@article{wang2026conditional,
      title={Conditional Rank-Rank Regression via Deep Conditional Transformation Models}, 
      author={Xiaoyi Wang and Long Feng and Zhaojun Wang},
      year={2026},
      journal={ArXiv},
      eprint={2603.07230},
      archivePrefix={arXiv},
      primaryClass={stat.ME},
      url={https://arxiv.org/abs/2603.07230}, 
}

@article{weymark1981generalized,
  title={Generalized Gini inequality indices},
  author={Weymark, John A},
  journal={Mathematical social sciences},
  volume={1},
  number={4},
  pages={409--430},
  year={1981},
  publisher={Elsevier}
}

\appendix

\section{Proofs of the Influence-Function and Robustness Results}\label{app:proofs}

\begin{proof}[Proof of \Cref{lemma:correctionterm}]
Fix $\theta$ and suppress it in the notation. Put $r_H=dH_X/dF_X$. For $0\leq\tau<1$, a version of the conditional law under $F_\tau$ is
\[
\gamma_\tau(dy,x)
=\frac{(1-\tau)\gamma_0(dy,x)+\tau r_H(x)H_{Y|X}(dy,x)}
       {1-\tau+\tau r_H(x)}.
\]
Consequently,
\[
\frac{\gamma_\tau-\gamma_0}{\tau}(dy,x)
=\frac{r_H(x)}{1-\tau+\tau r_H(x)}
\{H_{Y|X}-\gamma_0\}(dy,x).
\]
For $\tau\leq1/2$ the coefficient is bounded by $2r_H(x)$. The path integrability requirement therefore permits dominated convergence after integration against $\alpha_0$ and $F_X$. By \Cref{ass:linearization},
\[
\begin{split}
D_\theta(H)
&=\int r_H(x)\int\alpha_0(y,x)
                  \{H_{Y|X}-\gamma_0\}(dy,x)F_X(dx)\\
&=\int\left\{\alpha_0(y,x)
       -\int\alpha_0(t,x)\gamma_0(dt,x)\right\}H(dw).
\end{split}
\]
This is \Cref{eq:infdef}. Conditional centering gives mean zero and
$P\|\phi_0\|^2\leq P\|\alpha_0\|^2<\infty$.
\end{proof}

\begin{proof}[Proof of \Cref{lemma:correctiontermprime}]
For any conditional probability kernel $\gamma$, Fubini's theorem gives
\[
\begin{split}
\int\delta_0(t,x,\theta)\gamma(t,x)\gamma_0(dt,x)
&=\int\left\{\int\delta_0(t,x,\theta)\indic\{y\leq t\}
                         \gamma_0(dt,x)\right\}\gamma(dy,x)\\
&=\int\alpha_0(y,x,\theta)\gamma(dy,x).
\end{split}
\]
The equality holds coordinatewise. The bound $0\leq\gamma\leq1$ and integrability of $\delta_0$ justify Fubini under $F_X$. Apply the identity with $\gamma=\gamma_\tau$ and differentiate at zero to obtain \Cref{ass:linearization}. Moreover,
\[
\|\alpha_0(y,x,\theta)\|
\leq\mathbb E[\|\delta_0(Y,X,\theta)\|\mid X=x],
\qquad P\|\alpha_0\|^2\leq P\|\delta_0\|^2.
\]
The stipulated path integrability also holds for this $\alpha_0$. The result follows from \Cref{lemma:correctionterm}.
\end{proof}

\begin{proof}[Proof of \Cref{prop:controlfunction}]
Let $\dot\gamma_H$ denote the derivative at zero of the conditional CDF of $X$ given $Z$ along an admissible path. Differentiation under the expectation and conditioning on $(X,Z)$ yield
\[
\begin{split}
D_\theta(H)
&=\mathbb E[\partial_v\varrho(W,V,\theta)\dot\gamma_H(X,Z)]\\
&=\mathbb E[\delta_0(X,Z,\theta)\dot\gamma_H(X,Z)].
\end{split}
\]
This is the CDF linearization. Conditional Jensen's inequality gives
\[
P\|\delta_0\|^2\leq P\|\partial_v\varrho(W,V,\theta)\|^2<\infty.
\]
Applying \Cref{lemma:correctiontermprime} with $(Y,X)$ replaced by $(X,Z)$ gives the stated $\alpha_0$ and FSIF. A second application of conditional Fubini gives
\[
\mathbb E[\alpha_0(X,Z,\theta)\mid Z]
=\mathbb E[\partial_v\varrho(W,V,\theta)\gamma_0(X,Z)\mid Z],
\]
which also verifies \Cref{eq:phicontrolfunction} componentwise.
\end{proof}

\begin{proof}[Proof of \Cref{prop:functionalcounterfactuals}]
Let $\dot\gamma_H$ be the CDF derivative along an admissible path and let
$\dot F_H^*(t)=\int\dot\gamma_H(t,x)F_X^*(dx)$.
By assumption, this derivative belongs to $\mathcal T_C$. The chain rule for Hadamard derivatives and the assumed interchanges give
\[
\begin{split}
D_\theta(H)
&=\varphi'_{F_Y^*}[\dot F_H^*]
 =\int a_0(t)\int\dot\gamma_H(t,x)F_X^*(dx)\nu(dt)\\
&=\left.\frac{d}{d\tau}\right|_{\tau=0}
\int\int r_0(x)\left\{\int a_0(t)\indic\{y\leq t\}\nu(dt)\right\}
                         \gamma_\tau(dy,x)F_X(dx).
\end{split}
\]
Thus \Cref{ass:linearization} holds with the displayed representer. Writing $C_a=\int|a_0|\,d|\nu|$, we have
\[
|\alpha_0(y,x)|\leq r_0(x)C_a,
\qquad P\alpha_0(Y,X)^2\leq C_a^2P r_0(X)^2<\infty.
\]
This bound uses the total variation of $\nu$, so it also applies when $\nu$ is signed. The result follows by \Cref{lemma:correctionterm}.
\end{proof}

\begin{proof}[Proof of \Cref{prop:rankrank}]
Write $R_s=\gamma_{0,s}(Y_s,X)$ and let $\dot\gamma_{s,H}$ denote the CDF derivative along the path. The derivative of the product of ranks is
\[
\begin{split}
D_\theta(H)
&=\mathbb E[R_p\dot\gamma_{c,H}(Y_c,X)
             +R_c\dot\gamma_{p,H}(Y_p,X)]\\
&=\sum_{s\in\{c,p\}}
\mathbb E[\delta_{0,s}(Y_s,X)\dot\gamma_{s,H}(Y_s,X)].
\end{split}
\]
The mixture formula in the proof of \Cref{lemma:correctionterm} bounds each CDF difference quotient by $2r_H(X)$ for $\tau\leq1/2$. Since the ranks are bounded and $P r_H(X)=1$, dominated convergence justifies differentiation. Each $\delta_{0,s}$ and its tail-integral representer is between zero and one. Applying \Cref{lemma:correctiontermprime} to each component and adding the corrections gives the result.
\end{proof}

\begin{proof}[Proof of \Cref{prop:l-functionals}]
Put $q_0(s,x)=\gamma_0^{-1}(s,x)$ and let $\dot\gamma_H$ denote the path derivative. Under \Cref{ass:Lhadam-diff},
\[
\left.\frac{d}{d\tau}\gamma_\tau^{-1}(s,x)\right|_{\tau=0}
=-\frac{\dot\gamma_H(q_0(s,x),x)}{f_{Y|X}(q_0(s,x),x)}.
\]
The chain rule, conditioning on $X$, and the assumed interchanges imply
\[
D_\theta(H)
=-\mathbb E\left[b_0(X,\theta)\int_0^1
\frac{a'(q_0(s,X))}{f_{Y|X}(q_0(s,X),X)}
\dot\gamma_H(q_0(s,X),X)K(ds)\right].
\]
Using $\gamma_\tau(q,x)=\int\indic\{y\leq q\}\gamma_\tau(dy,x)$ gives the conditional-measure linearization. The term involving $s$ in the proposed representer integrates against a measure of total mass one and therefore has derivative zero. At each relevant quantile, continuity and positivity of the density imply $\gamma_0(q_0(s,x),x)=s$. Conditional Fubini then gives
$\mathbb E[\alpha_0(Y,X,\theta)\mid X]=0$.
Square-integrability is assumed. The result follows from \Cref{lemma:correctionterm}.
\end{proof}

\begin{proof}[Proof of \Cref{lemma:thm2CEINR}]
The two terms in the definition of $P\phi(W,\gamma_0,\alpha,\theta)$ are equal by iterated expectations. Their difference is zero.
\end{proof}

\begin{proof}[Proof of \Cref{thm:LR}]
For $H\in\mathcal H_0$, put $h=H_{Y|X}-\gamma_0$. The derivative of the original moment along $\gamma_0+\tau h$ is
$\mathbb E\int\alpha_0(y,X,\theta)h(dy,X)$ by \Cref{ass:linearization}. The derivative of the correction is negative. Their sum is zero. Linearity and continuity extend this equality from the generating directions to $\mathcal T$. The derivative in an integrable representer direction is zero by \Cref{lemma:thm2CEINR}.
\end{proof}

\begin{proof}[Proof of \Cref{thm:doublerobustness}]
For every admissible $\gamma$,
\[
P\phi(W,\gamma,\alpha_0,\theta)
=-\mathbb E\int\alpha_0(y,X,\theta)(\gamma-\gamma_0)(dy,X).
\]
Substitute \Cref{eq:exact-linear-representation} and use conditional centering for $\gamma=\gamma_0$. If \Cref{eq:exact-cdf-representation} holds, conditional Fubini gives
\[
\mathbb E\int\alpha_0(y,X,\theta)(\gamma-\gamma_0)(dy,X)
=P[\delta_0(Y,X,\theta)(\gamma-\gamma_0)(Y,X)],
\]
which proves the sufficient condition.
\end{proof}

\begin{proof}[Proof of \Cref{prop:DRcontrolfunction}]
Subtracting the moments at $\gamma$ and $\gamma_0$ gives
\[
P\{g(W,\gamma,\theta)-g(W,\gamma_0,\theta)\}
=P[h_1(W,\theta)(\gamma-\gamma_0)(X,Z)]
=P[\delta_0(X,Z,\theta)(\gamma-\gamma_0)(X,Z)].
\]
Apply the CDF-based sufficient condition in \Cref{thm:doublerobustness} with $(Y,X)$ replaced by $(X,Z)$.
\end{proof}

\begin{proof}[Proof of \Cref{prop:DRcounterfactuals}]
The stipulated exact expansion and Fubini give
\[
\varphi(F_\gamma^*)-\varphi(F_Y^*)
=\mathbb E\left[r_0(X)\int a_0(t)(\gamma-\gamma_0)(t,X)\nu(dt)\right]
=\mathbb E\int\alpha_0(y,X)(\gamma-\gamma_0)(dy,X).
\]
Absolute integrability follows from $\int|a_0|\,d|\nu|<\infty$, bounded CDF differences, and $P r_0=1$. Apply \Cref{thm:doublerobustness}.
\end{proof}

\begin{proof}[Proof of \Cref{prop:DRrankrank}]
With $\gamma_{-s}=\gamma_{0,-s}$, the original-moment difference is
\[
P[(\gamma_s-\gamma_{0,s})(Y_s,X)\gamma_{0,-s}(Y_{-s},X)].
\]
Conditioning on $(Y_s,X)$ identifies the exact CDF representer $\delta_{0,s}$. Thus the $s$ component satisfies \Cref{thm:doublerobustness}. The other correction is centered at its true CDF. Expanding both rank factors when both CDFs vary leaves the product of their errors after the two linear terms cancel, giving the stated remaining bias.
\end{proof}

\begin{proof}[Proof of \Cref{prop:representer-plugin}]
For part (i), work on the event
$\{\widehat\eta_{-\ell}\in\mathcal U\}$.
The assumed continuity bound and the Cauchy--Schwarz inequality give
\[
    \|\widehat\alpha_{-\ell}-\alpha_0\|_2^2
    \le
    C^2J\sum_{j=1}^J
    \|\widehat\eta_{j,-\ell}-\eta_{j,0}\|_j^{2\kappa_j}.
\]
The componentwise convergence rates and
$\mathbb P(\widehat\eta_{-\ell}\in\mathcal U)\to1$ therefore imply
\[
    \|\widehat\alpha_{-\ell}-\alpha_0\|_2^2
    =
    O_P\!\left(
        \sum_{j=1}^J r_{j,n}^{2\kappa_j}
    \right).
\]
The argument for direct estimation of $\delta_0$ is identical
under the corresponding continuity bound.

For part (ii), again work on $\{\widehat\eta_{-\ell}\in\mathcal U\}$ and suppress fold indices.
Decompose
\begin{align*}
    \widehat\alpha(y,x)-\alpha_0(y,x)
    &=
    \int_{[y,\infty)}
    (\widehat\delta-\delta_0)(t,x)\widehat\gamma(dt,x)\\
    &\quad+
    \int_{[y,\infty)}
    \delta_0(t,x)(\widehat\gamma-\gamma_0)(dt,x).
\end{align*}
Since $\widehat\gamma$ is a conditional probability measure,
\[
    \left\|
        \int_{[y,\infty)}
        (\widehat\delta-\delta_0)(t,x)\widehat\gamma(dt,x)
    \right\|
    \le
    \sup_t\|\widehat\delta(t,x)-\delta_0(t,x)\|.
\]

For the second integral, set
$H(t,x):=\widehat\gamma(t,x)-\gamma_0(t,x)$.
Componentwise Stieltjes integration by parts yields
\begin{align*}
    &\int_{[y,\infty)}
    \delta_0(t,x)(\widehat\gamma-\gamma_0)(dt,x)\\
    &=
    -\delta_0(y,x)H(y-,x)
    -\int_{(y,\infty)}H(t-,x)\,d_t\delta_0(t,x).
\end{align*}
To justify the unbounded interval, first apply integration by
parts on $[y,b]$ and then let $b\to\infty$.
The upper boundary term $\delta_0(b,x)H(b,x)$ vanishes because
$\delta_0$ is bounded and both conditional CDFs converge to one.
The remaining Stieltjes integral converges by the bounded-variation
assumption. The left limits and the open lower endpoint in this
integral account for possible atoms.

Since
\[
    \sup_t|H(t-,x)|\le\sup_t|H(t,x)|,
\]
the definitions of $B_0$ and $V_0$ imply
\[
    \left\|
        \int_{[y,\infty)}
        \delta_0(t,x)(\widehat\gamma-\gamma_0)(dt,x)
    \right\|
    \le
    (B_0+V_0)
    \sup_t|\widehat\gamma(t,x)-\gamma_0(t,x)|.
\]
Consequently, using
$\|a+b\|^2\le2\|a\|^2+2\|b\|^2$ and integrating with respect
to $P_{Y,X}$ gives
\begin{align*}
    \|\widehat\alpha-\alpha_0\|_2^2
    &\le
    2\mathbb E_X\!\left[
        \sup_t\|\widehat\delta(t,X)-\delta_0(t,X)\|^2
    \right]\\
    &\quad+
    2(B_0+V_0)^2
    \mathbb E_X\!\left[
        \sup_t|\widehat\gamma(t,X)-\gamma_0(t,X)|^2
    \right].
\end{align*}
On $\{\widehat\eta_{-\ell}\in\mathcal U\}$, the continuity bound gives
\[
    \mathbb E_X\!\left[
        \sup_t\|\widehat\delta(t,X)-\delta_0(t,X)\|^2
    \right]
    \le
    C^2J\sum_{j=1}^J
    \|\widehat\eta_j-\eta_{j,0}\|_j^{2\kappa_j}.
\]
Combining the assumed convergence rates with
$\mathbb P(\widehat\eta_{-\ell}\in\mathcal U)\to1$ therefore yields
\[
    \|\widehat\alpha-\alpha_0\|_2^2
    =
    O_P\!\left(
        \sum_{j=1}^J r_{j,n}^{2\kappa_j}
        +r_{\gamma,n}^2
    \right),
\]
which proves part (ii).
\end{proof}

\begin{proof}[Proof of \Cref{prop:automatic1}]
Let $R_n^*$ denote the coefficients of $\delta_n^*$, so that $GR_n^*=M$. From the first-order conditions of \Cref{eq:estimation-automatic1},
\[
\widehat R-R_n^*
=
(\widehat G+\lambda_n I)^{-1}
\left\{
\widehat M-M
-
(\widehat G-G)R_n^*
-
\lambda_nR_n^*
\right\}.
\]
The eigenvalue conditions and $\lambda_n\geq0$ bound the inverse in operator norm with probability approaching one. Also
$\operatorname{tr}(R_n^{*\prime}GR_n^*)=\|\delta_n^*\|_2^2\leq\|\delta_{\mathcal T_2}\|_2^2$,
so the lower eigenvalue bound implies a uniform Frobenius bound on $R_n^*$. Hence,
\[
\|\widehat\delta-\delta_n^*\|_2
\leq
C\left\{
\|\widehat M-M\|_F+
\|\widehat G-G\|_{\mathrm{op}}+\lambda_n
\right\}.
\]
The first conclusion follows by the triangle inequality.

For the second conclusion, observe that conditional Jensen's inequality gives
\[
\|\mathcal{K}\delta\|_2
\leq \|\delta\|_2
\]
for every $\delta\in L^2(P_{Y,X})$. Therefore,
\[
\begin{aligned}
\|\widehat\alpha-\alpha_{\mathcal T_2}\|_2
&\leq
\|(\widehat{\mathcal{K}}-\mathcal{K})
\widehat\delta\|_2
+
\|\mathcal{K}(\widehat\delta-\delta_{\mathcal T_2})\|_2\\
&\leq
\|(\widehat{\mathcal{K}}-\mathcal{K})
\widehat\delta\|_2
+
\|\widehat\delta-\delta_{\mathcal T_2}\|_2.
\end{aligned}
\]
The consistency statement follows.
\end{proof}

\section{Additional Conditions and Proofs for Asymptotic Theory}
\label{app:asymptotictheory}

\subsection{Consistency}\label{subsec:consistency}
\assumptionprefix{AR}

The sampling, cross-fitting, and weighting-matrix conditions of \Cref{sec:asymptotictheory} are maintained. The following standard conditions provide one route to consistency. They are not needed separately when consistency follows from an explicit estimator.

\begin{assumption}[Consistency]\label{ass:consistency}
(i) $\Theta$ is compact, $m(\theta)=P g(W,\gamma_0,\theta)$ is continuous and has the unique zero $\theta_0$, and $P\|g(W,\gamma_0,\theta)\|<\infty$ for each $\theta\in\Theta$.

(ii) For every $\theta\in\Theta$ and fold $\ell$,
\[
\begin{split}
P\|g(W,\widehat\gamma_{-\ell},\theta)-g(W,\gamma_0,\theta)\|&=o_P(1),\\
P\|d_\ell\|+P\|c_\ell\|+P\|\widehat\Delta_{-\ell}\|&=o_P(1).
\end{split}
\]

(iii) With probability approaching one, all $\widehat\gamma_{-\ell}$ belong to a deterministic class $\mathcal N_\gamma$ containing $\gamma_0$. For some $\zeta>0$, $C<\infty$, and measurable nonnegative $D_g$,
\[
\|g(W,\gamma,\theta)-g(W,\gamma,\theta')\|
\leq D_g(W,\gamma)\|\theta-\theta'\|^\zeta,
\qquad \sup_{\gamma\in\mathcal N_\gamma}P D_g(W,\gamma)\leq C,
\]
for every $\theta,\theta'\in\Theta$.
\end{assumption}

\begin{proposition}[Consistency]\label{prop:consistency}
Under \Cref{ass:consistency}, $\widehat\theta\to_P\theta_0$.
\end{proposition}

\begin{proof}
For fixed $\theta$, decompose the feasible moment around
$g(W,\gamma_0,\theta)+\phi(W,\gamma_0,\alpha_0,\theta_0)$ as in \Cref{eq:score-decomposition}. Conditional Markov's inequality and part~(ii) make the fold averages of all four differences $o_P(1)$. The oracle average converges to $m(\theta)$ by the law of large numbers; its correction has mean zero and is integrable. Thus $\widehat\psi(\theta)\to_P m(\theta)$ pointwise.

Since the fitted correction is fixed in $\theta$, part~(iii) gives
\[
\|\widehat\psi(\theta)-\widehat\psi(\theta')\|
\leq\left\{\frac1n\sum_{\ell=1}^L\sum_{i\in I_\ell}
D_g(W_i,\widehat\gamma_{-\ell})\right\}\|\theta-\theta'\|^\zeta.
\]
The expression in braces is $O_P(1)$ by conditional Markov's inequality and fixed $L$. Compactness and a finite-net argument therefore imply
$\sup_{\theta\in\Theta}\|\widehat\psi(\theta)-m(\theta)\|=o_P(1)$, as in \citet{newey1991uniform}. Together with $\widehat\Upsilon_n\to_P\Upsilon$, this gives uniform convergence of the GMM objective to $m(\theta)'\Upsilon m(\theta)$. Its unique minimum is $\theta_0$, so the standard extremum-estimator argument gives consistency.
\end{proof}

\subsection{Asymptotic linearity and Jacobian convergence}

\begin{proof}[Proof of \Cref{prop:asymptoticlinearity}]
Let $f_\ell=u_\ell-d_\ell+c_\ell$. By \Cref{ass:meansquareconsistency}, $P\|f_\ell\|^2=o_P(1)$. Conditional on $\mathcal W_\ell^c$, the observations in $I_\ell$ are i.i.d.\ and independent of the fitted functions. Hence
\[
\mathbb E\!\left[
\left\|\frac1{\sqrt{n_\ell}}\sum_{i\in I_\ell}
\{f_\ell(W_i)-Pf_\ell\}\right\|^2
\,\middle|\,\mathcal W_\ell^c\right]
\leq P\|f_\ell\|^2=o_P(1).
\]
Conditional Chebyshev's inequality implies that this centered sum is $o_P(1)$. Conditional probability bounds tending to zero in probability also give unconditional convergence, since probabilities are bounded by one.

Since $Pc_\ell=0$, $Pf_\ell=\mathcal R(\widehat\gamma_{-\ell})$.
Under \Cref{ass:rateandcontinuity}(i),
$\mathcal R(\widehat\gamma_{-\ell})=0$
with probability approaching one, so the conditional
bias term is identically zero on that event.
Under part~(ii), the quadratic bound and the stipulated rate imply $\sqrt n\,\|\mathcal R(\widehat\gamma_{-\ell})\|=o_P(1)$; part~(iii) assumes this directly. Since $n_\ell/n\to\pi_\ell$, in every case $\sqrt{n_\ell}\,Pf_\ell=o_P(1)$. Under \Cref{ass:rateremainder}(i), the centered sum of $\widehat\Delta_{-\ell}$ is $o_P(1)$ by the same conditional variance argument, and its mean is negligible at the $\sqrt{n_\ell}$ scale. Under part~(ii), conditional Markov's inequality yields
\[
\mathbb P\!\left(
\frac1{\sqrt{n_\ell}}\sum_{i\in I_\ell}
\|\widehat\Delta_{-\ell}(X_i)\|>\varepsilon
\,\middle|\,\mathcal W_\ell^c\right)
\leq\min\{1,\varepsilon^{-1}\sqrt{n_\ell}\,P\|\widehat\Delta_{-\ell}\|\}
\to_P0.
\]
Part~(iii) imposes the required convergence directly. Combining these facts with \Cref{eq:score-decomposition} gives
\[
\sqrt n\,\widehat\psi(\theta_0)
-\frac1{\sqrt n}\sum_{i=1}^n\psi_0(W_i)
=\sum_{\ell=1}^L\sqrt{\frac{n_\ell}{n}}\,o_P(1)=o_P(1),
\]
because $L$ is fixed.
\end{proof}

\begin{assumption}[Jacobian regularity]\label{ass:jacobian}
$P\|\partial_{\theta'}g(W,\gamma_0,\theta_0)\|<\infty$ and $J_0$ has full column rank. There is an open convex neighborhood $\mathcal N$ of $\theta_0$ such that:
(i) all fitted conditional distributions belong with probability approaching one to a deterministic class $\mathcal N_\gamma$ containing $\gamma_0$;
(ii) for every $\gamma\in\mathcal N_\gamma$, $g(W,\gamma,\theta)$ is continuously differentiable in $\theta\in\mathcal N$ almost surely, and, for some $\zeta_J>0$, $C_J<\infty$, and nonnegative $D_J$,
\[
\begin{split}
\|\partial_{\theta'}g(W,\gamma,\theta)-\partial_{\theta'}g(W,\gamma,\theta_0)\|
&\leq D_J(W,\gamma)\|\theta-\theta_0\|^{\zeta_J},\\
\sup_{\gamma\in\mathcal N_\gamma}P D_J(W,\gamma)&\leq C_J;
\end{split}
\]
(iii) for every fold,
\[
P\|\partial_{\theta'}g(W,\widehat\gamma_{-\ell},\theta_0)
-\partial_{\theta'}g(W,\gamma_0,\theta_0)\|=o_P(1).
\]
\end{assumption}

\begin{lemma}\label{lemma:Jacobian}
Under \Cref{ass:jacobian}, $\widehat J_n(\theta_n)\to_P J_0$ for every $\theta_n\to_P\theta_0$. If $\widehat\theta\to_P\theta_0$, then
\[
\overline J_n:=\int_0^1\widehat J_n\{\theta_0+t(\widehat\theta-\theta_0)\}\,dt
\to_P J_0.
\]
\end{lemma}

\begin{proof}
At $\theta_0$, conditional Markov's inequality applied to part~(iii), followed by the law of large numbers for the oracle Jacobian, gives $\widehat J_n(\theta_0)\to_P J_0$. On the event in parts~(i)--(ii),
\[
\|\widehat J_n(\theta)-\widehat J_n(\theta_0)\|
\leq D_{Jn}\|\theta-\theta_0\|^{\zeta_J},\qquad
D_{Jn}=\frac1n\sum_{\ell=1}^L\sum_{i\in I_\ell}D_J(W_i,\widehat\gamma_{-\ell})=O_P(1).
\]
The last conclusion follows by conditional Markov's inequality. This bound proves convergence at $\theta_n$ and, uniformly along the line segment joining $\theta_0$ and $\widehat\theta$, convergence of $\overline J_n$.
\end{proof}

\subsection{Normality and variance estimation}

\begin{proof}[Proof of \Cref{thm:asymptoticnormality}]
By consistency and interiority, the GMM first-order condition holds with probability approaching one:
$\widehat J_n(\widehat\theta)'\widehat\Upsilon_n\widehat\psi(\widehat\theta)=0$.
The integral mean-value identity gives
$\widehat\psi(\widehat\theta)=\widehat\psi(\theta_0)+\overline J_n(\widehat\theta-\theta_0)$.
By \Cref{lemma:Jacobian} and the weighting-matrix assumptions,
\[
\widehat J_n(\widehat\theta)'\widehat\Upsilon_n\overline J_n
\to_P J_0'\Upsilon J_0,
\]
whose limit is nonsingular. Solving the first-order condition, using \Cref{prop:asymptoticlinearity}, and applying the law of large numbers to the Jacobians gives \Cref{eq:parameter-linear-representation}. The i.i.d.\ multivariate central limit theorem applies to $\psi_0$ because $P\psi_0=0$ and $P\|\psi_0\|^2<\infty$. Slutsky's theorem yields \Cref{eq:asymptoticnormality}.
\end{proof}

\begin{proof}[Proof of \Cref{prop:variance-consistency}]
The score decomposition and \Cref{ass:meansquareconsistency,ass:varconsistency} give, for each fold,
\[
P\|\widehat\psi_\ell(W,\widetilde\theta_{-\ell})-\psi_0(W)\|^2=o_P(1).
\] Conditional Markov's inequality therefore gives
\[
R_n^2:=\frac1n\sum_{\ell=1}^L\sum_{i\in I_\ell}
\|\widehat s_{i\ell}-\psi_0(W_i)\|^2=o_P(1).
\]
Writing $B_n^2=n^{-1}\sum_i\|\psi_0(W_i)\|^2=O_P(1)$, Cauchy--Schwarz gives
\[
\left\|\widehat\Psi-\frac1n\sum_i\psi_0(W_i)\psi_0(W_i)'\right\|
\leq R_n^2+2R_n B_n=o_P(1).
\]
The oracle second-moment average converges to $\Psi$ by the law of large numbers. Thus $\widehat\Psi\to_P\Psi$; \Cref{lemma:Jacobian} and continuity of matrix inversion give $\widehat V\to_P V$. No fourth-moment assumption on the oracle score is required.
\end{proof}

\subsection{Automatic estimation and applications}

\begin{proof}[Derivation of \Cref{eq:automatic-interaction,eq:automatic-rate-bounds}]
By definition,
$e_\ell=\mathcal K(\widehat\delta_\ell-\delta_{\mathcal T_2})+\eta_\ell$.
Conditional Fubini gives
\[
\begin{split}
&\int[\mathcal K(\widehat\delta_\ell-\delta_{\mathcal T_2})](y,X)h_\ell(dy,X)\\
&\quad=\int(\widehat\delta_\ell-\delta_{\mathcal T_2})(t,X)
\left\{\int\indic\{y\leq t\}h_\ell(dy,X)\right\}\gamma_0(dt,X)\\
&\quad=\mathbb E[(\widehat\delta_\ell-\delta_{\mathcal T_2})(Y,X)h_\ell(Y,X)\mid X].
\end{split}
\]
The inner signed integral is precisely the CDF difference at $t$, including when the laws have atoms. This proves \Cref{eq:automatic-interaction}. The operator $\mathcal K$ is an $L^2(P_{Y,X})$ contraction: $|\mathcal K\delta(y,x)|\leq\mathbb E[\|\delta(Y,X)\|\mid X=x]$ and conditional Jensen's inequality applies. This gives the first bound in \Cref{eq:automatic-rate-bounds}. Ordinary Cauchy--Schwarz gives the second. For the third, conditional Jensen's inequality and $|h_\ell|\leq1$ imply
\[
\|\mathbb E[(\widehat\delta_\ell-\delta_{\mathcal T_2})h_\ell\mid X]\|_2
\leq\{P[\|\widehat\delta_\ell-\delta_{\mathcal T_2}\|^2h_\ell^2]\}^{1/2}
\leq r_{\delta,\ell}.
\]
The projection statement follows by the same Fubini identity applied to $\delta_0-\delta_{\mathcal T_2}$ and orthogonality of this difference to $\mathcal T_2$.
\end{proof}

\begin{proof}[Proof of \Cref{prop:ANcontrolfunction}]
Let $H_\ell(X,Z)=(\widehat\gamma_{-\ell}-\gamma_0)(X,Z)$ and let $\delta_0$ be the CDF-based representer at $\beta_*$. The tail representation gives
$d_\ell(Z)=\mathbb E[\delta_0(X,Z)H_\ell(X,Z)\mid Z]$.
Thus $\|d_\ell\|_2^2\leq P[\delta_0^2H_\ell^2]=o_P(1)$: for any $M>0$, the last expectation is at most $M^2P H_\ell^2+P[\delta_0^2\indic\{|\delta_0|>M\}]$, and one first lets $n\to\infty$ and then $M\to\infty$. The other mean-square conditions follow from part~(i) and conditional centering. Parts~(ii)--(iii) give the two remainder conditions, so \Cref{prop:asymptoticlinearity} applies. The unconstrained zero of the sample moment equals $\beta_*+\widehat\psi(\beta_*)$; it is consistent and belongs to $\Theta$ with probability approaching one by interiority. The Jacobian is $-1$, giving the stated variance. Parts~(iv)--(v) verify \Cref{ass:varconsistency}, so \Cref{prop:variance-consistency} applies.
\end{proof}

\begin{proof}[Proof of \Cref{prop:ANcounterfactuals}]
The original-moment difference is constant in $W$, and the true-representer centering difference equals
$d_\ell(X)=r_0(X)\int a_0(t)(\widehat\gamma_{-\ell}-\gamma_0)(t,X)\nu(dt)$.
Part~(i) therefore verifies \Cref{ass:meansquareconsistency}, using conditional centering for the representer error. Fubini yields the displayed expression for $\mathcal R$, so parts~(ii)--(iii) give \Cref{prop:asymptoticlinearity}. As the moment is affine in $\theta$ with derivative $-1$, its zero is $\theta_0+\widehat\psi(\theta_0)$, and is consistent and interior with probability approaching one. This proves normality with variance $P\psi_0^2$. Parts~(iv)--(v) give variance consistency.
\end{proof}

\begin{proof}[Proof of \Cref{prop:ANrank-rank}]
Since all CDFs take values in $[0,1]$, the difference of the products of child and parental ranks is bounded in absolute value by $|h_{c,\ell}(Y_c,X)|+|h_{p,\ell}(Y_p,X)|$. Also
$d_{s,\ell}(X)=\mathbb E[\delta_{0,s}(Y_s,X)h_{s,\ell}(Y_s,X)\mid X]$, with $0\leq\delta_{0,s}\leq1$. Part~(i) therefore implies mean-square continuity of the original moment and the true-representer correction, as well as the conditionally centered representer errors. Expanding the rank product cancels its two linear terms against the correction, leaving \Cref{eq:rankrank-remainder}. Its absolute value is at most $\|h_{c,\ell}\|_2\|h_{p,\ell}\|_2$. Parts~(ii)--(iii) thus give \Cref{prop:asymptoticlinearity}. The sample zero is $\theta_0+12\widehat\psi(\theta_0)$, which establishes consistency and the variance $144P\psi_0^2$. The preliminary-parameter contribution to variance continuity is $|\widetilde\theta_{-\ell}-\theta_0|/12$. Parts~(iv)--(v) and \Cref{prop:variance-consistency} give the final claim.
\end{proof}

\begin{proof}[Proof of \Cref{prop:ANl-functionals}]
Part~(i) verifies \Cref{ass:meansquareconsistency}, the third requirement by conditional centering. The displayed definition of $\mathcal R$ is \Cref{eq:population-remainder} for the general transformed moment, so parts~(ii)--(iii) verify \Cref{ass:rateandcontinuity,ass:rateremainder}. The remaining hypotheses permit direct application of \Cref{thm:asymptoticnormality}. Parts~(iv)--(v) verify \Cref{ass:varconsistency}, and \Cref{prop:variance-consistency} yields the variance result.
\end{proof}

\section{First-Step Convergence and the Moment Process}\label{app:empirical-process}

\subsection{Review of conditional-CDF estimators}\label{subsec:1st-steps-review}
We briefly review flexible conditional-CDF estimators and their convergence properties, which can help verify the mean-square consistency and remainder-rate conditions required by our asymptotic theory. Different results concern convergence rates, uniform consistency, pointwise limits, or process approximations, often in different topologies. Their applicability to the joint estimation-error process indexed by $(y,x)$ must be checked for the estimator at hand. These results should be distinguished from those concerning the nuisance-indexed moment process in \Cref{eq:empirical-process}, examined in the next subsection.

Classical kernel estimators of conditional distributions and conditional functionals are studied by \citet{stute1986almost} and \citet{hardle1988strong}. The usual local stochastic normalizations are $\sqrt{nh_n^d}$ for kernel estimators and $\sqrt{k_n}$ for nearest-neighbor estimators, subject to bias control. \citet{portier2025nearest} establishes weak convergence of a centered conditional-CDF process over outcome thresholds at a fixed covariate point, rather than jointly over $(y,x)$.
\citet{belloni2015some} develop pointwise normality, uniform convergence rates, and Gaussian approximations for series estimators. Their Comment~4.7 highlights that the standardized series process generally need not be asymptotically equicontinuous.

For machine-learning estimators, \citet{meinshausen2006quantile} establishes uniform consistency over thresholds at fixed covariate values using quantile regression forests, whereas \citet{wager2018estimation} derive pointwise normality for ``honest'' random forests under suitable assumptions.
For neural networks, \citet{farrell2021deep} provide mean-square bounds, including for binary-response models, and related regression-risk results are developed by \citet{schmidt2020nonparametric}.
\citet{liang2021how} studies convergence rates for distribution estimation using Generative Adversarial Networks (GANs) under adversarial losses.
At a fixed threshold $y$, estimation of the CCDF can be formulated as a regression of $\indic\{Y\le y\}$ on $X$, but applying such results uniformly over $y$ requires additional verification.

\citet{chernozhukov2014gaussian} develop Gaussian approximations for suprema of empirical processes whose complexity may increase with the sample size. While such approximations can support uniform inference without requiring weak convergence of the entire process to a fixed tight limit, they still require additional structure on the class of nuisance functions.

For averaged functionals, $\sqrt{n}$-inference can be obtained even with slow first-step convergence, as illustrated by counterfactual distributions \citep{rothe2010nonparametric} and weighted average conditional-quantile derivatives \citep{lee2022nonparametric}.

The convergence rates established for particular nuisance estimators can be sufficient for the remainder conditions of orthogonal estimation. However, these rates do not, by themselves, ensure the stochastic equicontinuity of the moment process in \Cref{eq:empirical-process}, as we demonstrate below.

\subsection{Consistency does not imply stochastic equicontinuity}\label{subsec:lack-SE}
For a full-sample nuisance estimator $\widehat\gamma$, an expansion of the plug-in moment separates a centered empirical-process term from a population term:
\[
\sqrt n\,P_n\{g(W,\widehat\gamma,\theta_0)-g(W,\gamma_0,\theta_0)\}
=\mathbb G_n(\widehat\gamma)-\mathbb G_n(\gamma_0)
 +\sqrt n\,P\{g(W,\widehat\gamma,\theta_0)-g(W,\gamma_0,\theta_0)\},
\]
where $P_n$ is the empirical average and
\begin{equation}\label{eq:empirical-process}
\mathbb G_n(\gamma)
=\frac1{\sqrt n}\sum_{i=1}^n
\{g(W_i,\gamma,\theta_0)-P g(W,\gamma,\theta_0)\}.
\end{equation}
Population expectations involving fitted functions hold the fitted function fixed. Stochastic equicontinuity and consistency can make the centered increment negligible \citep{andrews1994empirical,andrews1994asymptotics}. The population term requires a separate expansion and need not be negligible without correction \citep{newey1994asymptotic,chernozhukov2022locally}.

For a metric $d$, the relevant local equicontinuity condition is that, for every $\varepsilon,\eta>0$, there exists $\delta>0$ such that
\begin{equation}\label{eq:stoch-equic}
\limsup_{n\to\infty}P^*\left(
\sup_{d(\gamma,\gamma_0)<\delta}
\|\mathbb G_n(\gamma)-\mathbb G_n(\gamma_0)\|\geq\varepsilon
\right)\leq\eta,
\end{equation}
where $P^*$ denotes outer probability. The following construction shows that uniform CDF consistency, even at an $o(n^{-1/4})$ rate, does not ensure this condition on an unrestricted nuisance class.

As a counterexample, we now construct an estimator that is uniformly consistent at $o\left( n^{-1/4} \right)$ rate, but consider a nuisance class such that the moment process does not satisfy stochastic equicontinuity. Consider the following moment condition
$$
    g(W,\gamma,\theta) = \gamma(y_0,X) - \theta, \quad \theta_0=\frac{1}{2}, \quad X \sim U\left[0,1\right],
$$
with conditional distribution
$$
    \gamma_0(y,x) = \frac{1}{2}G_L(y) + \frac{1}{2}G_R(y)
$$
constant in $x$, where $G_L$ and $G_R$ are known continuous CDFs such that $G_L(y_0)=1$ and $G_R(y_0)=0$ for a fixed $y_0$.

For a Borel set $A \subset [0,1]$ and $0 < a \le \frac{1}{4}$, define
$$
    \gamma_{A,a}(y,x) := \left( \frac{1}{2} + a \indic\left\{ x \in A \right\} \right)G_L(y) + \left( \frac{1}{2} - a \indic\left\{ x \in A \right\} \right)G_R(y).
$$
Let $\mathcal G$ contain $\gamma_0$ and all the kernels $\gamma_{A,a}$ for Borel $A$ and $0<a\leq1/4$. Each is a measurable conditional CDF. Moreover, take $d(\gamma,\gamma_0)=\|\gamma-\gamma_0\|_\infty$, and consider the sequences
$$
    A_n := \bigcup_{i=1}^n \left( \left[ X_i-\frac{1}{2n^3}, X_i+\frac{1}{2n^3} \right]\cap[0,1] \right),
    \qquad
    a_n:=\min\left\{\frac14,n^{-1/3}\right\}.
$$

The set $A_n$ contains every observation $X_i$. Thus, $\gamma_{A_n,a_n}(y,x)$ differs from $\gamma_0(y,x)$ only for values $x \in A_n$, a region with probability at most $n^{-2}$. The sequence of weights $a_n$ is chosen so that the estimator converges uniformly at rate $o\left( n^{-1/4} \right)$, while $\sqrt{n}a_n\to\infty$.
Set the estimator to be $\widehat\gamma=\gamma_{A_n,a_n}$. Then, we obtain
$$
    \left\| \widehat\gamma - \gamma_0 \right\|_{\infty} = a_n= o\left( n^{-1/4} \right) \to 0.
$$
However, since eventually $a_n = n^{-\frac{1}{3}}$ and $\left\| \widehat\gamma-\gamma_0 \right\|_\infty < \delta$ for every fixed $\delta>0$ and all sufficiently large $n$, and evaluating the population integral with $A_n$ held fixed,
\begin{align*}
    \sup_{ \left\| \gamma-\gamma_0 \right\|_\infty < \delta} \left| \mathbb{G}_n(\gamma) - \mathbb{G}_n(\gamma_0) \right| & \ge
    \mathbb{G}_n(\widehat\gamma) - \mathbb{G}_n(\gamma_0)\allowdisplaybreaks\\[3pt]
    & = \sqrt{n} a_n \left[ \frac{1}{n}\sum_{i=1}^n \left( \indic\left\{ X_i \in A_n \right\} - \int_0^1 \indic\left\{ x \in A_n \right\}dx \right) \right] \allowdisplaybreaks\\[3pt]
    & = n^{\frac{1}{6}} \left( 1 - \int_0^1 \indic\left\{ x \in A_n \right\}dx \right) \allowdisplaybreaks\\[3pt]
    & = n^{\frac{1}{6}} \Big( 1 - \left| A_n \right| \Big) \allowdisplaybreaks\\[3pt]
    & \ge n^{\frac{1}{6}}\left( 1 - n^{-2} \right) \to \infty.
\end{align*}
Therefore, for every $\varepsilon>0$, the supremum in \Cref{eq:stoch-equic} exceeds $\varepsilon$ with probability one for all sufficiently large $n$, and asymptotic equicontinuity fails.

Without imposing additional structure on the model, such as smoothness of the CCDFs or on the complexity of the estimator class (sparsity, etc.), the moment process need not satisfy stochastic equicontinuity. In the counterexample above, it also fails to be asymptotically tight. Indeed, using $\mathbb{G}_n(\gamma_0)=0$,
$$
    \sup_{\gamma\in\mathcal{G}}|\mathbb{G}_n(\gamma)| \ge |\mathbb{G}_n(\widehat\gamma)| \ge n^{1/6}(1-n^{-2}) \to\infty.
$$
Consequently, the process cannot converge weakly to a tight random element \citep[Theorem~1.5.4]{van1996weak}. In such cases, a weak-limit argument for this unrestricted nuisance-indexed moment process is unavailable. This example does not rule out weak convergence for a particular conditional-CDF estimator under additional restrictions.

Cross-fitting controls the centered empirical-process increments by conditioning on each training sample. The FSIF correction removes the first-order population effect, leaving the nonlinear and interaction remainders analyzed in \Cref{sec:asymptotictheory}.

\section{Additional Simulation Results}\label{app:additionalresults}

\subsection{Diagnostics for Conditional Rank-Rank Regression}\label{app:diagCRRR}

Let $n_\ell=|I_\ell|$. For each generation $s\in\{c,p\}$, define
\[
\widehat\Delta_{s,-\ell}(x)
=
-\int
\bigl\{
    \widehat\alpha_{s,-\ell}(y,x)-\alpha_{0,s}(y,x)
\bigr\}
\bigl(\widehat\gamma_{s,-\ell}-\gamma_{0,s}\bigr)(dy,x),
\]
and set
\[
\widehat\Delta_{-\ell}(x)
=
\widehat\Delta_{c,-\ell}(x)
+
\widehat\Delta_{p,-\ell}(x).
\]
The conditional integrals entering these diagnostics are evaluated
numerically. The scaled interaction diagnostic is
\[
\frac{1}{L}\sum_{\ell=1}^L
\left|
    \frac{1}{\sqrt{n_\ell}}
    \sum_{i\in I_\ell}\widehat\Delta_{-\ell}(X_i)
\right|.
\]
Thus, the absolute value is taken after averaging within each evaluation fold, and the resulting quantities are averaged across folds.

The scaled asymptotic-linearity discrepancy is
\[
\left|
    \sqrt n\bigl(\widehat\theta^{LR}-\theta_0\bigr)
    -
    \frac{12}{\sqrt n}\sum_{i=1}^n\psi_0(W_i)
\right|,
\]
where $\psi_0$ is the oracle rank--rank score evaluated at $\theta_0$. Equivalently, this equals
$\sqrt n\,|\widehat\theta^{LR}-\widehat\theta^{\mathrm{or}}|$,
where
\[
\widehat\theta^{\mathrm{or}}
=
\theta_0+\frac{12}{n}\sum_{i=1}^n\psi_0(W_i).
\]

The oracle variance ratio is $\widehat V/V$, where
\[
V=144P\psi_0^2,
\qquad
\widehat V
=
\frac{144}{n}
\sum_{\ell=1}^L\sum_{i\in I_\ell}
\left\{
    \widehat\psi^{\mathrm{imp}}_{i\ell}
    (\widetilde\theta_{-\ell})
\right\}^2.
\]
Here $\widehat\psi^{\mathrm{imp}}_{i\ell}(\theta)$ denotes the implemented rank--rank score, including the numerical centering procedure, and $\widetilde\theta_{-\ell}$ is the preliminary training-sample plug-in estimator. The oracle variance $V$ is
evaluated by Gaussian quadrature.

\begin{sidewaystable}[p]
\centering
\begin{threeparttable}
\caption{First-step and asymptotic diagnostics: shared- and separate-sample nuisance estimation}
\label{tab:MLCRRR_diag}
\small
\setlength{\tabcolsep}{4pt}
\renewcommand{\arraystretch}{1.08}
\begin{tabular}{@{}rrl*{8}{r}@{}}
\toprule
$n$ & $c$ & Training
& \shortstack{$n^{1/4}$ RMSE\\$\gamma_c$}
& \shortstack{$n^{1/4}$ RMSE\\$\gamma_p$\tnote{a}}
& \shortstack{$n^{1/4}$ RMSE\\$\alpha_c$}
& \shortstack{$n^{1/4}$ RMSE\\$\alpha_p$\tnote{a}}
& \shortstack{$\sqrt n$ CDF--CDF\\product}
& \shortstack{Scaled\\$|\widehat\Delta|$}
& \shortstack{Scaled AL\\discrepancy}
& \shortstack{Oracle variance\\ratio} \\
\midrule
500 & 0.25 & Shared
& 0.9029 & 0.7741 & 0.5401 & 0.4445
& 0.6993 & 0.0883 & 1.6460 & 5.5716 \\
& & Separate
& 1.0794 & 0.9117 & 0.6201 & 0.5358
& 0.9804 & 0.0860 & 1.2781 & 5.4657 \\
\addlinespace[2pt]
500 & 0.50 & Shared
& 0.9071 & 0.7831 & 0.5538 & 0.4530
& 0.7107 & 0.0863 & 2.1705 & 7.9483 \\
& & Separate
& 1.0848 & 0.9136 & 0.6445 & 0.5523
& 0.9870 & 0.0928 & 1.2281 & 7.9743 \\
\addlinespace[2pt]
500 & 0.75 & Shared
& 0.9011 & 0.7627 & 0.5705 & 0.4686
& 0.6888 & 0.0911 & 3.2194 & 19.3203 \\
& & Separate
& 1.0810 & 0.9041 & 0.6579 & 0.5698
& 0.9741 & 0.0968 & 1.1776 & 20.0142 \\
\addlinespace[5pt]
1000 & 0.25 & Shared
& 0.8641 & 0.7417 & 0.5552 & 0.4364
& 0.6405 & 0.0561 & 1.3445 & 4.2236 \\
& & Separate
& 1.0686 & 0.9234 & 0.6427 & 0.5297
& 0.9842 & 0.0659 & 1.1413 & 4.5767 \\
\addlinespace[2pt]
1000 & 0.50 & Shared
& 0.8637 & 0.7535 & 0.5645 & 0.4499
& 0.6511 & 0.0703 & 2.2193 & 5.9961 \\
& & Separate
& 1.0686 & 0.9279 & 0.6523 & 0.5443
& 0.9892 & 0.0782 & 1.2339 & 6.6318 \\
\addlinespace[2pt]
1000 & 0.75 & Shared
& 0.8698 & 0.7502 & 0.5837 & 0.4656
& 0.6531 & 0.0866 & 3.1938 & 14.4694 \\
& & Separate
& 1.0738 & 0.9301 & 0.6757 & 0.5573
& 0.9965 & 0.0960 & 1.3218 & 16.4685 \\
\addlinespace[5pt]
5000 & 0.25 & Shared
& 0.7763 & 0.6197 & 0.6040 & 0.4415
& 0.4809 & 0.0413 & 0.7898 & 2.4422 \\
& & Separate
& 0.9713 & 0.8081 & 0.6930 & 0.5155
& 0.7841 & 0.0497 & 0.6122 & 2.7394 \\
\addlinespace[2pt]
5000 & 0.50 & Shared
& 0.7774 & 0.6223 & 0.6171 & 0.4516
& 0.4835 & 0.0494 & 1.3000 & 3.3514 \\
& & Separate
& 0.9707 & 0.8089 & 0.7037 & 0.5323
& 0.7846 & 0.0634 & 0.7322 & 3.8191 \\
\addlinespace[2pt]
5000 & 0.75 & Shared
& 0.7767 & 0.6214 & 0.6419 & 0.4734
& 0.4828 & 0.0516 & 2.2617 & 7.6701 \\
& & Separate
& 0.9644 & 0.8064 & 0.7290 & 0.5491
& 0.7772 & 0.0647 & 1.1431 & 8.9784 \\
\bottomrule
\end{tabular}

\begin{tablenotes}[flushleft]
\footnotesize
\item \textit{Notes:}
Entries are Monte Carlo means. Both procedures use XGBoost first steps with three-fold outer cross-fitting. Shared and separate training are implemented as described in the text. The nuisance-error, CDF--CDF-product, and interaction diagnostics use held-out fold means and 60 replications; the AL and variance diagnostics use all 300 replications. The CDF--CDF product provides a product-rate bound for the nonlinear remainder. The scaled interaction and AL diagnostics and the oracle variance ratio are defined in the preceding text. Individual $n^{1/4}$-scaled errors are descriptive rate benchmarks.
\item[a] For separate training, the parental scaled RMSEs are approximated
by multiplying the reported, rounded unscaled RMSEs by $n^{1/4}$.
\end{tablenotes}
\end{threeparttable}
\end{sidewaystable}

Table~\ref{tab:MLCRRR_diag} compares nuisance-error and first-order approximation diagnostics under shared and separate training. The individual $n^{1/4}$-scaled errors are larger under separate training, consistent with the smaller samples available for each nuisance estimator. Between $n=500$ and $n=5000$, the scaled child-CDF error falls from approximately $0.90$ to $0.78$ under shared training and from $1.08$ to $0.97$ under separate training. The corresponding parental-CDF errors fall from approximately $0.77$ to $0.62$ and from $0.91$ to $0.81$, respectively. The larger errors under separate training are consistent with the reduction in training-sample size, since each nuisance fit uses approximately half as many observations as under shared training. Under both procedures, the scaled child-representer error increases over these sample sizes, while the parental-representer error changes comparatively little. These finite-sample patterns do not establish individual $o(n^{-1/4})$ rates, and the measure-representer errors must be interpreted together with the interaction-remainder conditions in \Cref{sec:asymptotictheory}.

The reported CDF--CDF product diagnostic decreases between $n=500$ and $n=5000$, from approximately $0.69$--$0.71$ to $0.48$ under shared training and from $0.97$--$0.99$ to $0.78$ under separate training. The scaled interaction diagnostic also decreases between these sample sizes under both procedures, but is generally larger under separate training; for example, at $n=5000$ and $c=0.75$, it equals $0.0647$, compared with $0.0516$ under shared training. Thus, the bias reductions in Table~\ref{tab:MLCRRR_main} are not accompanied by smaller values of these diagnostics. Sample separation targets the dependence between nuisance-estimation errors entering second-order products. Reducing the covariance contribution to these products can improve bias even when individual errors and their norm-based bounds increase. The results are consistent with this mechanism, although the reported diagnostics do not isolate that covariance contribution directly.

The benefit of sample separation is particularly clear in the asymptotic-linearity diagnostic, which is smaller under separate training at every design point. At $n=5000$, the discrepancy falls from $0.7898$ to $0.6122$ for $c=0.25$, from $1.3000$ to $0.7322$ for $c=0.50$, and from $2.2617$ to $1.1431$ for $c=0.75$. The improvement is already substantial at $n=500$ when $c=0.75$, where the discrepancy falls from $3.2194$ to $1.1776$. These comparisons indicate closer agreement with the oracle first-order representation despite the larger individual nuisance errors. Within the separate-sample procedure, however, the discrepancy does not decline monotonically with $n$, particularly for $c=0.75$.

The oracle variance ratios decrease with the sample size under both procedures. At $n=5000$, the ratios under separate training are $2.7394$, $3.8191$, and $8.9784$ for $c=0.25$, $0.50$, and $0.75$, compared with $2.4422$, $3.3514$, and $7.6701$ under shared training. Thus, at these sample sizes the score-based variance estimator still exceeds the oracle variance. This is consistent with the conservative coverage under separate training and with the average estimated standard errors exceeding the empirical standard deviations in Table~\ref{tab:MLCRRR_main}. Taken together, the results show substantial gains in bias and RMSE from the locally robust correction, with additional sample separation bringing the estimator closer to its oracle first-order representation, despite the smaller training samples available for each nuisance fit. The remaining gap at $n=5000$ suggests that more accurate representer estimation would further improve finite-sample inference.
\subsection{Double Robustness of Conditional Rank-Rank Regression}\label{app:simulDRCRRR}

To illustrate the double robustness property, we run a smaller separate simulation using the baseline model of \citet{chernozhukov2025conditional}.
In this design,
$$
    X \sim N(0,1),
$$
and
$$
    \begin{pmatrix}
        Y_c\\
        Y_p
    \end{pmatrix}
    \big| X=x \sim N
    \left(
    \begin{pmatrix}
        x\\x
    \end{pmatrix},
    \begin{pmatrix}
        1 & c\\ c & 1
    \end{pmatrix}
    \right).
$$
The true parameter is again
$$
    \theta_0(c) = \frac{6}{\pi} \arcsin\left(\frac{c}{2}\right).
$$
The implementation combines local-linear kernel estimation with five-fold cross-fitting. It uses U-statistic centering. To apply the theory, fitted CDFs must define conditional probability measures; raw local-linear CDF estimates can require monotonicity and range correction. Any diagonal-removal modification of the centering must satisfy \Cref{rmk:computed-score}. We consider four different specifications for the child nuisance functions: (i) both models for $\gamma_0$ and $\alpha_0$ are correctly specified (CC); (ii) only the model for $\gamma_0$ is correctly specified (CM); (iii) only the model for $\alpha_0$ is correctly specified (MC); (iv) both models are misspecified (MM).
Misspecified child nuisances omit the dependence on $X$ and use their corresponding unconditional specifications. The parental nuisance functions $\gamma_p$ and $\alpha_p$ are estimated using their correct information sets in every configuration.

In this experiment we implement the estimator with U-statistic centering to remove the finite-sample diagonal self-interaction generated by the weighted empirical representations of the kernel nuisance estimators. This allows us to focus on the double-robustness property itself rather than on the finite-sample V-statistic centering effect.

Under \textsc{CC}, \Cref{thm:asymptoticnormality} applies if the fitted nuisances and the implemented centering satisfy its convergence and remainder conditions. Under \textsc{CM} and \textsc{MC}, the population moment remains valid by \Cref{prop:DRrankrank}; consistency additionally requires convergence of the sample moment and identification. Correct specification alone does not verify these conditions. However, the asymptotic expansion in \Cref{thm:asymptoticnormality} is formulated around the structural true nuisances $(\gamma_0,\alpha_0)$. Consequently, when one nuisance estimator converges to a misspecified probability limit, the mean-square conditions of that theorem are not satisfied as stated. We therefore interpret the \textsc{CM} and \textsc{MC} designs primarily as tests of double robustness and consistency, while their confidence-interval coverage is reported descriptively. Under \textsc{MM}, neither nuisance is correctly specified and double robustness no longer protects consistency.

This experiment uses
$$
    n \in\{625,2500,10000\}, \quad c\in\{0.25,0.50,0.75\},
$$
and 500 Monte Carlo replications for each design point.

\begin{table}[!htbp]
\centering
\begin{threeparttable}
\caption{Monte Carlo comparison of locally robust and plug-in estimators}
\label{tab:DRCRRR_main}
\scriptsize
\setlength{\tabcolsep}{2.4pt}
\begin{tabular}{@{}ccccccccccccc@{}}
\toprule
& & & \multicolumn{5}{c}{Locally robust} & \multicolumn{5}{c}{Plug-in} \\
\cmidrule(lr){4-8} \cmidrule(l){9-13}
$n$ & $c$ & $\theta_0$ & Bias & SD & Avg.\ SE & RMSE & Cov.\ 95\% & Bias & SD & Avg.\ SE & RMSE & Cov.\ 95\% \\
\midrule
\multicolumn{13}{@{}c@{}}{Panel A: \textsc{CC} ($\gamma_c$ and $\alpha_c$ correct)} \\
\midrule
625 & 0.25 & 0.2394 & -0.0033 & 0.0393 & 0.0406 & 0.0394 & 0.958 & 0.0038 & 0.0401 & 0.1179 & 0.0403 & 1.000 \\
625 & 0.50 & 0.4826 & -0.0078 & 0.0346 & 0.0345 & 0.0354 & 0.948 & 0.0061 & 0.0365 & 0.1272 & 0.0370 & 1.000 \\
625 & 0.75 & 0.7341 & -0.0136 & 0.0239 & 0.0232 & 0.0275 & 0.926 & 0.0115 & 0.0270 & 0.1363 & 0.0293 & 1.000 \\
\addlinespace[2pt]
2500 & 0.25 & 0.2394 & -0.0009 & 0.0190 & 0.0195 & 0.0191 & 0.960 & 0.0016 & 0.0192 & 0.0583 & 0.0193 & 1.000 \\
2500 & 0.50 & 0.4826 & -0.0031 & 0.0164 & 0.0164 & 0.0167 & 0.956 & 0.0008 & 0.0169 & 0.0630 & 0.0169 & 1.000 \\
2500 & 0.75 & 0.7341 & -0.0053 & 0.0104 & 0.0106 & 0.0117 & 0.930 & 0.0026 & 0.0113 & 0.0674 & 0.0116 & 1.000 \\
\addlinespace[2pt]
10000 & 0.25 & 0.2394 & 0.0002 & 0.0092 & 0.0096 & 0.0092 & 0.950 & 0.0007 & 0.0093 & 0.0290 & 0.0093 & 1.000 \\
10000 & 0.50 & 0.4826 & -0.0007 & 0.0081 & 0.0080 & 0.0081 & 0.948 & 0.0003 & 0.0081 & 0.0314 & 0.0081 & 1.000 \\
10000 & 0.75 & 0.7341 & -0.0019 & 0.0050 & 0.0051 & 0.0053 & 0.950 & -0.0000 & 0.0051 & 0.0336 & 0.0051 & 1.000 \\
\addlinespace
\midrule
\multicolumn{13}{@{}c@{}}{Panel B: \textsc{CM} ($\gamma_c$ correct and $\alpha_c$ misspecified)} \\
\midrule
625 & 0.25 & 0.2394 & -0.0004 & 0.0391 & 0.0489 & 0.0390 & 0.982 & 0.0038 & 0.0401 & 0.1179 & 0.0403 & 1.000 \\
625 & 0.50 & 0.4826 & -0.0011 & 0.0348 & 0.0453 & 0.0348 & 0.996 & 0.0061 & 0.0365 & 0.1272 & 0.0370 & 1.000 \\
625 & 0.75 & 0.7341 & -0.0019 & 0.0244 & 0.0394 & 0.0244 & 0.998 & 0.0115 & 0.0270 & 0.1363 & 0.0293 & 1.000 \\
\addlinespace[2pt]
2500 & 0.25 & 0.2394 & 0.0003 & 0.0191 & 0.0236 & 0.0191 & 0.990 & 0.0016 & 0.0192 & 0.0583 & 0.0193 & 1.000 \\
2500 & 0.50 & 0.4826 & -0.0009 & 0.0165 & 0.0218 & 0.0165 & 0.994 & 0.0008 & 0.0169 & 0.0630 & 0.0169 & 1.000 \\
2500 & 0.75 & 0.7341 & -0.0011 & 0.0105 & 0.0189 & 0.0105 & 1.000 & 0.0026 & 0.0113 & 0.0674 & 0.0116 & 1.000 \\
\addlinespace[2pt]
10000 & 0.25 & 0.2394 & 0.0006 & 0.0092 & 0.0117 & 0.0092 & 0.972 & 0.0007 & 0.0093 & 0.0290 & 0.0093 & 1.000 \\
10000 & 0.50 & 0.4826 & -0.0000 & 0.0081 & 0.0108 & 0.0081 & 0.984 & 0.0003 & 0.0081 & 0.0314 & 0.0081 & 1.000 \\
10000 & 0.75 & 0.7341 & -0.0007 & 0.0050 & 0.0093 & 0.0050 & 0.998 & -0.0000 & 0.0051 & 0.0336 & 0.0051 & 1.000 \\
\addlinespace
\midrule
\multicolumn{13}{@{}c@{}}{Panel C: \textsc{MC} ($\gamma_c$ misspecified and $\alpha_c$ correct)} \\
\midrule
625 & 0.25 & 0.2394 & -0.0045 & 0.0386 & 0.0491 & 0.0388 & 0.984 & -0.0649 & 0.0290 & 0.1145 & 0.0711 & 1.000 \\
625 & 0.50 & 0.4826 & -0.0106 & 0.0334 & 0.0460 & 0.0350 & 0.992 & -0.1383 & 0.0270 & 0.1213 & 0.1410 & 1.000 \\
625 & 0.75 & 0.7341 & -0.0195 & 0.0226 & 0.0413 & 0.0298 & 0.990 & -0.2173 & 0.0246 & 0.1277 & 0.2187 & 0.898 \\
\addlinespace[2pt]
2500 & 0.25 & 0.2394 & -0.0015 & 0.0187 & 0.0238 & 0.0187 & 0.988 & -0.0690 & 0.0136 & 0.0568 & 0.0704 & 1.000 \\
2500 & 0.50 & 0.4826 & -0.0046 & 0.0163 & 0.0222 & 0.0170 & 0.990 & -0.1434 & 0.0134 & 0.0602 & 0.1440 & 0.036 \\
2500 & 0.75 & 0.7341 & -0.0079 & 0.0103 & 0.0200 & 0.0130 & 0.998 & -0.2213 & 0.0116 & 0.0634 & 0.2217 & 0.000 \\
\addlinespace[2pt]
10000 & 0.25 & 0.2394 & -0.0001 & 0.0092 & 0.0117 & 0.0091 & 0.984 & -0.0697 & 0.0067 & 0.0283 & 0.0700 & 0.028 \\
10000 & 0.50 & 0.4826 & -0.0014 & 0.0082 & 0.0109 & 0.0083 & 0.988 & -0.1432 & 0.0064 & 0.0300 & 0.1434 & 0.000 \\
10000 & 0.75 & 0.7341 & -0.0030 & 0.0050 & 0.0098 & 0.0058 & 1.000 & -0.2221 & 0.0061 & 0.0316 & 0.2222 & 0.000 \\
\addlinespace
\midrule
\multicolumn{13}{@{}c@{}}{Panel D: \textsc{MM} (both child nuisances misspecified)} \\
\midrule
625 & 0.25 & 0.2394 & -0.0696 & 0.0284 & 0.0402 & 0.0752 & 0.594 & -0.0649 & 0.0290 & 0.1145 & 0.0711 & 1.000 \\
625 & 0.50 & 0.4826 & -0.1463 & 0.0265 & 0.0367 & 0.1487 & 0.000 & -0.1383 & 0.0270 & 0.1213 & 0.1410 & 1.000 \\
625 & 0.75 & 0.7341 & -0.2316 & 0.0239 & 0.0307 & 0.2328 & 0.000 & -0.2173 & 0.0246 & 0.1277 & 0.2187 & 0.898 \\
\addlinespace[2pt]
2500 & 0.25 & 0.2394 & -0.0705 & 0.0136 & 0.0197 & 0.0718 & 0.004 & -0.0690 & 0.0136 & 0.0568 & 0.0704 & 1.000 \\
2500 & 0.50 & 0.4826 & -0.1456 & 0.0132 & 0.0180 & 0.1462 & 0.000 & -0.1434 & 0.0134 & 0.0602 & 0.1440 & 0.036 \\
2500 & 0.75 & 0.7341 & -0.2257 & 0.0112 & 0.0151 & 0.2259 & 0.000 & -0.2213 & 0.0116 & 0.0634 & 0.2217 & 0.000 \\
\addlinespace[2pt]
10000 & 0.25 & 0.2394 & -0.0700 & 0.0066 & 0.0098 & 0.0703 & 0.000 & -0.0697 & 0.0067 & 0.0283 & 0.0700 & 0.028 \\
10000 & 0.50 & 0.4826 & -0.1438 & 0.0064 & 0.0089 & 0.1440 & 0.000 & -0.1432 & 0.0064 & 0.0300 & 0.1434 & 0.000 \\
10000 & 0.75 & 0.7341 & -0.2232 & 0.0060 & 0.0074 & 0.2233 & 0.000 & -0.2221 & 0.0061 & 0.0316 & 0.2222 & 0.000 \\
\bottomrule
\end{tabular}
\begin{tablenotes}[flushleft]
\footnotesize
\item \textit{Notes:} Results are based on 500 Monte Carlo replications. First steps are estimated by local-linear kernel methods with 5-fold cross-fitting, leave-one-out training pseudo-ranks, and U-statistic centering. Each panel corresponds to a different child-nuisance specification, while the parental nuisances $\gamma_p$ and $\alpha_p$ are correctly specified throughout. SD denotes the Monte Carlo standard deviation, Avg.\ SE the average estimated standard error, RMSE the root mean squared error, and Cov.\ 95\% the empirical coverage of nominal 95\% confidence intervals. Plug-in standard errors omit first-step estimation uncertainty and are reported only for comparison.
\end{tablenotes}
\end{threeparttable}
\end{table}

Table~\ref{tab:DRCRRR_main} illustrates the double-robustness property. Under \textsc{CC}, the locally robust estimator is approximately unbiased and its coverage is close to the nominal level. Under \textsc{CM}, where $\gamma_c$ is correctly specified but $\alpha_c$ is misspecified, the locally robust estimator remains approximately unbiased. The same occurs under \textsc{MC}, where $\alpha_c$ is correctly specified but $\gamma_c$ is misspecified. The latter case provides the clearest comparison with the plug-in estimator: misspecification of $\gamma_c$ produces a large and persistent plug-in bias, while the locally robust estimator remains close to the true parameter because the correction nuisance is correctly specified. Under \textsc{MM}, both child nuisance functions are misspecified and the bias of the locally robust estimator persists as the sample size increases, as expected when neither alternative of the double-robustness condition holds.

Coverage under \textsc{CM} and \textsc{MC} is reported only descriptively, since the asymptotic-normality result developed above is formulated around the structural true nuisances $(\gamma_0,\alpha_0)$ and does not cover cases in which one nuisance estimator converges to a misspecified probability limit. Asymptotic normality may still hold under additional conditions and an expansion around the corresponding probability limits, but this is outside the scope of the present analysis (see \citet{dukes2024doubly,vanderlaan2026doubly} for doubly robust inference under nuisance misspecification). The plug-in coverage results are likewise only descriptive because their standard errors omit first-step estimation uncertainty.

\begin{table}[!htbp]
\centering
\begin{threeparttable}
\caption{First-step and correction-term diagnostics}
\label{tab:DRCRRR_diag}
\scriptsize
\setlength{\tabcolsep}{2.1pt}
\begin{tabular}{@{}ccrrrrrrrrrrr@{}}
\toprule
$n$ & $c$ & \shortstack{RMSE\\$\gamma_c$} & \shortstack{RMSE\\$\gamma_p$} & \shortstack{RMSE\\$\alpha_c$} & \shortstack{RMSE\\$\alpha_p$} & \shortstack{$n^{1/4}$ RMSE\\$\gamma_c$} & \shortstack{$n^{1/4}$ RMSE\\$\gamma_p$} & \shortstack{$n^{1/4}$ RMSE\\$\alpha_c$} & \shortstack{$n^{1/4}$ RMSE\\$\alpha_p$} & \shortstack{$\sqrt{n}$ child\\product} & \shortstack{$\sqrt{n}$ parent\\product} & \shortstack{$\sqrt{n}$ CDF--CDF\\product} \\
\midrule
\multicolumn{13}{@{}c@{}}{Panel A: \textsc{CC} ($\gamma_c$ and $\alpha_c$ correct)} \\
\midrule
625 & 0.25 & 0.0595 & 0.0591 & 0.0355 & 0.0354 & 0.2975 & 0.2953 & 0.1773 & 0.1768 & 0.0531 & 0.0526 & 0.0880 \\
625 & 0.50 & 0.0592 & 0.0595 & 0.0357 & 0.0358 & 0.2959 & 0.2976 & 0.1784 & 0.1790 & 0.0532 & 0.0537 & 0.0882 \\
625 & 0.75 & 0.0593 & 0.0593 & 0.0357 & 0.0357 & 0.2964 & 0.2967 & 0.1785 & 0.1787 & 0.0533 & 0.0535 & 0.0884 \\
\addlinespace[2pt]
2500 & 0.25 & 0.0363 & 0.0366 & 0.0218 & 0.0219 & 0.2566 & 0.2591 & 0.1539 & 0.1551 & 0.0397 & 0.0404 & 0.0665 \\
2500 & 0.50 & 0.0362 & 0.0361 & 0.0218 & 0.0218 & 0.2558 & 0.2555 & 0.1542 & 0.1539 & 0.0396 & 0.0395 & 0.0654 \\
2500 & 0.75 & 0.0364 & 0.0365 & 0.0218 & 0.0219 & 0.2572 & 0.2581 & 0.1538 & 0.1546 & 0.0398 & 0.0401 & 0.0667 \\
\addlinespace[2pt]
10000 & 0.25 & 0.0218 & 0.0219 & 0.0131 & 0.0131 & 0.2185 & 0.2187 & 0.1306 & 0.1309 & 0.0286 & 0.0287 & 0.0478 \\
10000 & 0.50 & 0.0219 & 0.0218 & 0.0131 & 0.0131 & 0.2193 & 0.2181 & 0.1314 & 0.1310 & 0.0289 & 0.0287 & 0.0479 \\
10000 & 0.75 & 0.0218 & 0.0218 & 0.0130 & 0.0131 & 0.2183 & 0.2180 & 0.1305 & 0.1306 & 0.0286 & 0.0286 & 0.0477 \\
\addlinespace
\midrule
\multicolumn{13}{@{}c@{}}{Panel B: \textsc{CM} ($\gamma_c$ correct and $\alpha_c$ misspecified)} \\
\midrule
625 & 0.25 & 0.0595 & 0.0591 & 0.1145 & 0.0354 & 0.2975 & 0.2953 & 0.5727 & 0.1768 & 0.1704 & 0.0526 & 0.0880 \\
625 & 0.50 & 0.0592 & 0.0595 & 0.1156 & 0.0358 & 0.2959 & 0.2976 & 0.5778 & 0.1790 & 0.1710 & 0.0537 & 0.0882 \\
625 & 0.75 & 0.0593 & 0.0593 & 0.1178 & 0.0357 & 0.2964 & 0.2967 & 0.5892 & 0.1787 & 0.1747 & 0.0535 & 0.0884 \\
\addlinespace[2pt]
2500 & 0.25 & 0.0363 & 0.0366 & 0.1141 & 0.0219 & 0.2566 & 0.2591 & 0.8070 & 0.1551 & 0.2071 & 0.0404 & 0.0665 \\
2500 & 0.50 & 0.0362 & 0.0361 & 0.1152 & 0.0218 & 0.2558 & 0.2555 & 0.8149 & 0.1539 & 0.2085 & 0.0395 & 0.0654 \\
2500 & 0.75 & 0.0364 & 0.0365 & 0.1176 & 0.0219 & 0.2572 & 0.2581 & 0.8315 & 0.1546 & 0.2138 & 0.0401 & 0.0667 \\
\addlinespace[2pt]
10000 & 0.25 & 0.0218 & 0.0219 & 0.1141 & 0.0131 & 0.2185 & 0.2187 & 1.1405 & 0.1309 & 0.2492 & 0.0287 & 0.0478 \\
10000 & 0.50 & 0.0219 & 0.0218 & 0.1153 & 0.0131 & 0.2193 & 0.2181 & 1.1527 & 0.1310 & 0.2528 & 0.0287 & 0.0479 \\
10000 & 0.75 & 0.0218 & 0.0218 & 0.1176 & 0.0131 & 0.2183 & 0.2180 & 1.1757 & 0.1306 & 0.2567 & 0.0286 & 0.0477 \\
\addlinespace
\midrule
\multicolumn{13}{@{}c@{}}{Panel C: \textsc{MC} ($\gamma_c$ misspecified and $\alpha_c$ correct)} \\
\midrule
625 & 0.25 & 0.2281 & 0.0591 & 0.0355 & 0.0354 & 1.1407 & 0.2953 & 0.1773 & 0.1768 & 0.2023 & 0.0526 & 0.3369 \\
625 & 0.50 & 0.2279 & 0.0595 & 0.0357 & 0.0358 & 1.1397 & 0.2976 & 0.1784 & 0.1790 & 0.2033 & 0.0537 & 0.3392 \\
625 & 0.75 & 0.2277 & 0.0593 & 0.0357 & 0.0357 & 1.1384 & 0.2967 & 0.1785 & 0.1787 & 0.2032 & 0.0535 & 0.3378 \\
\addlinespace[2pt]
2500 & 0.25 & 0.2275 & 0.0366 & 0.0218 & 0.0219 & 1.6084 & 0.2591 & 0.1539 & 0.1551 & 0.2475 & 0.0404 & 0.4168 \\
2500 & 0.50 & 0.2273 & 0.0361 & 0.0218 & 0.0218 & 1.6071 & 0.2555 & 0.1542 & 0.1539 & 0.2477 & 0.0395 & 0.4105 \\
2500 & 0.75 & 0.2273 & 0.0365 & 0.0218 & 0.0219 & 1.6075 & 0.2581 & 0.1538 & 0.1546 & 0.2473 & 0.0401 & 0.4150 \\
\addlinespace[2pt]
10000 & 0.25 & 0.2273 & 0.0219 & 0.0131 & 0.0131 & 2.2734 & 0.2187 & 0.1306 & 0.1309 & 0.2970 & 0.0287 & 0.4972 \\
10000 & 0.50 & 0.2273 & 0.0218 & 0.0131 & 0.0131 & 2.2729 & 0.2181 & 0.1314 & 0.1310 & 0.2988 & 0.0287 & 0.4957 \\
10000 & 0.75 & 0.2273 & 0.0218 & 0.0130 & 0.0131 & 2.2727 & 0.2180 & 0.1305 & 0.1306 & 0.2965 & 0.0286 & 0.4954 \\
\addlinespace
\midrule
\multicolumn{13}{@{}c@{}}{Panel D: \textsc{MM} (both child nuisances misspecified)} \\
\midrule
625 & 0.25 & 0.2281 & 0.0591 & 0.1145 & 0.0354 & 1.1407 & 0.2953 & 0.5727 & 0.1768 & 0.6539 & 0.0526 & 0.3369 \\
625 & 0.50 & 0.2279 & 0.0595 & 0.1156 & 0.0358 & 1.1397 & 0.2976 & 0.5778 & 0.1790 & 0.6592 & 0.0537 & 0.3392 \\
625 & 0.75 & 0.2277 & 0.0593 & 0.1178 & 0.0357 & 1.1384 & 0.2967 & 0.5892 & 0.1787 & 0.6714 & 0.0535 & 0.3378 \\
\addlinespace[2pt]
2500 & 0.25 & 0.2275 & 0.0366 & 0.1141 & 0.0219 & 1.6084 & 0.2591 & 0.8070 & 0.1551 & 1.2984 & 0.0404 & 0.4168 \\
2500 & 0.50 & 0.2273 & 0.0361 & 0.1152 & 0.0218 & 1.6071 & 0.2555 & 0.8149 & 0.1539 & 1.3100 & 0.0395 & 0.4105 \\
2500 & 0.75 & 0.2273 & 0.0365 & 0.1176 & 0.0219 & 1.6075 & 0.2581 & 0.8315 & 0.1546 & 1.3370 & 0.0401 & 0.4150 \\
\addlinespace[2pt]
10000 & 0.25 & 0.2273 & 0.0219 & 0.1141 & 0.0131 & 2.2734 & 0.2187 & 1.1405 & 0.1309 & 2.5931 & 0.0287 & 0.4972 \\
10000 & 0.50 & 0.2273 & 0.0218 & 0.1153 & 0.0131 & 2.2729 & 0.2181 & 1.1527 & 0.1310 & 2.6202 & 0.0287 & 0.4957 \\
10000 & 0.75 & 0.2273 & 0.0218 & 0.1176 & 0.0131 & 2.2727 & 0.2180 & 1.1757 & 0.1306 & 2.6723 & 0.0286 & 0.4954 \\
\bottomrule
\end{tabular}
\begin{tablenotes}[flushleft]
\footnotesize
\item \textit{Notes:} Scenario definitions are as in Table~\ref{tab:DRCRRR_main}. Parent nuisances $\gamma_p$ and $\alpha_p$ are correctly specified in every reported configuration. First steps are estimated by local-linear kernel methods with 5-fold cross-fitting, leave-one-out training pseudo-ranks, the normal-reference $n^{-1/5}$ bandwidth rule, bandwidth scale equal to one, and a minimum of 30 neighbors. The $n^{1/4}$-scaled RMSE columns compare individual nuisance errors with an $n^{-1/4}$ benchmark. The last three columns report the child $\sqrt n\,\operatorname{RMSE}(\widehat\gamma_c) \operatorname{RMSE}(\widehat\alpha_c)$ product, the parental $\sqrt n\,\operatorname{RMSE}(\widehat\gamma_p) \operatorname{RMSE}(\widehat\alpha_p)$ product, and the CDF--CDF $\sqrt n\,\operatorname{RMSE}(\widehat\gamma_c) \operatorname{RMSE}(\widehat\gamma_p)$ product. The CDF--CDF product corresponds to the product-rate bound on the nonlinear remainder. The child and parental CDF--representer products are descriptive diagnostics and do not, by themselves, establish the interaction-remainder condition in \Cref{ass:rateremainder}.
\end{tablenotes}
\end{threeparttable}
\end{table}
\bigskip
Table~\ref{tab:DRCRRR_diag} reports the corresponding first-step diagnostics. Under \textsc{CC}, the nuisance errors and all the reported $\sqrt n$ product-rate diagnostics decrease with the sample size, consistently with the asymptotic conditions used for the correctly specified experiment. Under \textsc{CM}, the error of the misspecified $\alpha_c$ does not vanish, while under \textsc{MC} the same occurs for the misspecified $\gamma_c$. Consequently, the associated $n^{1/4}$ and product-rate diagnostics need not converge to zero. The population double-robustness identity continues to hold in these two configurations, provided the other generation is correctly specified. The observed bias patterns are consistent with this identity, but finite simulations do not prove consistency. Under \textsc{MM}, neither child nuisance converges to its true value and the corresponding diagnostics reflect the failure of the conditions required for double robustness. The parental nuisances remain correctly specified in all configurations and display decreasing estimation errors and product-rate diagnostics.

\end{document}